\documentclass[10pt,preprint]{aastex}

\newcommand{\NGC}[1]{\objectname[NGC]{NGC~{#1}}}
\newcommand{\SNJ}{\objectname[supernova]{SN~1993J}}

\slugcomment{Accepted for publication in \apj}

\shorttitle{LLAGN Radio Size}
\shortauthors{Anderson and Ulvestad}

\begin{document}


\title{The Size of the Radio-Emitting Region in Low-luminosity Active
  Galactic Nuclei}


\author{James M. Anderson\altaffilmark{1,2}}
\email{janderso@nrao.edu}
\and
\author{James S. Ulvestad\altaffilmark{1}}
\email{julvesta@nrao.edu}


\altaffiltext{1}{National Radio Astronomy Observatory, P.O. Box O,
  1003 Lopezville Road, Socorro, NM 87801.}
\altaffiltext{2}{Department of Physics, New Mexico Institute of Mining
  and Technology, Socorro, NM 87801.}
\altaffiltext{3}{Current address: Joint Institute for VLBI in Europe,
  Postbus 2, 7990 AA Dwingeloo, The Netherlands, anderson{@}jive.nl}


\begin{abstract}
  
  We have used the VLA to study radio variability among a sample
  of 18 low luminosity active galactic nuclei (LLAGNs), on time
  scales of a few hours to 10 days.  The goal was to measure or limit
  the sizes of the LLAGN radio-emitting regions, in order to use
  the size measurements as input to models of the radio emission
  mechanisms in LLAGNs.  We detect variability on typical time
  scales of a few days, at a confidence level of 99\%, in half of
  the target galaxies.  Either variability that is intrinsic to the
  radio emitting regions, or that is caused by scintillation in the
  Galactic interstellar medium, is consistent with the data.  For
  either interpretation, the brightness temperature of the emission is
  below the inverse-Compton limit for all of our LLAGNs, and has a mean
  value of about $10^{10}$~K.  The variability measurements plus
  VLBI upper limits imply that the typical angular size of the
  LLAGN radio cores at $8.5$~GHz is $0.2$~milliarcseconds, plus or
  minus a factor of two.  The $\sim 10^{10}$~K brightness temperature
  strongly suggests that a population of high-energy nonthermal
  electrons must be present, in addition to a hypothesized thermal
  population in an accretion flow, in order to produce the observed
  radio emission.
  

\end{abstract}



\keywords{ galaxies: active --- galaxies: jets --- galaxies:
  photometry --- radiation mechanisms: general --- radio continuum:
  galaxies }


\defcitealias{Fassnacht_T2001}{FT01}
\defcitealias{Barvainis_ea2004}{B04}
\defcitealias{Ho_ea1999}{H99}
\defcitealias{Anderson_UH2004}{A04}

\section{Introduction\label{sec:intro}}

Strong extragalactic radio sources generally are thought to be powered
by accretion onto massive black holes, resulting in the production of
powerful radio jets as well as self-absorbed radio cores.  The cores
and jets often undergo relativistic motion, resulting in a variety of
observed phenomena such as apparent superluminal motion and rapid
radio variability \citep[see][for reviews of these
topics]{Zensus_1997,Wagner_W1995,Ulrich_MU1997}.  Although the details
of the accretion and jet formation are not well understood, there
is still a consensus about the general physical processes that
dominate the cores of strong radio sources.

In recent years, it has become apparent that active galactic nuclei
(AGNs) are not restricted to massive black holes accreting at the
Eddington rate in galaxies whose luminosity is dominated by the AGN.
Indeed, it now appears that all galaxies with significant stellar
bulges harbor central black holes \citep{Kormendy_G2001}.  Careful
subtraction of template galaxy spectra reveals AGN-related emission
lines in roughly half of bright nearby galaxies
\citep{Ho_FS1997.0,Ho_FS1997.1}, and HST imaging reveals point-like
AGN cores in a number of these objects
\citep{Maoz_ea1996,Barth_ea1998,Ravindranath_ea2001}.  A question of
current interest is the mechanism by which energy and radiation are
produced in the centers of these low-luminosity AGNs (LLAGNs).
Although the LLAGNs are intrinsically quite weak at radio wavelengths,
their radio to optical ratio $R$ \citep[e.g.][]{Kellermann_ea1989}
often is found to be in the range of $10^2$--$10^6$
\citep{Ho_P2001,Ho_2002}, implying that the LLAGNs actually are
``radio-loud'' when their radio emission is considered as a fraction
of the overall AGN luminosity \citep[see also][]{Terashima_W2003}.  A
key discriminant among models which attempt to explain the origin of
the radio emission is the scale size of that emission.  If LLAGNs are
dominated by emission from low radiative efficiency accretion
\citep[e.g.][]{Narayan_Y1994,Mahadevan_1997,Narayan_MQ1998}, their
radio sources should be only tens of Schwarzschild radii in size,
whereas sources dominated by compact jets
\citep[e.g.][]{Falcke_B1999,Yuan_MF2002,Yuan_ea2002} should be
considerably larger.

The highest resolution imaging technique in astronomy is Very Long
Baseline Interferometry (VLBI), which can reveal the structures of
compact radio sources on milliarcsecond scales.  However, in LLAGNs,
recent imaging using the Very Long Baseline Array (VLBA\footnote{The
  VLBA and the Very Large Array (VLA) are operated by the National
  Radio Astronomy Observatory, a facility of the National Science
  Foundation operated under cooperative agreement by Associated
  Universities, Inc.})  shows that the galaxies often are dominated by
compact sources unresolved on scales near one milliarcsecond
\citep{Falcke_ea2000,Ulvestad_H2001.1,Nagar_ea2002,Anderson_UH2004}.
Even the nominally ``large'' jet models for LLAGNs, in fact, often
predict radio sizes smaller than a milliarcsecond
\citep[e.g.][]{Falcke_B1999}, so we are faced with the dilemma of
trying to measure radio sizes smaller than those that can be imaged by
normal interferometric techniques.

A possible solution to this dilemma is the investigation of intra-day
variable sources, or IDVs.  In the 1980s, Heeschen and collaborators
\citep{Heeschen_1984,Simonetti_CH1985,Heeschen_ea1987} discovered
radio flux ``flicker,'' whereby compact extragalactic radio sources
were seen to vary by a few percent on time scales of hours to days.
Reviews by \citet{Quirrenbach_1992} and \citet{Wagner_W1995}
summarized the state of studies of IDVs ten years ago; it was unclear
whether the IDV phenomenon was caused by intrinsic source variability
or by apparent variability caused by interstellar scintillation along
the line of sight to very compact sources.  More recently,
correlations of variability amplitude and time scale with Earth motion
relative to the Galactic interstellar medium have provided conclusive
evidence that at least some IDVs are caused by scintillation
\citep{Rickett_ea2001,Dennett-Thorpe_d2002}; an astonishing success of
scintillation models for short time scale radio variability is the
direct measurement of the expansion of gamma-ray bursts by virtue of
the short time scale variability impressed by the interstellar medium
\citep{Frail_ea1997}.  Since scintillation can occur only for radio
sources having size scales of tens of microarcseconds or less,
searches for scintillation provide a unique tool for investigating
radio emission on scales too small to image by conventional
interferometry.

In this paper, we report an exploration of radio variability of LLAGNs
on time scales ranging from a few hours to more than a week.  The
purpose of our observations is to determine the distribution of size
scales for the radio emission from a sample of LLAGNs either via
scintillation or intrinsic variability, and to use the results as a
discriminator among the various models for this emission.

\section{Source Selection\label{sec:source_selection}}

We have selected an LLAGN sample from the Palomar Seyfert Sample of
\citet{Ho_FS1997.0,Ho_FS1997.1}, which found emission lines
characteristic of LLAGNs in nearly half of the galaxies.  The Seyfert
galaxies in that sample were systematically observed in the radio by
\citet{Ho_U2001}, who detected nearly all of them using VLA snapshots
\citep[see also][]{Ulvestad_H2001.0}.  Because the Palomar Seyfert
galaxies were selected based only on the optical properties of the
galaxy nuclei, and because nearly all of the Seyfert galaxies were
detected in the radio, the orientation angles of any small-scale jets
in that sample are probably randomly oriented.  Therefore, we expect
that only $\sim 1$\% of the galaxies will have a jet pointed within
10\degr\ of us.  Furthermore, the components observed in Seyfert jets
are not generally relativistic \citep[see the discussion and
references in][]{Ulvestad_2003}, and thus the radio emission should
not be significantly Doppler boosted in any of our galaxies.  From the
Seyfert sample of \citet{Ho_U2001}, we selected the flat-spectrum
(defined as $\alpha > -0.35$, for $S_\nu \propto \nu^{+\alpha}$)
galaxies with peak 5~GHz flux densities of at least 2~mJy.  This flux
density limit was necessary to ensure that all objects would be
detectable by the VLBA, which provides upper limits to the source
sizes and constrains the interstellar scintillation calculations.

A model of the Galactic interstellar medium (ISM)
\citep[NE2001,][]{Cordes_L2002} and a simple model of advection
dominated accretion flow (ADAF) emission region size
\citep{Mahadevan_1997} were used to estimate minimum scintillation
timescales for our galaxies.  We excluded objects with timescales
longer than 7 days --- too long to be measured in our planned
variability program.

A large number of the Seyfert galaxies are gathered near the Virgo
cluster, so we further restricted our sample to contain those galaxies
within a few hours of 12$^\mathrm{h}$~RA, or those galaxies with high
declinations which could be observed with the VLA when the Virgo group
of galaxies was visible.  This permitted us to observe all of our
target sample galaxies within a reasonably small range of times at
the VLA.

Because we expected only a relatively small fraction of our galaxies
to show modest amplitude scintillation
\citep{Quirrenbach_ea1992,Kedziora_Chudczer_ea2001}, we needed
approximately 20 target galaxies in our sample in order to conclude at
a 99\% confidence level that the emission regions were too large to
scintillate if we found no variability.  Therefore, we added
additional target galaxies to our sample by including some LINER
galaxies from \citet{Falcke_ea2000}, two Seyfert galaxies with $\alpha
= -0.5$, and one Seyfert galaxy with $\alpha = -0.9$ which had right
ascensions that filled gaps in our RA coverage, in order to best
utilize the VLA time allocation.  The properties of our resulting
somewhat heterogeneous 18 sample galaxies are summarized in
Table~\ref{tab:sample}.

\section{VLBA Observations\label{sec:VLBA_obs}}

The probability that a radio source will show intraday variability due
to interstellar refraction is much higher for sources which are
compact or point-like on milliarcsecond scales
\citep[e.g.][]{Quirrenbach_ea1992}, so VLBI imaging of our galaxies is
necessary to correctly assess the number and flux densities of galaxies
in our sample that \emph{could} vary due to refractive scintillation.
Refractive scintillation variations are undetectable for sources
larger than about 100~$\mu$as at our observing frequency of $8.4$~GHz.


We used the VLBA to observe our ten sample galaxies with no previous
VLBI imaging at 8.4~GHz.  Details of the observations are presented in
Table~\ref{tab:VLBA_obs}.  Various on-source integration times were
used to achieve peak to RMS noise levels of at least 20 based on the
predicted peak flux density for each galaxy.  Observations were spread
over at least 4 hours of time to improve $(u,v)$ coverage and image
fidelity.  We applied an amplitude calibration using a priori gain
values together with system temperatures measured during the
observations; typically, this calibration is accurate to within 5\%.
Initial clock and atmospheric (phase) errors were derived from the
calibrator sources listed in Table~\ref{tab:VLBA_obs} using
phase-referencing \citep{Beasley_C1995}.  A data recording speed of
256~$\mathrm{Mbit~s^{-1}}$ was used to reduce switching cycle times
to further improve phase calibration.

This initial calibration was used to determine the galaxy core
positions shown in Table~\ref{tab:sample}.  Uncertainties in the
positions generally are dominated by the uncertainties in the phase
calibrator positions, but contributions from ionospheric and
tropospheric phase fluctuations and residual phase errors can be
important for some objects.  Another bright, nearby check source was
observed along with each target galaxy to test the effectiveness of
the phase calibration.  Excluding the check sources for
\NGC{777}\footnote{The check source \objectname[VLBA]{J0203+3041} was
  found to have a double-lobed structure with a separation of $\sim
  40$~mas.  The position listed in \citet{Beasley_ea2002} is
  approximately centered between the two lobes.  We believe that the
  observed source structure in the check source is real.} and
\NGC{2787}\footnote{The uncertainty in the position of the check
  source \objectname[JVAS]{J0853+6722} in the calibration archive was
  12~mas in RA and Dec, so we do not expect agreement at the
  milliarcsecond level.}, all check sources were measured
to be within 1.0~mas of their catalog positions, with a mean total
difference of 0.61~mas.  We therefore estimate that the
uncertainties in the measured positions of our target galaxy cores are
0.5~mas each in right ascension and in declination.

Eight of the ten galaxy cores were detected in this initial imaging
process.  For these galaxies, phase-only self-calibration was then
iteratively applied.  The resulting RMS noise levels in the images far
away from the galaxy cores are consistent with predictions based on
total integration time and vary from 30--40~$\mathrm{\mu
  Jy~beam^{-1}}$.  Beam widths are approximately 2~mas by 1~mas using
natural weighting.  Images of the detected galaxies are shown in 
Figure~\ref{fig:vlba_images}.

Similar processing steps were performed on the substantially brighter
check sources.  Peak flux densities were measured on images made with
the same self-calibration parameters as used on the target galaxies
(or with no self-calibration applied for \NGC{777} and \NGC{3227}).
Then, full self-calibration corrections were calculated for the
stronger check sources, and the peak flux densities measured again.
The ratio of these measurements indicates the amount of decoherence
remaining in the target galaxy measurements.  As shown in
Table~\ref{tab:VLBA_att}, this value is typically only a few percent
for the detected galaxies.  Target peak flux densities were estimated
from Gaussian fitting and corrected for decoherence.  Integrated flux
densities were calculated using Gaussian fitting for nearly unresolved
targets and hand-drawn regions for more complex sources.  For galaxies
which are well fit by a single Gaussian component, the 1-$\sigma$
estimates of the minimum and maximum size of the major axis of the
Gaussian component are shown in Table~\ref{tab:VLBA_att}.  The size
estimates are probably accurate to no better than 0.1~mas; sizes
less than 0.5~mas are highly unreliable and are consistent with the
sources being unresolved.
Estimated uncertainties from self-calibration and measurement errors
have been added in quadrature to the overall uncertainties in the
amplitude scale in order to derive the final flux-density
uncertainties.

Our results agree well with the 5~GHz observations of
\citet{Falcke_ea2000} for the galaxies in common with their study.
Although we have much higher signal to noise levels, we still find
unresolved emission where they found unresolved emission, and for
galaxies which were partially resolved, our position angles agree to
within 10\degr.  Further details for individual galaxies are given in
Appendix~\ref{sec:sources}.

\section{VLA Observations\label{sec:VLA_obs}}

Interstellar scattering properties change over long timescales because
the Earth's orbital velocity vector changes direction (and presumably
because different turbulent regions of the Galactic ISM move across
the line of sight) \citep{Rickett_ea2001,Dennett-Thorpe_d2002}.  In
order to improve the likelihood of detecting interstellar
scintillation, we observed our target sample during 2003 May and again
during 2003 September in order to allow the Earth's motion to change
substantially.  Observations were carried out using the VLA as shown
in Table~\ref{tab:VLA_obs}.  The time necessary to observe each of the
18 target galaxies once, plus supporting calibration observations, was
about 5~hours.  Each VLA run consists of $\sim$10~hour blocks on three
sequential days followed by $\sim$5~hour blocks on four nonsequential
days.  This observing strategy was designed to provide variability
information for changes over a few hours to changes over 24~hours to
changes over $\sim 10$ days.

Either intrinsic or scattering-induced variability on timescales of
days is expected to arise from emission within a few light-days of the
central black holes of our target galaxies.  This region is
unresolvable by the VLA.  Our target galaxies were selected to have
their 8.5~GHz radio emission on scales of a few arcseconds or less
dominated by an unresolved (or nearly unresolved) core.  Since the
Fourier transform of a point source has a constant amplitude in the
$(u,v)$-plane, changes in the observed $(u,v)$ coverage because of
hour angle changes or physical relocations of antennas should not
affect the measurement of the brightness of a point source.  This is
an important consideration because our snapshot observations of
individual objects were made at different hour angles and took place
during reconfiguration periods at the VLA, when antennas were moved
to different physical locations.  Although fewer than three antennas
were typically moved between any two neighboring observing blocks in
our program, the majority of the antennas had their physical locations
moved at some point during the observing runs in May and in September.

Our target galaxies are relatively nearby, as indicated by the
distances provided in Table~\ref{tab:sample}.  Therefore, some
emission from the host galaxy is expected to be present at spatial
scales from around 10~arcseconds to many arcminutes.  This galactic
emission typically has a steep spectral index \citep[$S_\nu \propto
\nu^{-0.75}$,][]{Condon1992}; we observed at 8.5~GHz to reduce the
contribution of the galactic background emission and decrease the
spatial scales to which the VLA was sensitive.  Our observations were
made with approximately A and B array configurations of the VLA, which
have the longest baselines \citep[see][]{Thompson_ea1980} and are
therefore least sensitive to extended emission.  By restricting the
VLA observations to only long baselines, the instrument is principally
sensitive to the unresolved core components in our target sample,
allowing us to study variability in the target cores despite changing
$(u,v)$ coverage.

Because the scientific results of this study depend critically upon
the data calibration and error analysis, extensive details of our VLA
observations and data reduction methods are presented in
Appendix~\ref{sec:VLA_obs_details}.  In brief, we made short snapshot
observations of our target galaxies using fast switching with a nearby
phase calibrator.  Compact symmetric objects \citep[CSOs,
see][]{Fassnacht_T2001} were used to provide accurate amplitude
calibration.  Standard software tools were used to process and
self-calibrate the data.  Data weighting in the $(u,v)$ plane was used
to isolate the core emission, and standard routines were used to
measure flux densities and uncertainties for target galaxies and phase
calibrators alike.

We used the resulting data to construct fully-calibrated flux-density
time series for each target galaxy and each calibration source.
Figure~\ref{fig:tar_time_0} presents this information in a graphical
form for \NGC{777}, and Figures \ref{fig:tar_time_1}
to~\ref{fig:tar_time_3} in the on-line paper present the series for
all target galaxies.  For convenience, the phase-calibrator time
series are shown immediately below their corresponding target galaxy
plots to facilitate comparison of brightness changes.

\section{Target Variability Statistics\label{sec:Target_statistics}}

In this section, we examine the statistical properties of the
short-term variations in the source flux densities.  We show that we
have detected significant variability in our sample, and compare the
rates of variability with other surveys.  And finally, we compute
structure functions from our target galaxy time series.  In
\S~\ref{sec:var_interp} we will discuss the physical interpretations
of the variability results.

\subsection{General Statistics\label{sec:gross_stat}}

We analyzed all of our objects, both target galaxies and calibrators,
to determine whether or not variability was present during the
observations.  Table~\ref{tab:var_stats} provides the results of our
simple statistical tests for variability on our sources.  Column~(2)
of Table~\ref{tab:var_stats} indicates the classification of the
structure of each object based on our VLA imaging.  The letter ``P''
indicates that the object is a target galaxy which is effectively a
point source to the VLA.  (All of the calibrator sources were
effectively point-like for the $(u,v)$ ranges used during their
calibration and measurement.)  ``D'' indicates that the emission is
dominated by a point source which contains at least 80\% of the flux
density measured with all $(u,v)$ spacings included.  ``E'' indicates
that the object is extended and has significant amounts of flux
density located outside the core component.  ``J'' indicates the
presence of a jet-like feature coming from the core.  ``C'' indicates
that the source was used as a phase calibrator, and ``S'' indicates
that the object is a CSO.

The mean flux density $<S_\nu>$ is calculated from an unweighted
average over all flux densities for each month.  The estimated
measurement error, $\sigma_\mathrm{e}$, indicates the \emph{expected}
scatter for a constant source, combining the random measurement noise
and the calibration uncertainties described in
Appendix~\ref{sec:VLA_obs_details}.  The observed RMS,
$\sigma_\mathrm{s}$, shows the actual scatter about the mean flux
level.  We also compute a de-biased RMS, $\sigma_\mathrm{d}$,
calculated as $\sqrt{\sigma_\mathrm{s}^2-\sigma_\mathrm{e}^2}$.  This
quantity provides an estimate of the true scatter in the data.  These
last three values have been divided by the mean flux density for each
month to scale each object to a common fractional variation scale.  In
addition, we calculate the probability that we would have observed a
scatter at least as large as $\sigma_\mathrm{s}$ assuming that the
estimated measurement errors are correct and normally distributed.
This probability is given by the $\chi^2$ distribution with $N-1$
degrees of freedom.  A probability value close to zero in
Table~\ref{tab:var_stats} is a good indication that an object
\emph{may} be variable, since we believe $\sigma_\mathrm{e}$
incorporates all instrumental and atmospheric error sources except
$(u,v)$ effects (see Appendix~\ref{sec:VLA_obs_details}).

\subsection{Is the Variability Real?\label{sec:is_var_real}}

We have carefully investigated the data for each object to assess the
reliability of the observations.  As indicated by
Table~\ref{tab:var_stats}, 7 of our 18 target galaxies have extended
emission in full-$(u,v)$ coverage images made from our data.  Of these
objects, \NGC{2273}, \NGC{2639}, \NGC{3227}, and \NGC{4472} have
brightness changes which appear to be correlated with changes
in $(u,v)$ coverage, either with changes in antennas locations or with
hour angle.  Because steep-spectrum AGNs typically contain emission
from extended jets, it is unsurprising that all three galaxies with
$\alpha \leq -0.5$ are included in this group.  These galaxies all
appear to have jet-like features at small spatial scales which
probably accounts for the observed changes with $(u,v)$ coverage, and
we will ignore these four target galaxies in our statistical analysis
of short-term variability.

We have labeled 8 of the target-month datasets as tentative for
measuring variability based on a close inspection of the data; these
are \NGC{3169}, \NGC{4450}, and \NGC{4579} in May and September, plus
\NGC{4203} in May and \NGC{5866} in September.  These data may have
problems with phase calibration or $(u,v)$ effects, but they could
also be perfectly valid.  For \NGC{5866} in September, the variability
classification depends on a single data-point.  Details are given in
Appendix~\ref{sec:sources}.  We classify the remaining galaxy datasets
as ``reliable'' for measuring variability.

One possible source of apparent variability could be rapid phase
fluctuations caused by the atmosphere which decrease the coherence by
differing amounts from observation to observation as the weather
changes.  Since the calibrators are strong enough to be
self-calibrated with very short solution intervals, this effect should
only impact the target galaxies, and should affect low elevation
sources the most, since the path lengths through the atmosphere are
the greatest.  Figure~\ref{fig:target_scatter_me} shows the observed
RMS level as a function of mean elevation angle for each month of
observing for our sources.  There is no significant increase in
scatter for target galaxies, whether extended or point-like (``P'' and
``D'' galaxies), at low elevation angles.  \NGC{777} has the two
highest scatter values, but no other point-like targets have
significantly high scatter at low elevation angles.

Similarly, large changes in elevation angle from one observation to
the next could be related to an increase in scatter caused by phase
decoherence differences at different elevations, subtle gain problems
depending on elevation, or even changes in the $(u,v)$ coverage of an
object.  Figure~\ref{fig:target_scatter_se} plots the RMS scatter as a
function of the RMS scatter in the elevation angle at which each
galaxy is observed.  Again, there is no significant trend visible in
the data.

Finally, it is expected that the scatter in the measured flux
densities should be highest for the weakest sources.  This could
include a systematic coherence loss for those galaxies weaker than
about 10~mJy where the signal to noise was potentially too low to use
a solution interval short enough to track the phase variations of the
atmosphere.  Figure~\ref{fig:target_scatter_fl} plots the source
flux-density scatter as a function of the mean flux density.  The
upper level of the scatter is uniformly about 4\% for sources stronger
than about 8~mJy.  The four extended galaxies with $(u,v)$ related
changes in flux density are all less than 8~mJy, creating the apparent
spike for these objects.  For point-like galaxies, only \NGC{777} and
\NGC{4565} have RMS scatter levels above 4\%.  Unfortunately, they are
also the two weakest sources, so we cannot clearly determine from this
plot whether the high scatter is due to poor self-calibration or
whether these objects are just more variable.

We adopt a $P_{\chi^2}$ upper limit cutoff of 0.01 to select sources
which are variable at our sensitivity level.  This cutoff corresponds
to a 99\% or greater confidence level in the variability.
Disregarding \objectname[COINS]{J0410$+$7656}, which is resolved and
has known problems with varying $(u,v)$ coverage, none of the CSO
sources has a $P_{\chi^2}$ value less than 0.01.  Combining the May
and September data, $10/28$ ($36\pm 9$\%) of the non-CSO phase
calibrators have $P_{\chi^2} < 0.01$.  Table~\ref{tab:var_fraction}
shows the variability fraction for various combinations of galaxy
classifications.  Although small number statistics prevent any
definite conclusions being drawn about the fraction of extended and
jet objects or core dominated objects, it seems that slightly more
than half of our target galaxies show significant variability.
Combining all of the reliable galaxies, $50\pm 11$\% are variable.
This fraction remains almost the same when the tentative galaxies are
included, rising slightly to $57\pm 9$\%.

The large fraction of target galaxies with $P_{\chi^2} \leq 0.01$ is
not simply caused by an underestimation of the true measurement errors
in the weak target galaxies.  Our estimation of the true measurement
errors appears to be valid, since several target galaxies actually have
$\sigma_\mathrm{e} < \sigma_\mathrm{s}$, including three cases where
the mean flux density was less than 10~mJy, and one case where some of
observed scatter was caused by $(u,v)$ changes.  Furthermore, the
galaxies which \emph{do} show variability often show coherent
variations over several-day time-periods (for example, \NGC{3031}
\NGC{3147} in September), something which would not be
expected if the measurement error was simply underestimated.

Our results are in reasonably good agreement with other surveys of
much stronger flat-spectrum ($\alpha > -0.5$) sources.
\citet{Heeschen_ea1987} found that $63\pm 8$\% of their flat spectrum
sources were variable at a $99.9$\% confidence level ($P_{\chi^2} <
0.001$) in their 1985 August and December observing sessions.  As they
point out, the amplitude of variability in their sample is very
low---the maximum modulation index (often written as $m$ or $\mu$,
with $m = \sigma/<S_\nu>$) they observed was only 3.4\% in their
flat-spectrum sample of 15 objects.  We find that only 35\% of our
galaxies are variable at that confidence level, but their measurement
error of about $0.27$\% is about five times lower than ours.
\citet{Quirrenbach_ea1992} found similar results for objects in their
sample, noting that sources with ``compact'' or ``very compact'' VLBI
structures show substantially larger amplitude variations than sources
extended on VLBI scales.  In their survey of 118 compact,
flat-spectrum sources, \citet{Kedziora_Chudczer_ea2001} found that
19\% of their sample showed variability above a 3-$\sigma$ level
(roughly corresponding to $P_{\chi^2} < 0.0005$).  The vast majority
of their sources have modulation indices at 8.6~GHz less than 2\%.  We
find a slightly higher fraction of galaxies at this confidence level,
even though our measurement errors are about a factor of two larger.
Our ``reliable'' galaxies are compact on milliarcsecond scales, whereas
the compact source sample of \citet{Kedziora_Chudczer_ea2001} was
taken from \citet{Duncan_ea1993}, which had a resolution of only $\sim
100$~mas, significantly larger than the VLBI scales used by
\citet{Quirrenbach_ea1992} to classify sources as ``compact'' or
``extended''.  Thus, our galaxies may show more variability because
they are angularly smaller, as would be expected either for
scintillation or intrinsic variability.  

Since the flat spectrum objects appear to have a large percentage of
objects varying at small modulation indices, this probably explains
most of the differences in the fraction of variable sources detected
by the different groups.  Given these constraints, our results fit
comfortably among these previous results, suggesting that it is
quite likely that the variability we have observed is actually real.
Other surveys looked at sources with flux densities $\gtrsim 1$~Jy,
while our target galaxies are about two orders of magnitude fainter,
so the results are not definitely conclusive as our weak target
galaxies may have different properties.

In a slightly different comparison, \citet{Lovell_ea2003} used the VLA
to search for intraday variability in 710 compact flat-spectrum
sources.  Their first epoch of observations found that 12\% of their
sources show RMS variations above 4\%.  For our reliable target
galaxies, 15\% of the target-months show scatter levels above 4\%, and
10\% of the target-months show debiased scatter levels above 4\%.
This is in excellent agreement with the \citet{Lovell_ea2003} results
for substantially brighter sources.  Again, this result highly
suggests that the variability in our target galaxies is real, but
cannot be conclusive.

\subsection{Structure Functions\label{sec:struc_func}}

As an alternative to examining our data as a set of flux density
changes as a function of time, we can transform our data into
amplitude changes as a function of time \emph{separations} (or lags)
by using the so-called ``structure functions'', a commonly used method
to investigate the time behavior of variable radio sources.  Since our
data were not sampled at regular time intervals, we create a pseudo
structure function as follows.  We start with the first order
structure function defined in the Appendix of
\citet{Simonetti_CH1985}, which we modify to become
\begin{equation}
  D^{(1)}(\tau) = \frac{1}{W^{(1)}(\tau)}
  \sum_{i=1}^{N}\sum_{j=i+1}^{N} w(i,j,\tau)\left[f(i)-f(j)\right]^2,
\end{equation}
where $f(i)$ is the fractional amplitude of the $i^\mathrm{th}$
observation, given by $f(i) = S_\nu(i)/<S_\nu>$, and
\begin{equation}
  W^{(1)}(\tau) = \sum_{i=1}^{N}\sum_{j=i+1}^{N} w(i,j,\tau).
\end{equation}
Since our observations were separated in a roughly logarithmic spacing
scheme, we calculate our structure functions at logarithmic intervals,
and use a weighting function given by
\begin{equation}
  w(i,j,\tau) = \left\{ \begin{array}{ll} 1 & \mbox{if\ }
        \log(\tau)-\zeta \leq 
        \log\left[\mbox{JD}(j) - \mbox{JD}(i)\right] <
        \log(\tau)+\zeta \\
        0 & \mbox{otherwise}
        \end{array} \right. .
\end{equation}
Here JD$(i)$ is the Julian Date of observation $i$, and $\tau$ is the
time lag in days.  The parameter $\zeta$ is an arbitrary number which
must be at least half of the logarithmic spacing in $\tau$ in order to
ensure that all data pairings are incorporated into the structure
function.  We set $\zeta$ equal to the spacing in $\log(\tau)$,
$\log(\tau_n) = n\zeta + \mbox{constant}$, $\zeta = 0.125$~dex, in
order to incorporate more data-points into each interval and partially
smooth the structure function.

The uncertainty in the structure function is given by
\begin{equation}
  \delta D^{(1)}(\tau) = \frac{2}{W^{(1)}(\tau)} \left\{
  \sum_{i=1}^{N}\sum_{j=i+1}^{N} w^2(i,j,\tau)\left[f(i)-f(j)\right]^2
  \left[ \delta f^2(i) + \delta f^2(j) \right] \right\}^{1/2},
\end{equation}
where $\delta f(i)$ is the measurement uncertainty in observation $i$.
For observations with nonzero measurement errors, the computed
structure function will be biased above the true level, as
$\left[f(i)-f(j)\right]^2$ does not average to zero.  This bias is
given by
\begin{equation}
  D_\mathrm{bias}^{(1)}(\tau) = \frac{1}{W^{(1)}(\tau)}
  \sum_{i=1}^{N}\sum_{j=i+1}^{N} w(i,j,\tau)
  \left[\delta f^2(i) + \delta f^2(j)\right].
\end{equation}
If the measurement errors are the same for all data-points, then these
equations reduce to the ones shown in \citet{Simonetti_CH1985}.

Figure~\ref{fig:target_struc_func} shows the structure functions for
our 18 target galaxies.  The structure function values are shown as
individual points with error-bars, but have not been corrected for
measurement bias.  The estimated bias levels are shown by the lines
going across each plot.  One can immediately see that the structure
functions are quite complicated.  This is partially a result of the
relatively low signal to noise of our faint target galaxies compared
with other intraday variability observations.

Galaxies such as \NGC{2273} and \NGC{3227} which have variability
caused by $(u,v)$ effects typically have structure function values an
order of magnitude above the bias level, while galaxies such as
\NGC{3079} and \NGC{4168} which have been classified as constant have
structure function values consistent within the error-bars with the
bias levels.  The majority of target galaxies have structure function
values at or below the estimated bias level for short time-lags, up to
a day or a few days.  This suggests that our flux-density uncertainty
levels are \emph{not} underestimated, and in fact they may be slightly
\emph{overestimated}.  We are therefore confident that the
$P_{\chi^2}$ probabilities and de-biased scatter levels are
trustworthy.

\section{Physical Interpretations of the Variability\label{sec:var_interp}}

Having established that variability \emph{is} present in many of our
target galaxies, we now investigate implications this variability has
for the physical properties of the radio emission regions.  We will
treat possible intrinsic and extrinsic variations in turn.

\subsection{Intrinsic Implications: Brightness Temperature\label{sec:T_b}}

One potentially important piece of information about the radio
emission is the brightness temperature of the source.  Some LLAGN
models such as the simple ADAF model suggest that the observed radio
emission is produced by synchrotron emission from thermal electrons.
In this case, the bulk of the observed emission comes from the region
where the optical depth reaches about unity, and the observed
brightness temperature should not be too different from the thermal
temperature of the electrons, which is expected to be a few times
$10^9$~K \citep[see, e.g.][]{Mahadevan_1997}.  However, if a
significant population of nonthermal electrons is present, as expected
for more complex ADAF and jet models, the brightness temperature could
easily exceed this value.  Results from VLBA imaging of our target
sample have found lower limits to the brightness temperature of
$10^8$--$ 10^9$~K (see Table~\ref{tab:VLBA_att} of this paper, Table~4
of \citealp{Anderson_UH2004}, Table~1 of \citealp{Falcke_ea2000}).
These VLBI measurement limits are unfortunately unable to discriminate
between current models.

Assuming that the variability we observed with the VLA is real and
intrinsic to the source, then the variability can improve our
understanding of the physical conditions in the source region smaller
than the VLBA observations can resolve.  The brightness temperature is given by
\begin{equation}
  T_\mathrm{b} = \frac{S_\nu \lambda^2}{2k\Omega} 
               = \frac{S_\nu c^2}{2k\Omega\nu^2},
\end{equation}
where $\Omega$ is the solid angle of the emission region.  Suppose
that a source changes in flux density\footnote{Both increases and
  decreases in emission can be used to calculate a brightness
  temperature.  A decrease in flux density simply indicates that an
  emission region which had a specified brightness temperature has
  been eliminated or absorbed.} by an amount $\left|\Delta
  S_\nu\right|$ in a time $\Delta t$.  The speed of light provides an
upper limit to the size of the variable radio source, with a maximum
solid angle for a distant, unbeamed, variable region of
\begin{equation}\label{eqn:omega}
  \Omega \leq \frac{\pi\left(c\Delta t\right)^2}{D^2},
\end{equation}
where $D$ is the distance to the source.  (This ignores relativistic
beaming effects --- probably a good assumption for most LLAGNs since
most Seyfert galaxy jets generally appear to be relatively slow; see
\citealp{Ulvestad_2003}.) The lower limit to the brightness
temperature of the variable component of emission is then
\begin{equation}
  T_\mathrm{b} \geq \frac{D^2}{2\pi k \nu^2} 
  \left(\frac{\left|\Delta S_\nu\right|}{{\Delta t}^2}\right),
\end{equation}
where the information in parentheses must be determined from the
observed time series.  As an example, during the first day of
observations in 2003 May, \NGC{777} went from 0.52~mJy to 0.87~mJy in
0.079~days (see Figure~\ref{fig:tar_time_0}), corresponding to
$T_\mathrm{b} \geq 5\times 10^{13}$~K, far in excess of the
inverse-Compton limit \citep[see, e.g.][]{Kellermann_P1968}.  Many
other target galaxies also have similar implied instances of brightness
temperatures above $10^{12}$~K.

However, because our target galaxies are relatively weak, the relative
uncertainty in $\left|\Delta S_\nu\right|$ is very significant.
Formal error analysis indicates that for the \NGC{777} data mentioned
above, the $\left|\Delta S_\nu\right|$ value is a 2.6-$\sigma$
difference.  If we assume that \NGC{777} was actually constant during
our observations, random \emph{error} in the flux density difference
would produce a brightness temperature measurement of at least
$10^{12}$~K 48\% of the time.  Two additional flux density differences
in \NGC{777} also suggest brightness temperatures above $10^{12}$~K,
but similar error analysis suggests that $10^{12}$~K temperatures
would be found 44\% and 41\% of the time for a constant source.  At
first glance, it seems unlikely that all three measurements would be
above $10^{12}$~K --- the combined probability that none of the three
are indeed above $10^{12}$~K is only 17\%.  Although this seemingly
suggests that it is likely that \NGC{777} has a variable emission
region with $T_\mathrm{b} \geq 10^{12}$~K, such an analysis is flawed,
because it does not take into account the statistics of all possible
data pairings.

There are 191 possible combinations of measurement points which can be
used to calculate brightness temperatures for \NGC{777} (although
there are actually 378 pairings of \NGC{777} data-points, differences
between May and September data-points are unlikely to yield
interesting brightness temperature limits).  We therefore expect
measurement noise to cause two 2.6-$\sigma$ or larger differences to
appear in the \NGC{777} dataset if \NGC{777} was \emph{actually
  constant} in flux density.  Thus, the modest signal to noise levels
of our data prevent us from simply relying on calculating the
brightness temperature limits for only those difference-pairs above
some threshold (such as 5-$\sigma$) to give an accurate estimate of
the brightness temperature of variations in the target galaxies.

In order to assess the effect of the multitude of statistical
opportunities to artificially generate high brightness-temperature
point differences, we created two slightly different Monte Carlo
simulations of our LLAGN datasets.  In the first simulation, a set of
random measurement errors is created from a normally-distributed
random number generator according to the actual measurement
uncertainties in the real dataset.  Then, using an \emph{assumed}
underlying brightness temperature and the observation dates from the
original dataset, a simulated flux density difference including the
random measurement errors is calculated for each pair of measurements.
This effectively assumes that \emph{every} pair of measurement points
has the equivalent of a flare starting at one point, and growing in
intensity to the second point.  (Increases or decreases in flux
density are mathematically equivalent in the $\left|\Delta
  S_\nu\right|$ model, so only rises are implemented to maintain
consistency.)  Next, the simulated difference pairs are numerically
analyzed, and the number of difference pairs which have a simulated
measurement difference above a minimum brightness temperature
threshold and a measurement flux density difference above a specified
sigma level are counted.  The results are recorded, and the same
process is repeated for a total of $10\,000$ trials using different
random measurement errors.

This simulation process is then repeated again for a slightly higher
\emph{assumed} brightness temperature, and so on, until a wide range
of assumed brightness temperatures has been covered.  Once these Monte
Carlo simulations are finished, the results are compared with number
counts from the \emph{real} dataset using the \emph{same} selection
criteria as for the simulated datasets.  For low \emph{assumed}
brightness temperatures, the simulated datasets have low simulated
brightness temperature measurements and most of the flux density
differences have small sigma levels, so that far fewer pairs are
counted above the threshold compared to the real dataset.  For high
\emph{assumed} brightness temperatures, the situation is reversed as
the random measurement errors are small compared to flux changes
required by the high \emph{assumed} brightness temperature and many
pairs are counted above the threshold.  We find the \emph{assumed}
brightness temperature at which the mean count number (above
threshold) from the Monte Carlo simulations matches the count number
from the real dataset, giving us the best estimate of the brightness
temperature implied by variations in the real dataset.  We also
determine the brightness temperature for which 10\% of the Monte Carlo
simulations have a count number at least as high as the real dataset,
giving us a 90\% confidence estimate that the actual brightness
temperature of the fluctuations in the real dataset is at least this
high.

Because this algorithm assumes that \emph{every} pair of data-points
contains a ``flare'' at the assumed brightness temperature, it will
tend to predict more data pairs which exceed our counting threshold
than would be seen in a real target having some quiescent periods.
Looking at the time series of the target galaxies for which we have
good confidence that the variability is real, the measured flux
density does not continuously rise or fall, but instead tends to both
rise \emph{and} fall over a characteristic timescale from a few hours
to a few days.  This suggests that intrinsic processes which lead to
variability in our target galaxies may be causing new sources of
emission to appear and fade away over the course of our observations.
Furthermore, it is possible that ``flares'' do not start at one of our
data-points, but begin somewhere between our observation times.  Thus,
not all pairs of data-points \emph{should} show variations of the
\emph{assumed} brightness temperature, as there should be periods when
the flux density is neither rising nor falling, but is roughly
constant, with other periods of somewhat rising and somewhat falling
values.

To make a first order correction for this effect, we have performed a
second Monte Carlo simulation.  We assume that each data pair contains
a random amount of emission up to the \emph{assumed} brightness
temperature.  The generating function for the physical brightness
temperature distribution for data pairs is probably a complicated
function of the prior history of the variations and the time
difference between the data points.  But for our first order
correction, we simply use a uniform random number generator to create
a variable emission temperature between zero and the \emph{assumed}
brightness temperature.  Otherwise, this second Monte Carlo simulation
follows the same prescription as the first simulation.

Table~\ref{tab:var_cat} shows the results of these simulations.  The
first simulation results are shown in Columns (6) and~(7), giving the
50\% (best) and 90\% (minimum) confidence estimates of $T_\mathrm{b}$,
respectively.  Columns (8) and~(9) give the results for the random
brightness temperature simulation.  Attempting to deal with all
biases, Columns (8) and~(9) are our best estimates of the variability
brightness temperatures.  They are always higher than those in Columns
(6) and~(7) because a higher brightness temperature is needed to
``compensate'' for the hypothesized quiescent periods.  Blank values
indicate that the simulation result for the brightness temperature was
less than the $10^5$~K minimum simulation temperature.  These results
are from simulations with a $2.5$-$\sigma$ flux-density difference
threshold and a temperature threshold of the \emph{assumed} brightness
temperature.  Simulations with other thresholds ($1.5$-$\sigma$ to
$3.5$-$\sigma$ and various temperature schemes) show essentially the
same results, having brightness temperatures generally within a factor
of 2 (0.3~dex) of the values in Table~\ref{tab:var_cat}.
(Calculations using different error distributions give essentially the
same results.  See Appendix~\ref{sec:Error_Distributions}.)  For
comparison, Columns (4) and~(5) show the results from a direct
analysis of the observed flux density variations in the target
datasets, showing the first highest and fourth highest apparent
brightness temperatures from all dataset pairings.

For the target galaxies that we are confident show real variability
(the ``var'' galaxies), the brightness temperatures from the Monte
Carlo simulations are typically somewhat larger than the lower limits
determined from VLBI observations, but they are still potentially
consistent with most accretion and emission models for LLAGNs.  The
best estimates of the brightness temperatures are between about
$10^{9}$ and $10^{11}$~K, with the 90\% confidence limits about half a
dex lower.  Although $10^{10}$~K is higher than the general
expectation from simple thermal electron ADAF models, physical
electron temperatures this high are possible when the accretion rate
approaches about $10^{-2}$~Eddington \citep[see, e.g., Figure~2
of][]{Mahadevan_1997}.

The brightness temperatures calculated here are also interesting in
terms of the inverse-Compton limit, which constrains the brightness
temperature of an incoherent-synchrotron emission source to a maximum
value of about $10^{12}$~K \citep{Kellermann_P1968}.  If the magnetic
field energy density is approximately in equipartition with the
particle energy density, the brightness temperature limit is about an
order of magnitude \emph{less}, as suggested by \citet{Readhead_1994},
\citet{Begelman_RS1994}, and \citet{Sincell_K1994} (these papers
suggest limits of about $6\times 10^{10}$~K, $10^{11}$~K, and $9\times
10^{10}$~K, respectively, for our LLAGNs).  Only \NGC{5866} (whose
variability we classify as tentative) approaches the $\sim 10^{12}$~K
inverse-Compton limit.  The remaining galaxies are all below the
equipartition limit of about $10^{11}$~K, and for the 90\% confidence
limits to the brightness temperature, \emph{none} of the variable
galaxies exceed the equipartition limit.

We would like to reiterate that these brightness temperatures are
probably lower limits to the actual brightness temperatures of the
emission regions responsible for the variability.
Equation~\ref{eqn:omega} assumes that the variable region of emission
expands spherically at a velocity of $c$.  Since we have no way to
measure the actual size of the emission region if the variability is
intrinsic to the target galaxy, the emission region could well be
smaller than this maximum limit, the actual brightness temperatures
thereby being substantially higher than those listed in
Table~\ref{tab:var_cat}.  Furthermore, if there are multiple,
independent regions of variable emission, the sum of all of the
emission from the target galaxy core would tend to have smaller
fluctuations as the individual variations would tend to cancel one
another out.  Finally, we would like to emphasize that the brightness
temperatures determined in this section are for the regions of
\emph{variable} emission, and do not give any information about
regions of constant emission.  This constant emission could have a
brightness temperature significantly different than the variable
emission, depending on the physical processes leading to the constant
and variable emission.

\subsection{Extrinsic Variability: Interstellar Scintillation\label{sec:ISS}}

An alternative variability explanation is refractive interstellar
scintillation of the radio emission, caused by density perturbations
in the Galactic interstellar medium, which distort the wavefront of
radio waves passing through the medium.  As the interstellar medium
appears to move across the line of sight toward a source (because of
the relative motion between the Sun and the ISM and the orbital motion
of the Earth), the observer sees alternating regions of increased or
decreased apparent brightness of the source.  Reviews of this process
can be found in \citet{Rickett_1990}, \citet{Narayan_1992}, and
references therein.  The variability in some extragalactic sources has
now conclusively been attributed to refractive scintillation
\citep[see, e.g.][]
{Rickett_ea2001,Dennett-Thorpe_d2001,Dennett-Thorpe_d2002,Jauncey_ea2003.0}
If significant amounts of the core flux come from a region smaller
than several tens of microarcseconds across on the sky, our target
galaxies have a good possibility of showing refractive scintillation
in our measurements.  Using properties of the scatter behavior in the
observational data and making some assumptions about the Galactic
interstellar medium in the direction of target galaxies gives
estimates of the solid angles subtended by the sources on the sky for
extrinsic variability.


In our relatively simple analysis, we follow the discussions in
\citet[see also the errata notice in
\citealt{Walker_2001}]{Walker_1998} and \citet{Rickett_2002}.  For
weak scattering caused by a single, thin region of the ISM, the
refractive medium will cause waves propagating toward the observer to
constructively and destructively interfere on a characteristic angular
size scale.  This scale is related to the size of the first Fresnel
zone,\footnote{The first Fresnel zone is the surface bounded by a
  circle on a plane perpendicular to the direction of the source, on
  which the geometric path from the source to the observer is
  \onehalf\ radian longer than the direct path.  In weak (strong)
  scattering, density perturbations cause additional phase changes of
  $<$\onehalf\ ($>$\onehalf) radian across this zone.} which is given
by
\begin{equation}\label{eqn:iss_Fresnel_angle}
  \theta_\mathrm{F} = 1/\sqrt{Lk},
\end{equation} where $k$ is the wavenumber of the radio waves and
$L$ is the distance to the ISM screen.  Converting to frequency in
units of gigahertz and distance in kiloparsecs, the angular size in
microarcseconds is $\theta_\mathrm{F} \approx 8.11/\sqrt{L\nu}$.  As
the ISM screen moves across the line of sight toward the source with a
transverse velocity relative to the observer of $v$, the screen will
appear to move a distance equal to the characteristic angular size
scale in a time
\begin{equation}\label{eqn:iss_Fresnel_time}
  t_\mathrm{F} = v^{-1} \sqrt{L/k}.
\end{equation}
With the velocity in
kilometers per second, frequency in gigahertz, and distance in
kiloparsecs, the timescale in days is given by $t_\mathrm{F} \approx
14.0 \,v^{-1} \sqrt{L/\nu}$.  

Following \citet{Rickett_ea1995}, we have adopted a transverse
velocity of $50$~km~s$^{-1}$ to account for the relative motions of
the Earth about the Sun, the motion of the Sun with respect to the
local standard of rest, and the velocities of plasma clouds within the
ISM.  Using the NE2001 software from \citet{Cordes_L2002}, which
contains a model of electron densities and fluctuations in the ISM, we
have calculated effective distances to phase screens for the lines of
sight to our extragalactic sources, and hence values for
$\theta_\mathrm{F}$ and $t_\mathrm{F}$.  We also used NE2001 to
calculate the transition frequency ($\nu_0$) between strong and weak
interstellar scattering \citep[see, e.g.,][]{Rickett_2002}.  These
quantities are shown in Table~\ref{tab:var_scintillation}.  Our values
are in basic agreement with the plots in \citet{Walker_2001} which are
based on the older Galactic electron density model of
\citet{Taylor_C1993}.

It is important to recognize that we do not have exact knowledge about
the ISM screens which might be causing scintillation of our target
galaxies.  Detailed studies of individual objects undergoing strong
scintillation have shown that the transverse velocities and distances
of the screens are often substantially different from the
``expected'' values.  However, we hope that the results we obtain
using our NE2001-based predictions will be reasonably correct \emph{on
  average}.  The angular size information is only proportional to the
square root of the screen distance, so distance errors of a factor of
ten will result in size errors of a factor of three.  The screen
velocity could easily have errors of a factor of a few.  Also, since
the transition frequency is close to the observation frequency, it is
not clear whether the scintillation is actually in the weak or strong
regime, so another error factor of about two may be possible.  Added
in quadrature, the size estimate error for any specific object
should be less than about a factor of five, assuming that the
variability is entirely due to interstellar scattering.

We expect that the intrinsic sizes of the emission regions of our
target galaxies will be substantially larger than the Fresnel angle
($\theta_\mathrm{F}$).  In this case, the integrated brightness of the
target emission will be the sum over many different regions of the
plasma screen.  The scintillation effects of each area of the screen
are relatively independent, so the brightness variations caused by each
Fresnel angular region will tend to cancel one another.  This causes
the amplitude of the variation to \emph{decrease} and the timescale of
the variation to \emph{increase} roughly as the square root of the
number of independent Fresnel zones for a uniformly illuminated
source.  Since the number of Fresnel zones will go as the square of
the source angular size, the variations will depend on the source size
to roughly the first power.  Taking the equations for weak scattering
from \citet{Walker_1998}, the angular size of the source based on a
measurement of the variability timescale $\tau_\mathrm{max}$ is
\begin{equation}\label{eqn:iss_size_tau}
  \theta_\tau =
  \frac{\tau_\mathrm{max}}{t_\mathrm{F}}\theta_\mathrm{F}.
\end{equation}
Similarly, an alternate estimate of the source angular size is given
by the modulation index, which is just our de-biased scatter from
\S~\ref{sec:gross_stat}.  Scintillation theory predicts
\begin{equation}\label{eqn:iss_size_m}
  \theta_m = \left(\frac{m_\mathrm{p}}{m}\right)^{6/7}\theta_\mathrm{F},
\end{equation}
where $m_\mathrm{p}$ is the expected modulation index for a point
source; $m_\mathrm{p} = \left(\nu_0/\nu\right)^{17/12} \sim 100\%$
since our observing frequency is close to the predicted transition
frequency.  

The variability timescales were estimated by eye from the structure
function plots in Figure~\ref{fig:target_struc_func}.  For
interstellar scintillation, the structure function at small time lags
is expected to rise as a function of the time lag to some power (a
straight line in a log-log plot), rising to a maximum at the
characteristic timescale of variability for the object, followed by a
plateau region at a value $2m^2$ \citep[see, e.g.,][]{Beckert_ea2002}.
Our target galaxies generally have noisy structure functions which do
not allow for a simple deduction of the timescale of the initial
maximum.  In Table~\ref{tab:var_scintillation} we report the lag
($\tau_\mathrm{max}$) for the initial peak in the structure function
with values clearly above the observational bias levels.  Because the
structure function values for timescales longer than 10~days are
generally poorly sampled, peaks occurring at timescales larger than
10~days have been indicated by a lower limit symbol.

Using the scintillation model predictions and our $\tau_\mathrm{max}$
and $m$ measurements, we calculated angular sizes of our target
galaxies.  Table~\ref{tab:var_scintillation} shows the results of these
calculations for the targets which show significant variability
($P_{\chi^2} < 0.01$).  Our calculations
assume that \emph{all} of the emission comes from a compact core, as
suggested by VLA and VLBA measurements which have similar flux
densities.  If \emph{part} of the emission is from an extended
emission region (say more than about 1~mas in size), the interstellar
scattering would only affect the remaining compact core, which would
\emph{decrease} the measured modulation index, and therefore
\emph{increase} the angular size of the compact core calculated from
the modulation index.  The angular size calculated from the
variability timescale would remain unchanged.

If weak interstellar scattering is responsible for the variability in
our target galaxies, then $\theta_\tau$ and $\theta_m$ should have
approximately the same value.  The two methods generally agree with
one another within a factor of 10, although there is one case where
$\theta_m \sim 40\,\theta_\tau$.  Equations \ref{eqn:iss_Fresnel_angle}
and~\ref{eqn:iss_Fresnel_time} have inverse dependencies on the
distance $L$ to the ISM screen, so in principle the distance to the
screen could be found if one believes both the $\tau_\mathrm{max}$ and
$m$ measurements.  Of the 16 variability instances in
Table~\ref{tab:var_scintillation}, 9 have $\theta_\tau < \theta_m$, so
there is no significant tendency for either $\theta_\tau$ or
$\theta_m$ to be too large or small.  This suggests that the distance
estimates from the NE2001 software are probably not significantly
biased high or low.

Assuming that the angular size calculated from the modulation index is
approximately correct, we have also calculated the equivalent
\emph{radius} of the emission region, assuming that the source is
circular on the sky.  The linear radius is given in units of
Schwarzschild radii in Table~\ref{tab:var_scintillation}, as this
quantity is more likely to be of interest for comparison with LLAGN
models.  The radii have a mean of 540~$R_\mathrm{S}$, and range from
38 to 1200~$R_\mathrm{S}$.  Given the range of black hole masses,
target galaxy distances, and calculated angular sizes, the range in
radius seems relatively small, but we have not performed a
careful analysis of observational biases which might limit the range
in radius we could measure.  A typical radius of
500~$R_\mathrm{S}$ is in reasonable agreement with predictions for
ADAF sizes at 8~GHz.  However, as described above, the variability
timescale and modulation index resulting from interstellar
scintillation are actually most dependent on the \emph{area} (solid
angle) of the source on the sky, rather than a one-dimensional angle.
The exact shape of the emission on the sky is not constrained by our
measurements, so the target galaxies could also have an elongated
structure, as would be expected from jets.  Therefore, the targets
could also have a major to minor axis ratio of $\sim 5$ and still be
within the angular size limits from our VLBA observations.  This means
that jet-like structures are also still allowed from the size
restrictions from our interstellar scintillation calculations.

Since the brightness temperature of the emission depends on the solid
angle of the source, the brightness temperature derived from the
interstellar scattering size estimate is valid whether the source is
circular \emph{or} elongated.  Table~\ref{tab:var_scintillation} shows
the brightness temperatures derived in this way.  Remarkably, the
brightness temperatures are also reasonably constant, with a mean of
$10^{10.6}$~K and a standard deviation of only a factor of 3.  There
are no galaxies which appear to have brightness temperatures above the
inverse Compton limit of $\sim 10^{12}$~K.  (Brightness temperatures
as high as $10^{14}$~K \emph{could} have been detected in our sample
galaxies.) Although these brightness temperatures are moderately high
for the standard ADAF models, they can probably be accommodated with
the inclusion of a small non-thermal electron population.
Alternatively, these brightness temperatures are quite reasonable for
jet models.

The differences of up to a factor of 40 between $\theta_\tau$ and
$\theta_m$ seem rather larger than we would expect for good agreement
between the two methods of calculating angular size.  The structure
function plots for our targets are very noisy, which is partially due
to the small flux density changes we are measuring.  The structure on
short timescales may also be affected by other things besides
interstellar scattering, such as problems in the instrumentation or
intrinsic flicker in the target galaxies.  Whatever the cause, the
$\tau_\mathrm{max}$ estimates for the structure function plots in
Figure~\ref{fig:target_struc_func} are probably not very reliable.  We
have somewhat better faith that the modulation index calculations for
the variability are closer to being correct, as the modulation index
contains information from \emph{all} of the data-points, while the
individual structure function points depend on only a few measurements
each.  The modulation-index-based angular sizes in
Table~\ref{tab:var_scintillation}, which assume that all of the
variability is caused by interstellar scattering, have a mean angular
size of 76~$\mu$as.  However, the structure functions for our targets
do not show the expected power law rise at small time lags.  This may
be caused by at least some of the variability being intrinsic to the
target emission regions.  In that case, the fractional variability due
to scintillation would be smaller than $\sigma_\mathrm{d}$, which
would \emph{increase} the sizes of the emission regions.  Therefore,
sizes in Table~\ref{tab:var_scintillation} should be treated as
probable lower limits, and brightness temperatures as upper limits.

\subsection{Extrinsic or Intrinsic Variability?\label{sec:extrinsic_or_intrinsic}}

Intrinsic variability is a viable explanation for the variability we
observed in the nuclear emission regions of our galaxies.  The
inferred brightness temperatures are in the range
$10^{9}$--$10^{11}$~K --- less than the inverse Compton limit.  The
roughly few days timescale of the variability as measured from the
structure functions implies a variable emission region size smaller
than about 50~$\mu$as for the typical distance of our sample galaxies,
which is well within the size limits from our VLBA observations.  The
total extent of the radio emitting region can, however, be
significantly larger than this size.

Scintillation-induced extrinsic variability is also consistent with
our observations.  Most of our ``reliable'' galaxy datasets show
variability in at least one epoch, and we use this information to
estimate the sizes for all of our galaxies.  (Because the
scintillation seen in some IDVs is known to be transient, we expect
some galaxies which \emph{could} scintillate to actually remain
constant, see
\citealp{Kedziora_Chudczer_ea1997,Kedziora_Chudczer_ea2001}.)
Assuming all of the variability is caused by scintillation, the mean
angular size is 76~$\mu$as, and the brightness temperature is
$10^{10.6}$~K.  Once again, the brightness temperatures are below the
inverse Compton limit.

Thus, whether the variability is extrinsic or intrinsic, the
implications for the physical conditions of the emission regions are
approximately the same.  Furthermore, large Doppler boosting factors
are unnecessary to explain the brightness temperatures, which are
comfortably within even the equipartition inverse-Compton limit of
\citet{Readhead_1994} and others.  This is in stark contrast to
intraday variability observations of bright, distant AGNs which
occasionally imply apparent brightness temperatures of $10^{19}$~K or
higher \citep[see, e.g.,][]{Wagner_W1995}. In such cases, interstellar
scintillation is normally preferred over intrinsic variations to
explain the origin of the variability, as it predicts dramatically
larger angular sizes and reduces the necessary Doppler factors.

As discussed in \S~\ref{sec:ISS}, we believe that at least some
intrinsic variability is present in our galaxies based on their
structure function behavior.  We therefore combine our information to
come up with our best estimates of the physical parameters of the LLAGN
emission at $8.5$~GHz for our VLBA unresolved nuclei.  The VLBA
imaging typically gives upper limits of $0.5$--$0.7$~mas for the major
axis of the emission region.  The scintillation results give a mean
lower limit of $0.076$~mas.  Correcting for a reasonable amount of
contamination by intrinsic variability, we estimate that the mean
actual size of the emission regions is $0.2$~mas, plus or minus a
factor of two for individual galaxies.  In terms of mean linear size,
this is about 1400~$R_\mathrm{S}$ plus or minus a factor of two.

This corresponds to an estimated solid angle of about $0.04$~mas$^2$,
which is potentially more meaningful, since the scintillation results
are shape-insensitive.  The emission can therefore be circular on the
sky or in a narrow strip.  For jet models, this allows the emission to
extend out to a distance of $0.5$~mas (or $1.0$~mas in the case of
\NGC{5866}) so long as the emission is only $0.08$~mas ($0.04$~mas)
wide.  This suggests that the mean brightness temperature is about
$10^{10.1}$~K, with individual galaxies being perhaps plus or minus
$0.6$~dex different.  This well matches the brightness temperature
findings from intrinsic variability and is consistent with the lower
limits from our VLBA imaging.

A brightness temperature of $10^{10.1}$~K is close to an order of
magnitude higher than expected from the traditional ADAF model with
electrons in a thermal energy distribution.  Thus, our results require
an additional population of high-energy non-thermal electrons in
order to produce the observed brightness temperatures.  Whether this
non-thermal electron population is present in the accretion disk, or
in an outflow from the accretion disk, or in a jet remains an open
question.

Our results compare favorably with other results for the sizes of
compact LLAGN.  VLBA measurements of the size of \NGC{3031} are
discussed in Appendix~\ref{sec:sources}.  For Sgr~A$^\ast$,
\citet{Bower_ea2004} claim to have measured the size of the emission
region at high frequencies using the VLBA.  Their fit to the angular
size of Sgr~A$^\ast$ predicts a major axis size of 160~$R_\mathrm{S}$
in radius at our observing frequency of 8.46~GHz.  Sgr~A$^\ast$ has a
radio luminosity about $10^5$ times weaker than the median target
galaxy in our sample, the black hole mass for Sgr~A$^\ast$ is about 10
times smaller than the typical black hole mass in our sample, and the
emitting region is $10^3$--$10^4$ times closer.  Therefore, the
brightness temperature of Sgr~A$^\ast$ at 8.46~GHz is still close to
the brightness temperature of our objects.  It is possible that the
size of the radio emission region scales with black hole mass and
radio luminosity over an extremely wide range of luminosities and mass
accretion rates.

\section{Medium-Term Variability\label{sec:long-term}}

Active galactic nuclei are well known for having variable emission at
all observed frequencies \citep[see, for example, the review
by][]{Ulrich_MU1997}.  In addition to the short intraday variability,
AGNs typically also show relatively large amplitude variations over
longer timescales, from months to years.  As with other wavelength
regimes, long-term radio variability in low-luminosity AGNs is less
well studied, but is known to be present, as indicated by the studies
of \citet{Nagar_ea2002} who found that almost half of their LLAGN
sample observed with the VLA at 8.5~GHz changed in flux density by at
least 20\% after a 15~month time interval.

Many of our own LLAGN target galaxies also changed significantly from
the May to September variability runs.  To measure this effect, we
calculated the average flux density from the \emph{last} three days of
our 2003~May observations and the average flux density from the
\emph{first} three days of our 2003~September observations.  These
days had the most similar $(u,v)$ coverage between the two months,
with most of the VLA antennas in their A~configuration locations.
Therefore, these three day averages should minimize $(u,v)$
differences, and provide reliable flux density estimates to compare
\emph{all} of our target galaxies.  Averaging over several days for
each month should also minimize variations caused by short-term
variability, whether intrinsic or extrinsic in origin. 

Fractional variation values were calculated from
\begin{equation}
  F = \frac{2\left(<S_\mathrm{Sep}> - <S_\mathrm{May}>\right)}
  {<S_\mathrm{Sep}> + <S_\mathrm{May}>},
\end{equation}
along with associated uncertainty estimates.  These values are
listed in Table~\ref{tab:var_stats}, and plotted as a function of mean
flux density in Figure~\ref{fig:long_term_var_plot}.  The CSO
calibrators all have fractional variations less than 5\%, and are
centered about $F=0$, suggesting that the absolute flux density
calibration of both months was performed properly.  The other phase
calibrators tend to be more varied.  Roughly equal numbers of
calibrators increased (positive $F$) and decreased (negative $F$) in
flux density from May to September.

The fractional variations of the target galaxies are rather different.
Seven of the eighteen target galaxies have flux density changes of
more than 10\% from May to September.  Of these galaxies, only one
(\NGC{3226}) increased in brightness, while six decreased
substantially over the time period.  The random probability that at
least six of seven galaxies would all increase or decrease is 13\%, so
the apparent excess of decreasing flux densities is not statistically
significant.  \emph{None} of these seven galaxies have short-term
variability related to $(u,v)$ effects, and the target galaxy changes
do not appear related to changes in the flux densities of their phase
calibrators.  After carefully examining the data, and remembering that
the numbers of calibrators with increasing and decreasing flux
densities are about equal, we are confident in the measured fractional
variations of these galaxies.

The probability that a target galaxy would change dramatically between
the two observation periods does not seem to be related to the
observed flux density, as the large fractional variations occur over
essentially the entire target flux-density range.  However,
Figure~\ref{fig:long_term_var_lum} shows that most of the
large fractional variations occur in the galaxies with the weakest
8.5~GHz luminosities in our sample.  We are unsure whether this result
is a real effect or just coincidence, although it might be possible
that the weakest galaxies are able to vary more easily over the $\sim
4$~month timespan between our observations.

\citet[hereafter \citetalias{Barvainis_ea2004}]{Barvainis_ea2004} have
completed a study of the long-term variability of AGNs to search for
possible differences between radio-loud and radio-quiet AGNs.  Radio
loudness is defined here as the ratio $R$ between radio and optical
flux densities \citep{Kellermann_ea1989}, with $R < 3$ defining a
radio-quiet quasar (RQQ), $3 < R < 100$ defining a radio-intermediate
quasar (RIQ), and $R>100$ defining a radio-loud quasar (RLQ).
The AGNs used in \citetalias{Barvainis_ea2004} range from ``classical''
Seyfert galaxies to bright quasars.  They find no significant
differences between the radio core variability and spectral index
properties as a function of the radio-loudness parameter $R$.

Our study extends the results in \citetalias{Barvainis_ea2004} by
identifying the variability statistics for flat-spectrum LLAGNs, which
are significantly lower in luminosity, extending the range of radio
luminosities examined by about 1.5 orders of magnitude to weaker
sources.  Although the radio powers of our LLAGN cores are weak, the
ratios of radio to optical flux densities classify our galaxies as
radio-loud or radio-intermediate \citep[see][]{Ho_2002}.  Our galaxies
are some four orders of magnitude fainter than the weakest radio loud
AGN of \citetalias{Barvainis_ea2004}.  The combined range of
radio-loud radio luminosities is therefore about 8 orders of
magnitude, from $10^{20}$--$10^{28}$~W~Hz$^{-1}$.

The amplitude of medium-term variability seen in Figures
\ref{fig:long_term_var_plot} and~\ref{fig:long_term_var_lum} is quite
similar to the variability found by \citet{Barvainis_ea2004},
accounting for the differences between our fractional variation $F$
and their de-biased RMS variability.
Figure~\ref{fig:long_term_var_hist} presents a histogram of the number
of sources as a function of the absolute fractional variability.
Combining our extended and point-like results, our medium-term sample
variability results are quite similar to the results of
\citetalias{Barvainis_ea2004}, given the small number statistics.  Our
variability results for our non-CSO calibrators are also similar to
their results.

The quasars investigated in \citetalias{Barvainis_ea2004} tend to be
located at cosmological distances (redshifts in the range $0 < z <
4$), while our most distant galaxy is only 66.5~Mpc away ($z =
0.017$).  Thus, our VLA observations measure emission on much smaller
physical scales than the \citetalias{Barvainis_ea2004} observations.
However, large-scale jet emission tends to have a steep spectral
index, so the \citetalias{Barvainis_ea2004} study should be dominated
by core emission at 8.5~GHz.  Furthermore, \citet{Ulvestad_AB2004}
find that most of the radio-quiet quasars they have investigated with
the VLBA are consistent with all of the arcsecond (VLA) scale flux
density being located in the milliarcsecond (VLBA) scale core.  Thus,
our VLA observations \emph{are} actually measuring emission from the
same size scales as the \citetalias{Barvainis_ea2004} observations.

Although a more detailed long-term study of variability in LLAGNs is
needed, our results indicate that both the medium- and short-term
variability properties of LLAGNs are similar to those of far more
luminous quasars.  The radio spectral indices of the nuclear regions
of Seyfert LLAGNs \citep{Ulvestad_H2001.0} are also similar to the
quasar sample in \citetalias{Barvainis_ea2004}, with approximately
equal numbers of sources with spectral indices above and below $\alpha
= 0$ and concentrated in the range $-1 < \alpha < +1$.  These results
suggests that the physical processes responsible for producing the
observed radio emission from AGN cores \emph{may} be the same for all
AGNs, despite a luminosity range of some 8 orders of magnitude.
However, the luminous quasars have combined optical and X-ray
luminosities which are close to the Eddington limit \citep[for
radio-quiet as well as radio loud quasars, see][]{Ulvestad_AB2004},
while our low-luminosity AGNs have bolometric luminosities orders of
magnitude \emph{below} the Eddington limit.  It is not clear that the
accretion processes which lead to the optical and X-ray emission are
necessarily related to the processes which lead to the radio emission
ubiquitously seen in AGNs.

\section{Conclusions\label{sec:conclusions}}

We have conducted a VLA variability study of a sample of 18
predominantly flat-spectrum LLAGNs to investigate the sizes of the
radio emission regions in these objects. Our analysis included new and
published milliarcsecond-scale VLBA imaging of all 18 objects.  

The majority of our sample galaxies have essentially all of their
large (arcsecond) scale flux confined within a single,
sub-milliarcsecond core.  \NGC{2273} and \NGC{3227}, which have steep
radio spectral indices ($\alpha \leq -0.5$) on VLA scales, have
extended structure on VLBA scales, as does \NGC{3079}.  Three of our
galaxies have measured sizes from our VLBA observations of $\sim
1.5$~mas in the major axis, and are effectively unresolved in the
minor axis.  The 11 remaining galaxies with VLBA detections all have
sizes less than 1~mas and are probably smaller than $0.5$~mas.

We have detected short-term variability in our LLAGN sample on time
scales from slightly less than a day to 10~days.  The fraction of
galaxies which are variable at or above the 4\% level agrees very well
with a larger sample of far more luminous AGNs performed by
\citet{Lovell_ea2003}.  The fraction of galaxies with smaller
variability amplitudes also agrees quite well with other existing
studies of more luminous AGNs.

The observed variability is consistent with intrinsic variability, but
is also partially consistent with scintillation caused by the Galactic
interstellar medium.  Both intrinsic and extrinsic (scintillation)
explanations for the variability yield consistent results for the
radio source sizes and brightness temperatures.  We estimate that the
mean brightness temperature of the emission regions is about
$10^{10.1}$~K, and that the mean angular size of the emission regions
is about $0.2$~mas, which corresponds to a mean radial size in
Schwarzschild radii is about 1400~$R_\mathrm{S}$.

Our medium-term variability measurements are also consistent with the
variability of far more luminous quasars, analyzed by
\citet{Barvainis_ea2004}.  This result suggests that the physical
processes which control the regions emitting the bulk of the radio
emission in AGNs \emph{may} remain the same from the highest
luminosity quasars to the low-luminosity AGNs, spanning some 8 orders
of magnitude.



\acknowledgments We thank Greg Taylor for useful discussions about
using CSOs for amplitude calibration at the VLA.  We thank Dale Frail
and Barney Rickett for advice on scintillation in the ISM.  We would
also like to thank Jean Eilek, Sera Markoff, and Feng Yuan for
valuable discussions on the physics of accretion and jet models.
J.M.A.  gratefully acknowledges support from the predoctoral
fellowship program from NRAO.  This research has made use of the
SIMBAD database, operated at CDS, Strasbourg, France, and NASA's
Astrophysics Data System Bibliographic Services.



Facilities: \facility{NRAO/VLA()}, \facility{NRAO/VLBA()}.



\appendix

\section{VLA Observational Details\label{sec:VLA_obs_details}}

\subsection{Observing Strategy\label{sec:VLA_strategy}}

We used observations of the phase calibrator sources listed in
Table~\ref{tab:sample} to determine initial phase and amplitude
corrections for each VLA antenna.  The calibrator and target pairs
were observed in fast switching mode to minimize the observing time
lost due to overhead in the telescope control software.  A cycle time
of slightly over four minutes was used for all targets, with a
correlator integration time of 3\onethird~s.  For most galaxies, three
fast switching cycles were performed for each snapshot to bring the
total integration time on the target galaxies to 10~minutes, resulting
in a noise level of about 50~$\mu$Jy~beam$^{-1}$.  For galaxies
stronger than 50~mJy, this was normally reduced to one or two fast
switching cycles as signal to noise ratios far greater than 200 were
not needed for this study.

The source \objectname[3C]{3C~147}\footnote{Located at a large right
  ascension gap in our source list, this calibrator selection meshed
  with our target snapshot sequence better than the canonical VLA
  calibrator, \objectname[3C]{3C~286}, which is near a relatively
  crowded right ascension.} was observed once per day to set the
amplitude scale.  However, this calibration method is typically only
accurate at the 1 or 2 percent level \citep[VLA Calibrator
Manual\footnote{\url{http://www.aoc.nrao.edu/$\sim$gtaylor/calib.html}};]
[hereafter \citetalias{Fassnacht_T2001}]{Fassnacht_T2001}.  In order
to improve the relative calibration of the VLA from day to day during
our variability campaign, we also observed a set of 9 compact
symmetric objects (CSOs) to serve as stable relative calibrators,
following the suggestions in \citetalias{Fassnacht_T2001}.  CSOs are
compact on VLA scales, which is ideal for instrument calibration, but
their milliarcsecond scale emission is dominated by steep-spectrum
radio lobes on both sides of the core.  The fraction of emission
coming from the core is only a few percent, and little Doppler
boosting is present, so short-term variations resulting from ejections
of new jet components or wobbling of the jet orientation angle should
be minimal.  In their study, \citetalias{Fassnacht_T2001} found the
mean variation of their CSO sample to be only 0.7\% using VLA
observations in A and B~configurations.  Five of our CSOs overlap the
\citetalias{Fassnacht_T2001} sample; the other four were selected from
\citet{Peck_T2000} to extend our calibrator list to smaller right
ascensions.

Two of the CSOs (\objectname[COINS]{J0650$+$6001} and
\objectname[COINS]{J1035$+$5628}) were able to serve as phase
calibrators for our target galaxies.  The remaining CSOs were observed
in 60~s snapshot observations.  \objectname[COINS]{J1244$+$4048} was
observed at least five times during each day at a variety of elevation
angles to check for gain variations with time or elevation that are
not corrected by the standard gain curves provided by the VLA.

For our $\sim 5$~hour observing blocks, we cycled through each object
in our target sample and calibrator list in an order which minimized
antenna slew times, generally going in order of right ascension.  For
the $\sim 10$~hour blocks, each target galaxy was observed at least
twice.  We attempted to have observations of the same target separated
by at least 3~hours.  Additional target and CSO observations were made
to fill out remaining observing time.

\subsection{Initial Flux Density Calibration\label{sec:VLA_initial_red}}

A careful process of data reduction and analysis was undertaken.  The
VLA data were reduced using the AIPS software package from NRAO
\citep{van_Moorsel_KG1996}.  Since our observations were made during
reconfiguration periods at the VLA, the antenna positions for each
observing run were updated to reflect better estimates of the
positions made from calibration observations made by NRAO staff.

Next, the data were flagged to eliminate known problems.
Records of malfunctions and other problems by the telescope operators
were used to flag specific antennas and/or time intervals.  The first
couple of integrations of each source scan are frequently corrupted at
the VLA, especially during fast switching, so we flagged the first
6\twothirds~s (2 integration times) of each scan.

Next, data for \objectname[3C]{3C~147} and
\objectname[COINS]{J1244$+$4048} were flagged and reduced using
standard methods in AIPS.  Calibration steps used to determine the
absolute calibration of the VLA using \objectname[3C]{3C~147} followed
the guidelines in the VLA Calibrator Manual.  These guidelines
restrict the $(u,v)$ range allowed to 0--40~kilowavelengths
(k$\lambda$) at 8.5~GHz.  For antennas locations similar to
A~configuration, this restricts the number of useful antennas on each
arm to only 2~antennas.  Solution intervals of 10 or 20~s were using
during calculations of the gain solutions, depending on atmospheric
conditions.  Antennas gains for \objectname[COINS]{J1244$+$4048} were
calculated in a similar manner, except that only baselines greater
than 10~k$\lambda$ were allowed.  Short baselines were removed to
prevent any extended emission in the primary beam from affecting the
observations and to reduce problems resulting from shadowing and
crosstalk between neighboring antennas --- such problems were most
severe during the first few days of our May observing run when most of
the antennas were still in the D~configuration.

The flux density of \objectname[COINS]{J1244$+$4048} was calculated
from the antennas gain calibrations for each day.  The results had a
scatter of about 2\%, as expected from the analysis of
\citetalias{Fassnacht_T2001}.  Nevertheless,
\objectname[COINS]{J1244$+$4048} appeared to be roughly constant in
brightness during our observations in both May and September.  We
formed average flux densities for the May and September periods
separately in each intermediate frequency channel of the VLA.

\subsection{Full Amplitude and Phase Calibration\label{sec:VLA_secondary_red}}

Next, the entire dataset was calibrated using
\objectname[COINS]{J1244$+$4048} for absolute gain calibration.
Because \objectname[COINS]{J1244$+$4048} is a point source to the VLA,
\emph{all} antennas could be used in the gain calibration of target
and CSO snapshot observations (only a few antennas typically could be
calibrated using \objectname[3C]{3C~147}), dramatically increasing the
precision of the calibration, as the error is reduced approximately as
the inverse of the square root of the number of antennas.  This
painstaking process was iteratively repeated in a process of
examination of the data, flagging, calibration, and re-examination of
data.

After careful flagging, the antenna gains for each CSO and phase
calibrator were calculated with the task CALIB.  An initial
calibration of only phase changes was used with short solution
intervals (10 to 20~seconds) to remove rapid changes caused by the
atmosphere, and then a second calibration was performed for both
amplitude and phase over the entire scan length.  The results from
these calculations were used to search for additional problems with
the data.  All of the phase calibrators and CSOs were imaged to verify
that they were point-like for the $(u,v)$ ranges used for determining
antenna gains.  When all detectable errors in the data were found,
final antennas gain solutions were calculated and applied to the data,
with the target galaxy gains interpolated from their phase calibrator
information.

\subsection{CSO Analysis\label{sec:CSO_analysis}}

Figure~\ref{fig:cso_var} shows the brightness time series for our
fully calibrated CSO data.  The data for each CSO are divided into
separate plots for May (left) and September (right).  All of the plots
are shown with the same fractional deviation scale from $-4$\% to
$+4$\% from the mean brightness for each month, as indicated by the
right-hand vertical axis of each plot.  

The CSO data were analyzed to determine improved amplitude
calibrations for the VLA observations.  We performed a least squares
fit to the CSO data to find the mean brightness of each individual CSO
for each month plus multiplicative amplitude corrections for each
observing day.  The CSO \objectname[COINS]{J0410$+$7656} was resolved
significantly by the VLA, so it was excluded from the least squares
fit.  No significant variation was seen as a function of elevation
angle as shown in Figure~\ref{fig:cso_elev}.  This suggests that the
elevation gain curves used by the VLA software have properly corrected
for elevation effects in the data.  Similarly, no trends were seen for
hour angle or time of day.  Interestingly, the data for individual
CSOs appear remarkably stable for measurements made on any single day,
as seen in Figure~\ref{fig:cso_var} for objects like
\objectname[COINS]{J0204$+$0903}, \objectname[COINS]{J0650$+$6001},
and \objectname[COINS]{J1035$+$5628}.  These objects show a modest
scatter in brightness level from day to day, but within any given day
the scatter is typically only a few tenths of a percent.  These
observations were made many hours apart, with widely varying elevation
and azimuth angles and $(u,v)$ coverage, but the measured brightness
remains extremely constant, with a sub-day RMS scatter of only 0.5\%
averaged over all CSOs.

The CSOs show day-to-day variations of a percent or more in amplitude,
and there is more scatter present in the May data than is in the
September data.  We believe that this higher level of variability is
\emph{not} caused by the extreme change in $(u,v)$ coverage, as the
high level of scatter can be independently seen in the first three
days of observations during May, for which the positions of the
antennas are identical.  Rather, the higher level of scatter is
probably caused by more variable conditions in the atmosphere or
ionosphere during 2003 May.  Weather conditions during September were
generally clear with surface winds calm or less than 3~m~s$^{-1}$,
whereas the May weather was generally overcast with low clouds, winds
greater than 5~m~s$^{-1}$, and a severe thunderstorm on one date.
Self-calibration solution intervals as small as 10~seconds were
necessary to adequately track phase fluctuations during May, whereas
the phase coherence time for September was normally several times
longer.  The greater phase stability during September leads to more
reliable measurements for that month.  Results for individual CSOs are
shown in Table~\ref{tab:CSO_list}.  The mean RMS scatter for CSOs in
May was 1.38\%, while for September it was only 0.71\%.  This residual
scatter after the least squares adjustment is assumed to indicate the
minimum level of uncertainty in the gain calibration of individual
objects, and it has been added in quadrature to the noise-induced
measurement error determined for each snapshot observation.  This
total error for the CSOs is shown as 1-$\sigma$ error-bars in
Figure~\ref{fig:cso_var}.  The error due to random noise is typically
less than 0.1\% for the CSO objects, so it generally is negligible
relative to the measured amplitude scatter.

Figure~\ref{fig:cso_hist} shows a histogram of the residuals from the
CSO fitting process.  Although the profile is not exactly symmetric,
the deviations are not dramatically different from a Gaussian
distribution.  These deviations are probably representative of the
deviations to be expected in the galaxy observations.

\subsection{Snapshot Imaging\label{sec:VLA_snapshots}}

In order to retain as much similarity between the calibrator and
target data as possible, the processing steps to extract brightness
information from individual snapshots were kept as similar as possible
for all objects.  

Preliminary imaging of the snapshot data was used to investigate the
structure of the sources for use in self-calibration.  For those
sources which were unresolved, a point source model was used for
self-calibration.  This model was used for the calibrator sources as
well as for several of the target galaxies.  For target galaxies with
apparent structure, data from multiple snapshot observations were
combined (improving $(u,v)$ coverage) to produce a
high-fidelity template image of the galaxy using an iterative process
of imaging and self-calibration.  Then, the clean components from this
template image were used as the source model for self-calibration of
the individual snapshot observations.  For those target galaxies with
substantial differences between their May and September appearances
(because of a significant change in the core flux level or because of
large amounts of extended flux detected in the numerous short
baselines in the May data), different templates were used for May and
September data.  

As a fringe benefit, the self-calibration process was also used to
move the core emission of the galaxy to the phase center.  Errors in
the a priori calibration from the phase calibrator result in images
which wobble around from snapshot to snapshot.  Self-calibration was
used to shift all of the snapshot images to the same apparent
position, allowing for easier comparisons between images and reducing
measurement scatter resulting from flux falling in different pixel
distributions.  The scatter in the position of the source peak in the
resulting images was generally less than 0.03~pixels for typical
sources in our sample.

For target galaxies weaker than 15~mJy, we averaged the two
polarizations and two intermediate frequency channels to improve the
signal to noise of self-calibration; tests showed a coherence loss
less than 0.2\% due to this averaging.  Solution intervals for target
galaxies were kept as short as possible while still providing good
fits to the entire snapshot dataset.  These solution intervals were
around 30~s for most of the galaxies, with the strongest galaxies able
to use solution intervals of only 10~s and the weakest galaxy
(\NGC{777}) requiring solution intervals of 4~minutes.

The AIPS task IMAGR was used to produce deconvolved images of our
sources using the CLEAN algorithm.  All snapshot images were
individually examined to verify that there were no significant
problems with individual datasets.  However, human interaction was
eliminated in the final processing stages using a common set of CLEAN
parameters to ensure that all snapshot imaging and deconvolution was
treated in the same way.

For our variability study, we are only concerned with measuring the
flux density in the core of the galaxy, which is unresolved by the
VLA.  We explored many different schemes to weight data in the $(u,v)$
plane to study the effects of possible extended emission on our
measurement of the core flux density.  We selected natural weighting
to improve sensitivity.  Minimum baseline lengths of 0--40~k$\lambda$
were normally sufficient to prevent extended emission from corrupting
the core flux density measurement, so we conservatively restricted the
$(u,v)$ range for nearly all objects to baselines longer than
100~k$\lambda$, removing sensitivity to regions of emission larger
than about 2\arcsec.  (We used $(u,v)$ ranges of 0--200~k$\lambda$ for
\objectname[COINS]{J0410$+$7656}, and 0--400~k$\lambda$ for
\objectname[VLA_Cal]{J1035$+$5628},
\objectname[VLA_Cal]{J1215$+$3448}, and
\objectname[COINS]{J1400$+$6210}, which have small scale structure
resolved by the VLA.)  Further concerns regarding the effects of
extended emission and jet features on the measurements of the core
flux densities are discussed in \S~\ref{sec:Target_statistics}.

Source flux densities and error estimates were obtained through a
combination of methods.  The task JMFIT fit a two-dimensional Gaussian
to the source core to measure the peak and integrated flux densities.
The maximum pixel flux density was recorded as an alternative measure
of the peak flux density.  (Self-calibration shifted the core to the
same pixel position, so this maximum is a reliable estimate of the
flux density of the unresolved core.)  In addition to error estimates
from JMFIT, the noise in each image was measured in a symmetric region
surrounding the source.  Finally, the task UVFIT was used to fit a
point source of fixed location at the phase center of the image to the
self-calibrated $(u,v)$ data, as an alternate method of measuring the
core flux density.  The variability analysis presented in this paper
was independently performed on each type of flux density measurement.
Although the results varied slightly, no significant differences among
measurement techniques were found.  For simplicity, we only use
the JMFIT peak flux density results in this paper.

The error estimates for the measured brightnesses were modified by
adding the mean CSO residual scatter appropriate for the month of
observation in quadrature to the random measurement error.  For the
target galaxies, we also added a 0.5\% uncertainty in the gain
transfer from the phase calibrators to allow for errors introduced in
the time interpolation process and for the spatial extrapolation of
the phase calibration from the calibrator to the target positions.  We
also added an uncertainty ($\sim 1$\% for a 5~mJy source) to account
for errors introduced by the phase-only self-calibration process
\cite[see][equation 10--15]{Cornwell_F1999}.  Although this error is
insignificant for bright sources such as the phase calibrators, it can
be very significant for our weak galaxies with only a few millijanskys
of flux density.  For the phase calibrators and CSO gain calibrators,
only the CSO statistical gain uncertainty was added to the measurement
uncertainty.

\section{Non-Normal Error Distributions\label{sec:Error_Distributions}}

The statistical analysis made in the main body of our paper, including
the $\chi^2$ distribution probability and the Monte Carlo simulations,
assumed a normal error distribution.  The random and systematic
errors in our VLA observations are not guaranteed to be normally
distributed.  However, careful data flagging removed significant
outliers in amplitude and phase, eliminating large error sources in
the data.  Small problems near the noise level do remain in the data,
but we believe these to be small and infrequent enough to not cause
significant departures from a normal error distribution.  Mathematical
operations on the interferometry data (such as the conversion of real
and imaginary numbers to amplitudes and phases) do not strictly
preserve the normal error distribution shape, but we also believe that
this effect is not critically important to our results.  However, the
nature of the systematic errors is not well understood, and
significant non-normal errors could possibly have been introduced by
such terms.

We have examined the residual deviations for our galaxies to search
for signs of significant deviations from a normal error distribution,
but have found none.  Figure~\ref{fig:NGC4168_hist} shows the residual
histogram for \NGC{4168}, one of our ``constant'' galaxies.  Given the
small number of data-points, a normal distribution agrees relatively
well with the measured distribution of errors.  Histograms for our
other ``constant'' galaxies show similar results.  The distributions
are not significantly skewed toward positive or negative errors, and
they do not show unexpectedly large numbers of errors at either very
large or very small deviations.  Sources which we classify as variable
tend to have structure functions (\S~\ref{sec:struc_func}) showing
larger flux density differences on longer time-scales, and frequently
show coherent variations in the time series plots, suggesting that
large outliers to the normal error distribution are \emph{not}
responsible for the variability that we detect.  We believe that the
errors in our galaxy data are reasonably well approximated by a normal
error distribution, and are therefore confident that the standard
statistical analysis presented in this paper is reasonably accurate.

To investigate the dependence of our Monte Carlo simulations on the
exact form of the error distribution, we performed additional
simulations using the measured error distribution of \NGC{4168} and,
alternatively, the CSO error distribution (see
Figure~\ref{fig:cso_hist}) instead of a normal error distribution.
The \NGC{4168} distribution is similar to the normal distribution, but
there are no errors at very large $\sigma$ values because of the small
number of data-points.  Since \NGC{4168} is not a strong source, the
deviations in its error distribution are probably dominated by the
``random'' errors in the $(u,v)$ data.  The CSO distribution is a
composite of many different strong sources, each with different
individual scatter levels, and probably gives a better prediction of
the possible deviations for the systematic errors.

The suggested brightness temperatures from the alternative error
distributions were not substantially different from the normal
distribution results.  The implied brightness temperatures were
generally higher, but only by about $0.1$~dex (a factor of $1.26$) on
average for the 50\% confidence level calculations.  For the 90\%
confidence level results, about half of the galaxies had implied
brightness temperatures increase by less than $0.1$~dex, and about
half of the galaxies had increases between $0.2$ and $0.3$~dex (that
is, less than a factor of two), with a maximum increase of only
$0.33$~dex.  The small brightness temperature increases are probably
caused by the lack of high-$\sigma$ errors in the \NGC{4168} and CSO
error distributions as the absence of high-$\sigma$ errors requires
larger intrinsic changes in brightness to yield the same observed
measured flux density changes.  These changes are within the $0.3$~dex
simulation scatter using different selection criteria as described in
\S~\ref{sec:T_b}, and do not significantly change our results.

\section{Individual Sources\label{sec:sources}}

\begin{figure}
\plotone{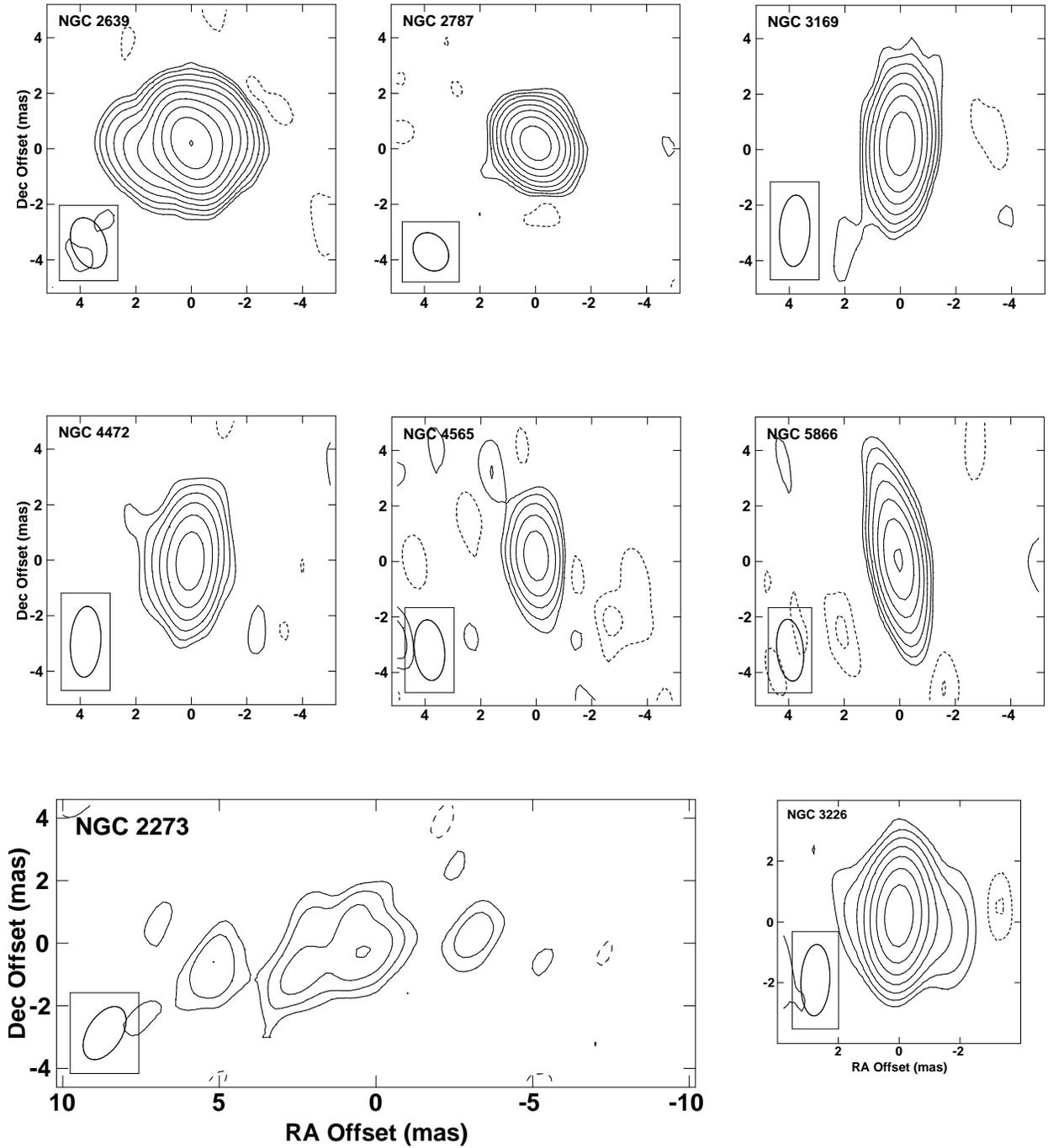}
\caption{VLBA 8.4~GHz images of the eight
  detected LLAGNs are shown as contour plots.  Contours start at 2
  times the RMS noise level (see Table~\ref{tab:VLBA_att}) and
  increase by factors of 2.  Negative contours are indicated by dashed
  lines.  The restoring beam for each image is shown in the bottom
  left-hand corner of each individual image.
  \label{fig:vlba_images} }
\end{figure}

\begin{figure}
\epsscale{0.8}
\plotone{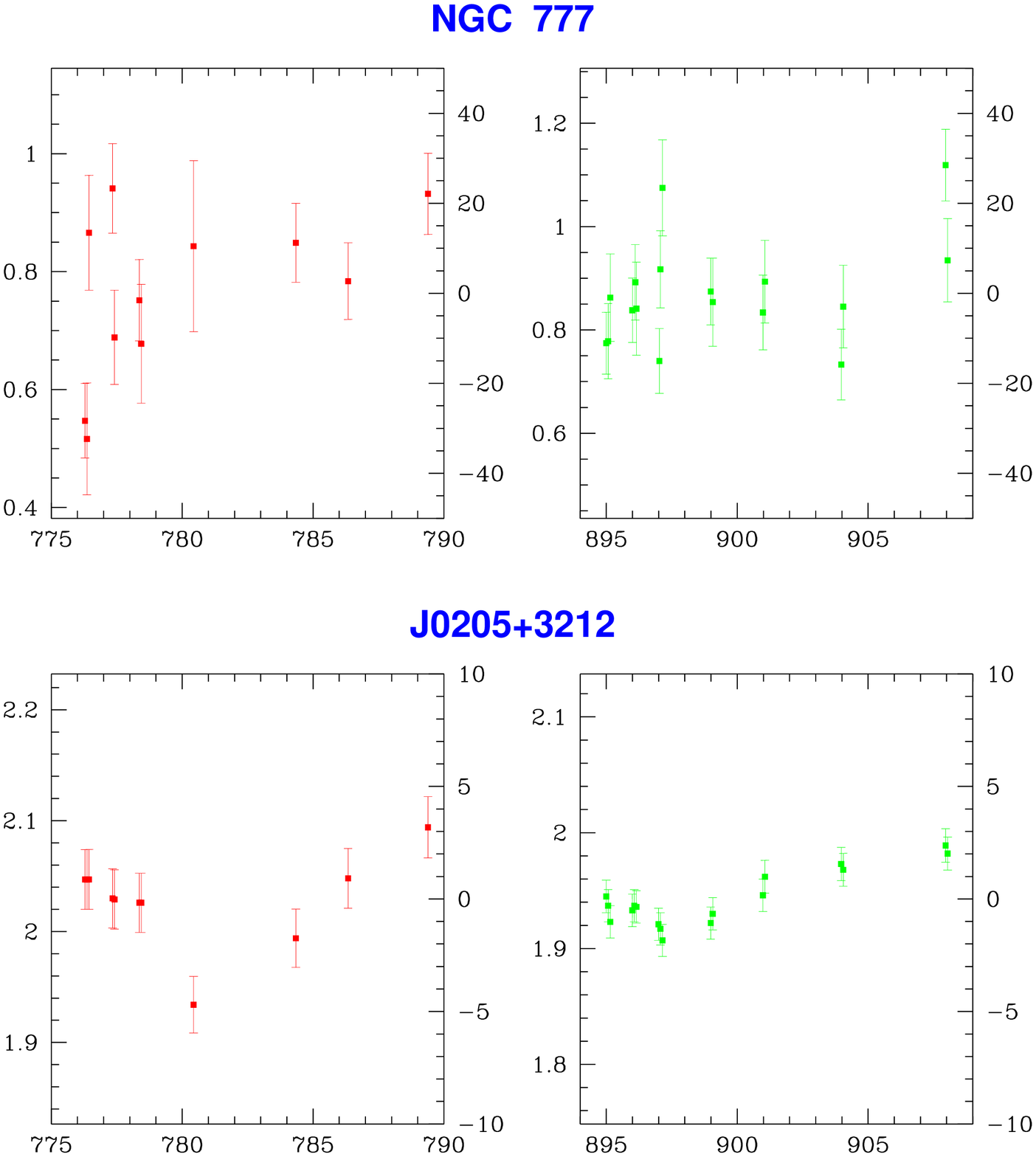}
\caption{
  Example target time series for NGC~{777}.  The peak flux density is
  shown as a function of time, with the left-hand plot of each object
  (red points) displaying 2003 May data, and the right-hand plot
  (green points) showing 2003 September data.  The corresponding phase
  calibrator time series is shown immediately below NGC~{777}.  The
  horizontal axis indicates the number of days since Julian Date
  $2\,452\,000$.  The vertical axis on the left-hand side of each
  individual plot shows the peak flux density of the measurement, in
  units of millijanskys per beam for NGC~{777} and in units of janskys
  per beam for the phase calibrator.  The vertical axis on the
  right-hand side of each plot gives the relative difference from the
  mean value for the month.  Error-bars show the 1-$\sigma$
  uncertainty in the measurements, including random errors and the
  estimated systematic errors.  (See the on-line paper for color time
  series plots for all target galaxies.)
  \label{fig:tar_time_0} }
\end{figure}

\begin{figure}
\epsscale{0.8}\vspace{-0.75in}
\figurenum{\ref{fig:tar_time_0}a}
\plotone{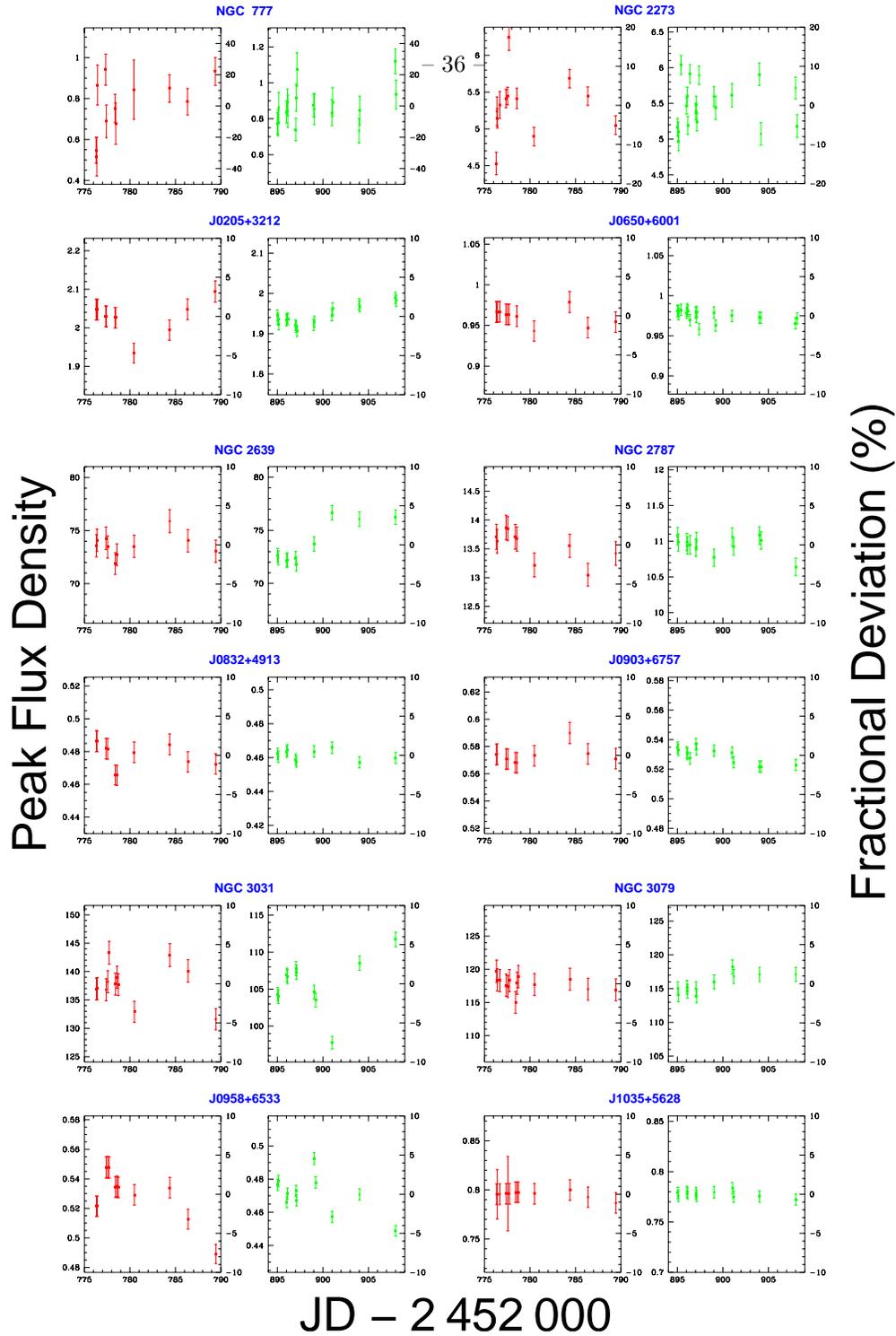}
\caption{\footnotesize Target time series for NGC~{777}, NGC~{2273}, NGC~{2639},
  NGC~{2787}, NGC~{3031}, and NGC~{3079}.  The peak flux density is
  shown as a function of time for six target galaxies, with the
  left-hand plot of each object (red points) displaying 2003 May data,
  and the right-hand plot (green points) showing 2003 September data.
  For each target galaxy (NGC objects), the corresponding phase
  calibrator time series is shown immediately below the target plots.
  The horizontal axis indicates the number of days since Julian Date
  $2\,452\,000$.  The vertical axis on the left-hand side of each
  individual plot shows the peak flux density of the measurement, in
  units of millijanskys per beam for the target galaxy and in units
  of janskys per beam for the phase calibrators.  The vertical axis on
  the right-hand side of each plot gives the relative difference from
  the mean value for the month.  The fractional scale extends from
  $-10$\% to $+10$\% for all objects except NGC~{777} and NGC~2273.
  Error-bars show the 1-$\sigma$ uncertainty in the measurements,
  including random errors and the estimated systematic errors.
  \label{fig:tar_time_1} }
\end{figure}

\begin{figure}
\epsscale{0.8}
\figurenum{\ref{fig:tar_time_0}b}
\plotone{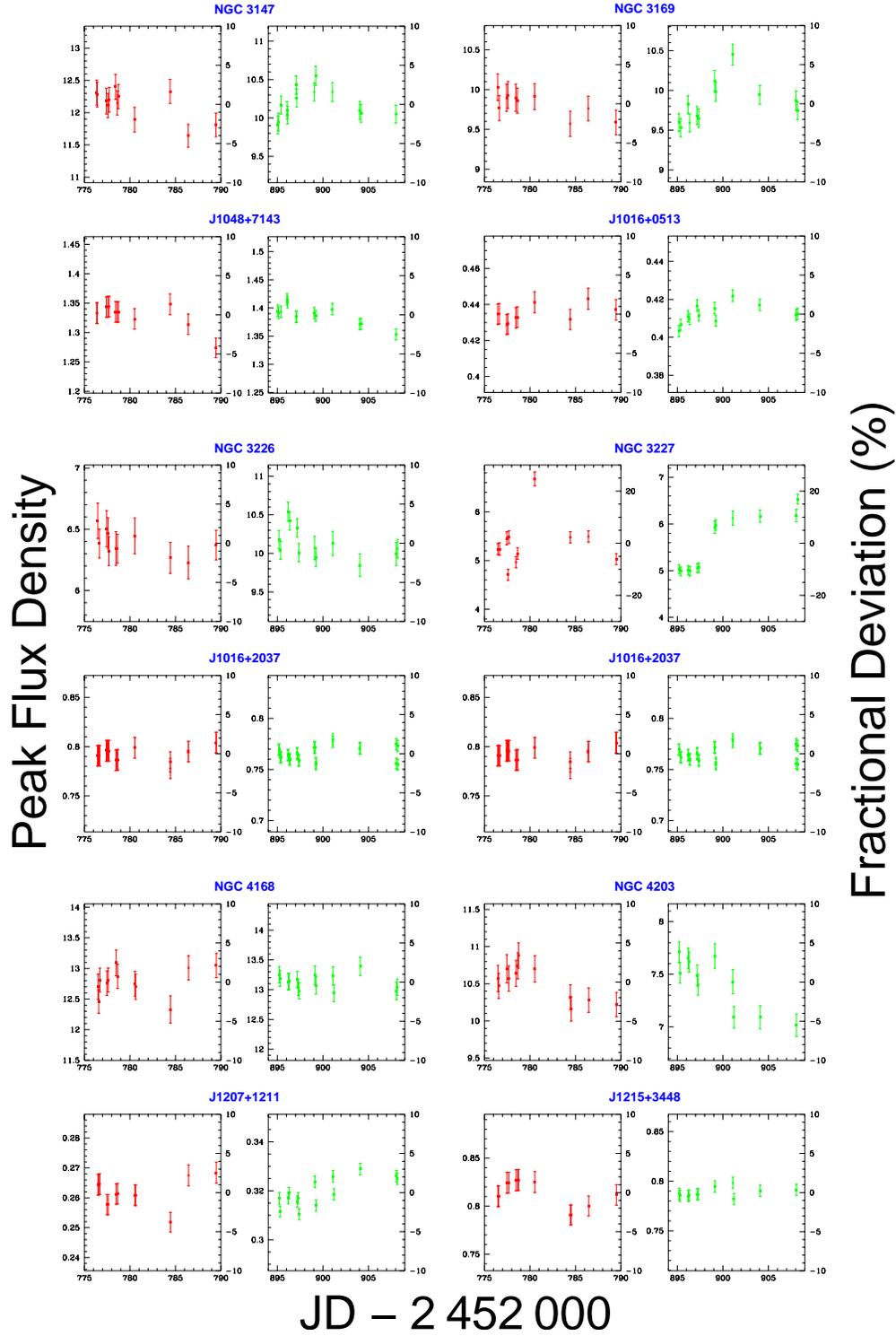}
\caption{Target time series for NGC~{3147}, NGC~{3169}, NGC~{3226},
  NGC~{3227}, NGC~{4168}, and NGC~{4203}.  Details are the same as
  Figure~2a.  The fractional scale extends from
  $-10$\% to $+10$\% for all objects except NGC~{3227}.
  \label{fig:tar_time_2} }
\end{figure}

\begin{trivlist}
  
\item[\textbf{CSO \objectname[COINS]{J0410$+$7656}}] This CSO is very
  resolved by our VLA observations.  Although following the $(u,v)$
  restrictions suggested by the VLA Calibrator
  Manual
  results in measurements which are approximately constant with
  varying hour angle, large changes from day to day with antennas
  moves prevented this source from being useful as an amplitude
  calibration tool.

\item[\textbf{CSO \objectname[COINS]{J0427$+$4133}}] The time series
  of this CSO during our September run suggests that variability may
  be present in this object.  The $\chi^2$ analysis suggests that the
  probability of seeing the measured scatter is less than 5\% for a
  constant source, and the coherent nature of the variation further
  supports this.  VLBA observations by \citet{Peck_T2000} at 8.4~GHz
  show that the CSO is dominated by a bright central component with a
  peak flux density 72\% of the flux density measured by our VLA
  observations, suggesting that variability may indeed be likely in
  this object.
  
\item[\textbf{CSO \objectname[COINS]{J1823$+$7938}}]
  \citetalias{Fassnacht_T2001} found that this CSO had an RMS scatter
  of $1.3$\%, double the scatter of most of the other CSOs in
  their survey.  Our measurements place
  \objectname[COINS]{J1823$+$7938} in the middle range of scatter,
  with two CSOs better and five CSOs worse than
  \objectname[COINS]{J1823$+$7938}.  However, Figure~3 of
  \citetalias{Fassnacht_T2001} shows that
  \objectname[COINS]{J1823$+$7938} had extended periods of stability
  during their long time series, and our short time series may have
  fortuitously taken place during similar periods of minimal
  variability.

\item[\textbf{\NGC{777}}] VLA observations of \NGC{777} show a point
  source with position RA $02^\mathrm{h}~00^\mathrm{m}~14\fs907$ Dec
  $31\degr~25\arcmin~45\farcs90$, with an uncertainty of about 25~mas
  in each dimension.  No emission was detected in the VLBA data which
  was at least $5$-$\sigma$ above the RMS noise level within 200~mas
  of this position.  No object could be identified with imaging
  performed out to 1\arcsec\ from this position.  Corrected for
  decoherence, this corresponds to an upper limit of 180~$\mathrm{\mu
    Jy~beam^{-1}}$.  Various weighting schemes were applied to the
  $(u,v)$ data to search for emission on scales up to 6~mas in
  diameter without success.  VLA measurements were made on the day
  before and the day after the September VLBA observations, but showed
  rapid variability on the prior day, with peak flux densities varying
  from $0.7$ to $1.1$~mJy~beam$^{-1}$.  Measurements in 2003 May were
  as low as $0.5$~mJy~beam$^{-1}$.  We are confident that the
  variability detected by the VLA is real, and therefore that a
  compact source with brightness temperature of more than $10^{10}$~K
  is normally present in this galaxy.  We suspect that this LLAGN was
  simply too weak to be detected by the VLBA during our observing run.

\item[\textbf{\NGC{2273}}] This Seyfert~2 galaxy has a steep spectrum
  at VLA scales ($\alpha^{5.0}_{8.5} = -0.5$).  Our VLBA imaging shows
  an elongated multiple component structure aligned East-West on the
  sky.  The area shown in Figure~\ref{fig:vlba_images} contains an
  integrated flux density of 2.4~mJy, but analysis from running CLEAN
  suggests that as much as 10~mJy of emission may be present on large
  scales within 40~mas of the region shown, with the surface
  brightness lying below the noise level of the image.  The peak
  brightness temperature of $4\times 10^6$~K is calculated from the
  peak flux density value using the area of the synthesized beam.
  \citet{Lal_SG2004} show results from a 5~GHz VLBI observation made
  on 1998 February~18.  In contrast to our observations, they find
  only a single, slightly resolved component with a peak of
  7.5~mJy~beam$^{-1}$.  It is possible that an outburst has occurred
  since the \citeauthor{Lal_SG2004} observations --- in the $\sim
  5.6$~years between observations, a jet component traveling at $c$ in
  the plane of the sky could reach 11~mas from the core, well within
  the limits from our VLBA image.  The VLA flux density changes appear
  to be correlated with $(u,v)$ changes, and we do not include this
  object in our short-term variability statistics.

\item[\textbf{\NGC{2639}}] This Seyfert~1.9 galaxy is known to be
  variable on timescales of several years.  \citet{Ho_U2001} found a
  spectral index $\alpha^{1.4}_{5.0} = +0.47$ (1.4~GHz data from 1999
  August 29, 5.0~GHz data from 1999 October 31).  The \citet{Ho_U2001}
  measurements were significantly brighter than previous measurements,
  consistent with possible resolution effects and/or variability in
  the core.  \NGC{2639} was also observed with the VLBA by
  \citet{Wilson_ea1998} on 1996 May~31 at 1.7, 5.0, and 15~GHz,
  finding $\alpha^{1.7}_{5.0} = +1.8$ and $\alpha^{5.0}_{15} = -0.04$.
  These observations found the emission to be unresolved at 1.7 and
  5.0~GHz, but somewhat resolved at 15~GHz, with a deconvolved size of
  $0.70$~mas $\times < 0.15$~mas with PA 111\degr.  Our 8.4~GHz
  results are consistent with a two-component source, with a 34~mJy
  unresolved core and a 34~mJy resolved component with size $1.94$~mas
  $\times < 0.1$~mas at PA 110\degr.  Taken at face value, this size
  implies a brightness temperature of at least $10^{9.8}$~K, and
  suggests that an expanding jet-like feature is present in this
  galaxy (but is also marginally consistent with our $0.04$~mas$^2$
  solid angle result).  However, the $0.1$~mas upper limit to the
  width is almost a factor of 20 smaller than the beam width; a more
  realistic upper limit is $0.5$~mas, which corresponds to a minimum
  brightness temperature of $10^{9.1}$~K.  \citet{Lal_SG2004} show
  results from a 5~GHz VLBI observation made on 1998 February~18 which
  also find a slightly resolved source at position angle $\sim
  90\degr$.  The VLA flux density changes appear to be correlated with
  $(u,v)$ changes, and we do not include this object in our short-term
  variability statistics.

\item[\textbf{\NGC{2787}}]  \citet{Falcke_ea2000} found a peak flux
  density of 11.2~mJy~beam$^{-1}$ in their 5.0~GHz VLBA observations
  from 1997 June 16.  Combined with 15~GHz VLA observations by
  \citet{Nagar_ea2000} from 1996 October, they calculated a spectral
  index of $\alpha^{5.0}_{15} = -0.45$.  Our 8.4~GHz VLBA measurement
  is brighter than both of these previous observations.  Our
  observation has a peak to RMS noise ratio of almost 400, but shows
  no sign of any extended emission.  The JMFIT deconvolved size is
  only about 0.20~mas.
  
\item[\textbf{\NGC{3031}}] This galaxy is more commonly known as
  \object[Messier]{M81}.  The compact VLBA core contains virtually all
  of the radio flux seen on VLA scales.  \citet{Bietenholz_ea1996} and
  \citet{Bietenholz_BR2000} use these observations to measure the size
  of the emission region of \NGC{3031}, finding a mean result of $530
  \pm 100$~$\mu$as $\times$ $180 \pm 40$~$\mu$as (at position angle
  50\degr) at 8.4~GHz for their single component model.  (The two
  component model is very similar.)  This measurement suggests a
  brightness temperature of about $10^{10.3}$~K.  Given the factor of
  5 uncertainty in any individual galaxy, our interstellar
  scintillation angular size result of about 100~$\mu$as is in
  remarkably good agreement with their findings.  If some fraction of
  the variability of \NGC{3031} is \emph{intrinsic} to the source,
  this would \emph{increase} the interstellar scattering size
  estimate, bringing our value into even closer agreement with
  \citeauthor{Bietenholz_BR2000}
  
  This nearby \citep[3.55~Mpc,][]{Freedman_ea2001} galaxy has been
  known to contain a compact, variable radio nucleus for many years.
  \citet{Crane_GC1976} and \citet{deBruyn_ea1976} present radio
  measurements from 1967 to 1975 showing that \NGC{3031} had varied by
  up to a factor of $\sim 2$ over that time.  Following the \SNJ\ 
  explosion, the nucleus of \NGC{3031} was repeatedly observed as part
  of the \SNJ\ monitoring campaign, as it was located within the
  primary beam area of the VLA and VLBA telescopes targeting \SNJ.
  \citet[hereafter \citetalias{Ho_ea1999}]{Ho_ea1999} present flux
  densities for \NGC{3031} at 1.4, 4.9, 8.4, and 15.2~GHz starting
  3~days after the supernova explosion and extending to almost
  1400~days after the explosion.  Since the supernova is located about
  170\arcsec\ away from the nucleus, there is no problem with
  confusion.  However, changing array configurations, uncertainty in
  the flux density calibrations (CSOs were generally not observed),
  and other problems cause the uncertainty levels in the individual
  measurements to be far higher than the measurements presented in
  this work.  Variability statistics are presented in
  Table~\ref{tab:M81}; a $P_{\chi^2}$ analysis gives a probability of
  only $10^{-13}$ that the nuclear emission was constant.  The 8.4~GHz
  radio light-curves in \citetalias{Ho_ea1999} suggest that \NGC{3031}
  has periodic ``outbursts'' which can double the emission levels,
  separated by periods of relative ``quiet''.  The high scatter in
  Table~\ref{tab:M81} reflects the strong outburst periods.  The
  observations of \NGC{3031} presented in this work show far lower
  variability levels, but are consistent with the ``quiet'' period
  scatter levels.  Assuming that the variability in
  \citetalias{Ho_ea1999} is intrinsic, the brightness temperature
  limits are higher than the values in Table~\ref{tab:var_cat}, but
  still consistent with inverse-Compton limits.  The 90\% confidence
  estimate to the brightness temperature lower limit is actually
  consistent with the equipartition brightness temperature limit of
  $\sim 10^{11}$~K.
  
  Figure~\ref{fig:M81_struc_func} shows structure function plots for
  the \citetalias{Ho_ea1999} data.  The 1.4 and 4.9~GHz data show no
  significant intraday variability.  The 8.4~GHz data show significant
  variability down to $\sim 0.5$~days.  If the 8.4~GHz data are
  interpreted in terms of interstellar scintillation, the source
  angular size predictions from both the initial maximum in the
  structure function and the de-biased measurement scatter are about
  12~$\mu$as at 8.4~GHz.  However, we agree with the conclusion by
  \citetalias{Ho_ea1999} that the large amplitude variations seen in
  their data are almost certainly intrinsic to \NGC{3031}, as the
  ``outbursts'' are visible in all frequencies and are nearly
  coincident in time.  Furthermore, the flux density measurements of
  \SNJ, only 170\arcsec\ away, show no evidence for large amplitude
  fluctuations over a period of almost four years.  We find it very
  unlikely that any phase screen covering \NGC{3031} for this time
  period would never affect the supernova emission as well.  However,
  the cause of the few percent, ``quiet'' intraday variability
  detected in this work could still be either intrinsic or extrinsic
  to the source itself.
  
\item[\textbf{\NGC{3079}}] This galaxy contains complicated structure
  on milliarcsecond scales along with water maser emission.
  \citet{Sawada_Satoh_ea2000} present VLBA images of this galaxy from
  1996 October~20 observations showing a resolved object with multiple
  components.  They found over 18~mJy in three milliarcsecond scale
  components at $8.4$~GHz, which is far less than the $\sim 120$~mJy
  we observed on arcsecond scales with the VLA.  The VLBI nuclear
  component has a sharply peaked spectrum, with $\alpha^{8.4}_{15}
  \approx +0.9$ at low frequencies and $\alpha^{15}_{22} = -1.8$ at
  high frequencies.  \citet{Ho_U2001} found a peak spectral index of
  $\alpha^{1.4}_{4.9} = +0.2$ on arcsecond scales.  The VLA scale
  emission is also extended and shows a small jet feature in our
  full-$(u,v)$ images.  The extended emission in this galaxy did not
  cause any $(u,v)$ problems, as we found \NGC{3079} to remain
  constant with an observed scatter of only $\sim 1$\%.
  
\item[\textbf{\NGC{3147}}] This galaxy was unresolved in the VLBA
  imaging of \citet{Ulvestad_H2001.1} and \citet[hereafter
  \citetalias{Anderson_UH2004}]{Anderson_UH2004}.
  
\item[\textbf{\NGC{3169}}] \citet{Falcke_ea2000} found a peak flux
  density of 6.2~mJy~beam$^{-1}$ in their 5.0~GHz VLBA observations
  from 1997 June 16, while \citet{Nagar_ea2000} found
  6.8~mJy~beam$^{-1}$ in their 15~GHz VLA A-array image.  Our VLBA
  image had a peak of $8.6$~mJy with a peak to RMS noise ratio of
  $\sim 200$, giving a 3-$\sigma$ upper limit to the size of the
  unresolved core of 0.65~mas.  The variability seen in our VLA
  observations has some features which correlate with the phase
  calibrator behavior, and we classify this galaxy as possibly
  variable.
  
\item[\textbf{\NGC{3226}}] This LINER~1.9 galaxy was previously
  observed by \citet{Falcke_ea2000} with the VLBA at 5.0~GHz, finding
  a peak flux density of 3.5~mJy~beam$^{-1}$.  Our 8.4~GHz VLBA
  measurement is more than 2 times brighter.  Note that our measured
  VLBA position is almost 0\farcs2 north of the position found by
  \citet{Falcke_ea2000}, but is in agreement with our own 8.5~GHz VLA
  position and is reasonably close to the 15~GHz VLA position of
  \citet{Nagar_ea2000}.  The self-calibrated image shows a
  core-dominated structure with extensions to the East and West.  The
  structure can be modeled as an unresolved core with a peak flux
  density of 7.9~mJy~beam$^{-1}$ and an elongated structure 1.5~mas
  away at PA 74\degr\ with a peak flux density of 0.4~mJy~beam$^{-1}$.
  \citet{Falcke_ea2000} list a very uncertain position angle of
  64\degr.  Thus, all of the emission is contained within a region
  $1.5$~mas $\times < 0.5$~mas in size, corresponding to a brightness
  temperature of at least $10^{8.3}$~K.  Extended emission seen in our
  full-$(u,v)$ VLA images is removed by our $(u,v)$ weighting, and
  does not affect our variability measurements.
  
\item[\textbf{\NGC{3227}}] This Seyfert~1.5 galaxy has the steepest
  radio spectral index in our galaxy sample, with $\alpha^{1.7}_{5.0}
  =-0.9$ as measured by \citet{Mundell_ea1995} using MERLIN.  Their
  data show a resolved structure aligned at PA $-10\degr$, with a flux
  density of 8~mJy at 5~GHz.  Although the 5~GHz peak flux density
  measured by MERLIN was greater than 2~mJy~beam$^{-1}$, our 8.4~GHz
  measurements find no detectable emission on VLBI scales.  The
  5-$\sigma$ upper limit to any compact emission is
  260$\mu$Jy~beam$^{-1}$.  Our VLA measurements are severely
  handicapped by brightness changes caused by $(u,v)$ effects, and we
  do not include this object in our short-term variability study.

\item[\textbf{\NGC{4168}}]  This galaxy appears unresolved in the VLBA
  imaging of \citetalias{Anderson_UH2004}, and appears to be constant
  in our short-term variability analysis.  
  
\item[\textbf{\NGC{4203}}] This galaxy was also essentially unresolved
  by \citet{Ulvestad_H2001.1} and \citetalias{Anderson_UH2004} in
  their VLBA imaging.  We classify the May VLA measurements as
  tentative as they have some similarity to the phase calibrator
  behavior.  However, the September observations show a significant
  decline with time which we believe is reliable.

\item[\textbf{\NGC{4235}}]  Essentially unresolved in the VLBA imaging
  of \citetalias{Anderson_UH2004}, this galaxy shows modest
  variability in the September VLA data.
  
\item[\textbf{\NGC{4450}}] Although essentially unresolved in the VLBA
  study of \citetalias{Anderson_UH2004}, this galaxy has about 10~mJy
  of extended emission in our VLA images, about twice as much as is
  present in the core.  Because of the extensive extended emission, we
  classify the variability as tentative, although we find no signs of
  $(u,v)$ effects in the flux variations.

\item[\textbf{\NGC{4472}}] The imaging of this Seyfert~2 galaxy is
  consistent with an unresolved point source.  The source is extended
  on VLA scales, but the peak flux density in our VLA imaging is
  similar to the peak flux density in our VLBA image.  Variations in
  our VLA observations appear highly correlated with $(u,v)$ changes,
  and we do not include this galaxy in our short-term variability
  analysis.  
  
\item[\textbf{\NGC{4565}}] This Seyfert~1.9 galaxy is relatively weak
  in our VLBA imaging with a peak of only $1.9$~mJy; self-calibration
  only partially corrected residual phase errors from the original
  calibration steps.  At most we can only say that the object appears
  to be compact on VLBI scales.  \citet{Falcke_ea2000} found a peak
  brightness of 3.2~mJy~beam$^{-1}$ in their 1997 June 16 observations
  at 5.0~GHz, and 15~GHz VLA A-array imaging by \citet{Nagar_ea2000}
  found 3.7~mJy~beam$^{-1}$.  Although the September $P_{\chi^2}$
  value of $0.06$ is above our variability cutoff level,
  Figure~\ref{fig:tar_time_3} (in the on-line paper) clearly shows
  variations on a $\sim 4$~day timescale.  This is borne out in
  Figure~\ref{fig:target_struc_func} which shows that the structure
  function values for time lags of several days are well above the
  noise level, even if the entire variability does not meet meet our
  $\chi^2$ test.
  
\item[\textbf{\NGC{4579}}] This galaxy was essentially unresolved by
  the VLBA study of \citet{Ulvestad_H2001.1} and
  \citetalias{Anderson_UH2004}.  We have classified the short-term
  variability as tentative since the May data could possible have
  $(u,v)$ effects and the September data could possibly be influenced
  by the phase calibration.
  
\item[\textbf{\NGC{5866}}] This galaxy was classified as having a
  transition nucleus (intermediate between an \ion{H}{2} nucleus and a
  LINER nucleus) by \citet{Ho_FS1997.0}.  \citet{Falcke_ea2000} found
  a slightly resolved core with a peak flux density of
  7.0~mJy~beam$^{-1}$ and a very uncertain position angle of 11\degr\ 
  in their 5.0~GHz VLBA observations.  Our 8.4~GHz VLBA observations
  show a similar result, with a deconvolved core size of about
  $1.9$~mas $\times < 0.2$~mas with a position angle of 12\degr.
  Unfortunately, the beam size for our observations was $2.2$~mas
  $\times 1.0$~mas at PA 6\degr\ --- nearly aligned with the possible
  source structure.  The galaxy is at high declination, so a full
  12$^\mathrm{h}$ observing run could better resolve the core.  The
  May short-term variability appears to be reliable.  However, the
  September variability classification effectively depends on the
  single low point on the second day of observations, and we classify
  this target-month dataset as tentative.

\end{trivlist}




\bibliography{jma_astro}

\begin{thebibliography}{94}
\expandafter\ifx\csname natexlab\endcsname\relax\def\natexlab#1{#1}\fi

\bibitem[{{Anderson} {et~al.}(2004){Anderson}, {Ulvestad}, \&
  {Ho}}]{Anderson_UH2004}
{Anderson}, J.~M., {Ulvestad}, J.~S., \& {Ho}, L.~C. 2004, \apj, 603, 42,
  ({A04})

\bibitem[{{Barth} {et~al.}(1998){Barth}, {Ho}, {Filippenko}, \&
  {Sargent}}]{Barth_ea1998}
{Barth}, A.~J., {Ho}, L.~C., {Filippenko}, A.~V., \& {Sargent}, W.~L.~W. 1998,
  \apj, 496, 133

\bibitem[{{Barth} {et~al.}(2002){Barth}, {Ho}, \& {Sargent}}]{Barth_HS2002}
{Barth}, A.~J., {Ho}, L.~C., \& {Sargent}, W.~L.~W. 2002, \aj, 124, 2607

\bibitem[{{Barvainis} {et~al.}(2004){Barvainis}, {Leh\'ar}, {Birkinshaw},
  {Falke}, \& {Blundell}}]{Barvainis_ea2004}
{Barvainis}, R., {Leh\'ar}, J., {Birkinshaw}, M., {Falke}, H., \& {Blundell},
  K.~M. 2004, \apj, submitted, ({B04})

\bibitem[{{Beasley} \& {Conway}(1995)}]{Beasley_C1995}
{Beasley}, A.~J., \& {Conway}, J.~E. 1995, in ASP Conf. Ser. 82: Very Long
  Baseline Interferometry and the VLBA, ed. J.~A. {Zensus}, P.~J. {Diamond}, \&
  P.~J. {Napier} (San Francisco: ASP), 328

\bibitem[{{Beasley} {et~al.}(2002){Beasley}, {Gordon}, {Peck}, {Petrov},
  {MacMillan}, {Fomalont}, \& {Ma}}]{Beasley_ea2002}
{Beasley}, A.~J., {Gordon}, D., {Peck}, A.~B., {Petrov}, L., {MacMillan},
  D.~S., {Fomalont}, E.~B., \& {Ma}, C. 2002, \apjs, 141, 13

\bibitem[{{Beckert} {et~al.}(2002){Beckert}, {Fuhrmann}, {Cim{\` o}},
  {Krichbaum}, {Witzel}, \& {Zensus}}]{Beckert_ea2002}
{Beckert}, T., {Fuhrmann}, L., {Cim{\` o}}, G., {Krichbaum}, T.~P., {Witzel},
  A., \& {Zensus}, J.~A. 2002, in Proceedings of the 6th EVN Symposium, ed.
  E.~{Ros}, R.~W. {Porcas}, A.~P. {Lobanov}, \& J.~A. {Zensus} (Bonn:
  Max-Planck-Institut fuer Radioastronomie), 79

\bibitem[{{Begelman} {et~al.}(1994){Begelman}, {Rees}, \&
  {Sikora}}]{Begelman_RS1994}
{Begelman}, M.~C., {Rees}, M.~J., \& {Sikora}, M. 1994, \apjl, 429, L57

\bibitem[{{Bietenholz} {et~al.}(2000){Bietenholz}, {Bartel}, \&
  {Rupen}}]{Bietenholz_BR2000}
{Bietenholz}, M.~F., {Bartel}, N., \& {Rupen}, M.~P. 2000, \apj, 532, 895

\bibitem[{{Bietenholz} {et~al.}(1996)}]{Bietenholz_ea1996}
{Bietenholz}, M.~F., {et~al.} 1996, \apj, 457, 604

\bibitem[{{Bower} {et~al.}(2004){Bower}, {Falcke}, {Herrnstein}, {Zhao},
  {Goss}, \& {Backer}}]{Bower_ea2004}
{Bower}, G.~C., {Falcke}, H., {Herrnstein}, R.~M., {Zhao}, J., {Goss}, W.~M.,
  \& {Backer}, D.~C. 2004, Science, 304, 704

\bibitem[{{Condon}(1992)}]{Condon1992}
{Condon}, J.~J. 1992, \araa, 30, 575

\bibitem[{{Cordes} \& {Lazio}(2002)}]{Cordes_L2002}
{Cordes}, J.~M., \& {Lazio}, T.~J.~W. 2002, (astro-ph/0207156)

\bibitem[{{Cornwell} \& {Fomalont}(1999)}]{Cornwell_F1999}
{Cornwell}, T., \& {Fomalont}, E. 1999, in ASP Conf. Ser. 180: Synthesis
  Imaging in Radio Astronomy II, ed. G.~B. {Taylor}, C.~L. {Carilli}, \& R.~A.
  {Perley} (San Francisco: ASP), 187

\bibitem[{{Crane} {et~al.}(1976){Crane}, {Giuffrida}, \&
  {Carlson}}]{Crane_GC1976}
{Crane}, P.~C., {Giuffrida}, T.~S., \& {Carlson}, J.~B. 1976, \apjl, 203, L113

\bibitem[{{de Bruyn} {et~al.}(1976){de Bruyn}, {Crane}, {Price}, \&
  {Carlson}}]{deBruyn_ea1976}
{de Bruyn}, A.~G., {Crane}, P.~C., {Price}, R.~M., \& {Carlson}, J.~B. 1976,
  \aap, 46, 243

\bibitem[{{de Vaucouleurs} {et~al.}(1991){de Vaucouleurs}, {de Vaucouleurs},
  {Corwin}, {Buta}, {Paturel}, \& {Fouque}}]{deVaucouleurs_ea1991}
{de Vaucouleurs}, G., {de Vaucouleurs}, A., {Corwin}, H.~G., {Buta}, R.~J.,
  {Paturel}, G., \& {Fouque}, P. 1991, {Third Reference Catalogue of Bright
  Galaxies} (New York: Springer-Verlag)

\bibitem[{{Dennett-Thorpe} \& {de Bruyn}(2001)}]{Dennett-Thorpe_d2001}
{Dennett-Thorpe}, J., \& {de Bruyn}, A.~G. 2001, \apss, 278, 101

\bibitem[{{Dennett-Thorpe} \& {de Bruyn}(2002)}]{Dennett-Thorpe_d2002}
---. 2002, \nat, 415, 57

\bibitem[{{Duncan} {et~al.}(1993){Duncan}, {White}, {Wark}, {Reynolds},
  {Jauncey}, {Norris}, \& {Savage}}]{Duncan_ea1993}
{Duncan}, R.~A., {White}, G.~L., {Wark}, R., {Reynolds}, J.~E., {Jauncey},
  D.~L., {Norris}, R.~P., \& {Savage}, L.~T.~A. 1993, \pasa, 10, 310

\bibitem[{{Falcke} \& {Biermann}(1999)}]{Falcke_B1999}
{Falcke}, H., \& {Biermann}, P.~L. 1999, \aap, 342, 49

\bibitem[{{Falcke} {et~al.}(2000){Falcke}, {Nagar}, {Wilson}, \&
  {Ulvestad}}]{Falcke_ea2000}
{Falcke}, H., {Nagar}, N.~M., {Wilson}, A.~S., \& {Ulvestad}, J.~S. 2000, \apj,
  542, 197

\bibitem[{{Fassnacht} \& {Taylor}(2001)}]{Fassnacht_T2001}
{Fassnacht}, C.~D., \& {Taylor}, G.~B. 2001, \aj, 122, 1661, ({F}T01)

\bibitem[{{Frail} {et~al.}(1997){Frail}, {Kulkarni}, {Nicastro}, {Feroci}, \&
  {Taylor}}]{Frail_ea1997}
{Frail}, D.~A., {Kulkarni}, S.~R., {Nicastro}, S.~R., {Feroci}, M., \&
  {Taylor}, G.~B. 1997, \nat, 389, 261

\bibitem[{{Freedman} {et~al.}(2001)}]{Freedman_ea2001}
{Freedman}, W.~L., {et~al.} 2001, \apj, 553, 47

\bibitem[{{Gallimore} {et~al.}(1999){Gallimore}, {Baum}, {O'Dea}, {Pedlar}, \&
  {Brinks}}]{Gallimore_ea1999}
{Gallimore}, J.~F., {Baum}, S.~A., {O'Dea}, C.~P., {Pedlar}, A., \& {Brinks},
  E. 1999, \apj, 524, 684

\bibitem[{{H{\' e}raudeau} \& {Simien}(1998)}]{Heraudeau_S1998}
{H{\' e}raudeau}, P., \& {Simien}, F. 1998, \aaps, 133, 317

\bibitem[{{Heeschen}(1984)}]{Heeschen_1984}
{Heeschen}, D.~S. 1984, \aj, 89, 1111

\bibitem[{{Heeschen} {et~al.}(1987){Heeschen}, {Krichbaum}, {Schalinski}, \&
  {Witzel}}]{Heeschen_ea1987}
{Heeschen}, D.~S., {Krichbaum}, T., {Schalinski}, C.~J., \& {Witzel}, A. 1987,
  \aj, 94, 1493

\bibitem[{{Ho}(2002)}]{Ho_2002}
{Ho}, L.~C. 2002, \apj, 564, 120

\bibitem[{{Ho} {et~al.}(1997{\natexlab{a}}){Ho}, {Filippenko}, \&
  {Sargent}}]{Ho_FS1997.0}
{Ho}, L.~C., {Filippenko}, A.~V., \& {Sargent}, W.~L.~W. 1997{\natexlab{a}},
  \apjs, 112, 315

\bibitem[{{Ho} {et~al.}(1997{\natexlab{b}}){Ho}, {Filippenko}, \&
  {Sargent}}]{Ho_FS1997.1}
---. 1997{\natexlab{b}}, \apj, 487, 568

\bibitem[{{Ho} \& {Peng}(2001)}]{Ho_P2001}
{Ho}, L.~C., \& {Peng}, C.~Y. 2001, \apj, 555, 650

\bibitem[{{Ho} \& {Ulvestad}(2001)}]{Ho_U2001}
{Ho}, L.~C., \& {Ulvestad}, J.~S. 2001, \apjs, 133, 77

\bibitem[{{Ho} {et~al.}(1999){Ho}, {van Dyk}, {Pooley}, {Sramek}, \&
  {Weiler}}]{Ho_ea1999}
{Ho}, L.~C., {van Dyk}, S.~D., {Pooley}, G.~G., {Sramek}, R.~A., \& {Weiler},
  K.~W. 1999, \aj, 118, 843, ({H99})

\bibitem[{{Jauncey} {et~al.}(2003){Jauncey}, {Bignall}, {Lovell},
  {Kedziora-Chudczer}, {Tzioumis}, {Macquart}, \& {Rickett}}]{Jauncey_ea2003.0}
{Jauncey}, D.~L., {Bignall}, H.~E., {Lovell}, J.~E.~J., {Kedziora-Chudczer},
  L., {Tzioumis}, A.~K., {Macquart}, J.-P., \& {Rickett}, B.~J. 2003, in ASP
  Conf. Ser. 300: Radio Astronomy at the Fringe, ed. J.~A. {Zensus}, M.~H.
  {Cohen}, \& E.~{Ros} (San Francisco: ASP), 199

\bibitem[{{Jim{\' e}nez-Benito} {et~al.}(2000){Jim{\' e}nez-Benito},
  {D{\'{\i}}az}, {Terlevich}, \& {Terlevich}}]{Jimenez-Benito_ea2000}
{Jim{\' e}nez-Benito}, L., {D{\'{\i}}az}, A.~I., {Terlevich}, R., \&
  {Terlevich}, E. 2000, \mnras, 317, 907

\bibitem[{{Kedziora-Chudczer} {et~al.}(1997){Kedziora-Chudczer}, {Jauncey},
  {Wieringa}, {Walker}, {Nicolson}, {Reynolds}, \&
  {Tzioumis}}]{Kedziora_Chudczer_ea1997}
{Kedziora-Chudczer}, L., {Jauncey}, D.~L., {Wieringa}, M.~H., {Walker}, M.~A.,
  {Nicolson}, G.~D., {Reynolds}, J.~E., \& {Tzioumis}, A.~K. 1997, \apjl, 490,
  L9+

\bibitem[{{Kedziora-Chudczer} {et~al.}(2001){Kedziora-Chudczer}, {Jauncey},
  {Wieringa}, {Tzioumis}, \& {Reynolds}}]{Kedziora_Chudczer_ea2001}
{Kedziora-Chudczer}, L.~L., {Jauncey}, D.~L., {Wieringa}, M.~H., {Tzioumis},
  A.~K., \& {Reynolds}, J.~E. 2001, \mnras, 325, 1411

\bibitem[{{Kellermann} \& {Pauliny-Toth}(1968)}]{Kellermann_P1968}
{Kellermann}, K.~I., \& {Pauliny-Toth}, I.~I.~K. 1968, \araa, 6, 417

\bibitem[{{Kellermann} {et~al.}(1989){Kellermann}, {Sramek}, {Schmidt},
  {Shaffer}, \& {Green}}]{Kellermann_ea1989}
{Kellermann}, K.~I., {Sramek}, R., {Schmidt}, M., {Shaffer}, D.~B., \& {Green},
  R. 1989, \aj, 98, 1195

\bibitem[{{Kormendy} \& {Gebhardt}(2001)}]{Kormendy_G2001}
{Kormendy}, J., \& {Gebhardt}, K. 2001, in AIP Conf. Proc. 586: 20th Texas
  Symposium on relativistic astrophysics, ed. J.~C. {Wheeler} \& H.~{Martel}
  (Melville, NY: AIP), 363

\bibitem[{{Lal} {et~al.}(2004){Lal}, {Shastri}, \& {Gabuzda}}]{Lal_SG2004}
{Lal}, D.~V., {Shastri}, P., \& {Gabuzda}, D.~C. 2004, \aap, in press
  (astro-ph/0406597)

\bibitem[{{Lovell} {et~al.}(2003){Lovell}, {Jauncey}, {Bignall},
  {Kedziora-Chudczer}, {Macquart}, {Rickett}, \& {Tzioumis}}]{Lovell_ea2003}
{Lovell}, J.~E.~J., {Jauncey}, D.~L., {Bignall}, H.~E., {Kedziora-Chudczer},
  L., {Macquart}, J.-P., {Rickett}, B.~J., \& {Tzioumis}, A.~K. 2003, \aj, 126,
  1699

\bibitem[{{Ma} {et~al.}(1998)}]{Ma_ea1998}
{Ma}, C., {et~al.} 1998, \aj, 116, 516

\bibitem[{{Mahadevan}(1997)}]{Mahadevan_1997}
{Mahadevan}, R. 1997, \apj, 477, 585

\bibitem[{{Maoz} {et~al.}(1996){Maoz}, {Filippenko}, {Ho}, {Macchetto}, {Rix},
  \& {Schneider}}]{Maoz_ea1996}
{Maoz}, D., {Filippenko}, A.~V., {Ho}, L.~C., {Macchetto}, F.~D., {Rix}, H., \&
  {Schneider}, D.~P. 1996, \apjs, 107, 215

\bibitem[{{McElroy}(1995)}]{McElroy_1995}
{McElroy}, D.~B. 1995, \apjs, 100, 105

\bibitem[{{Mundell} {et~al.}(1995){Mundell}, {Holloway}, {Pedlar}, {Meaburn},
  {Kukula}, \& {Axon}}]{Mundell_ea1995}
{Mundell}, C.~G., {Holloway}, A.~J., {Pedlar}, A., {Meaburn}, J., {Kukula},
  M.~J., \& {Axon}, D.~J. 1995, \mnras, 275, 67

\bibitem[{{Nagar} {et~al.}(2000){Nagar}, {Falcke}, {Wilson}, \&
  {Ho}}]{Nagar_ea2000}
{Nagar}, N.~M., {Falcke}, H., {Wilson}, A.~S., \& {Ho}, L.~C. 2000, \apj, 542,
  186

\bibitem[{{Nagar} {et~al.}(2002){Nagar}, {Falcke}, {Wilson}, \&
  {Ulvestad}}]{Nagar_ea2002}
{Nagar}, N.~M., {Falcke}, H., {Wilson}, A.~S., \& {Ulvestad}, J.~S. 2002, \aap,
  392, 53

\bibitem[{{Narayan}(1992)}]{Narayan_1992}
{Narayan}, R. 1992, Phil. Trans. R. Soc. London, 341, 151

\bibitem[{{Narayan} {et~al.}(1998){Narayan}, {Mahadevan}, \&
  {Quataert}}]{Narayan_MQ1998}
{Narayan}, R., {Mahadevan}, R., \& {Quataert}, E. 1998, in Theory of Black Hole
  Accretion Disks, ed. M.~A. {Abramowicz}, G.~{Bjornsson}, \& J.~E. {Pringle}
  (Cambridge: Cambridge University Press), 148

\bibitem[{{Narayan} \& {Yi}(1994)}]{Narayan_Y1994}
{Narayan}, R., \& {Yi}, I. 1994, \apjl, 428, L13

\bibitem[{{Nelson} \& {Whittle}(1995)}]{Nelson_W1995}
{Nelson}, C.~H., \& {Whittle}, M. 1995, \apjs, 99, 67

\bibitem[{{Peck} \& {Taylor}(2000)}]{Peck_T2000}
{Peck}, A.~B., \& {Taylor}, G.~B. 2000, \apj, 534, 90

\bibitem[{{Prugniel} {et~al.}(2001)}]{Prugniel_ea2001}
{Prugniel}, P., {et~al.} 2001, in Mining the Sky, ed. A.~J. Banday, S.~Zaroubi,
  \& M.~Bartelmann (Heidelberg: Springer), 683

\bibitem[{{Quirrenbach}(1992)}]{Quirrenbach_1992}
{Quirrenbach}, A. 1992, Reviews of Modern Astronomy, 5, 214

\bibitem[{{Quirrenbach} {et~al.}(1992)}]{Quirrenbach_ea1992}
{Quirrenbach}, A., {et~al.} 1992, \aap, 258, 279

\bibitem[{{Ravindranath} {et~al.}(2001){Ravindranath}, {Ho}, {Peng},
  {Filippenko}, \& {Sargent}}]{Ravindranath_ea2001}
{Ravindranath}, S., {Ho}, L.~C., {Peng}, C.~Y., {Filippenko}, A.~V., \&
  {Sargent}, W.~L.~W. 2001, \aj, 122, 653

\bibitem[{{Readhead}(1994)}]{Readhead_1994}
{Readhead}, A.~C.~S. 1994, \apj, 426, 51

\bibitem[{{Rickett}(2002)}]{Rickett_2002}
{Rickett}, B. 2002, \pasa, 19, 100

\bibitem[{{Rickett}(1990)}]{Rickett_1990}
{Rickett}, B.~J. 1990, \araa, 28, 561

\bibitem[{{Rickett} {et~al.}(1995){Rickett}, {Quirrenbach}, {Wegner},
  {Krichbaum}, \& {Witzel}}]{Rickett_ea1995}
{Rickett}, B.~J., {Quirrenbach}, A., {Wegner}, R., {Krichbaum}, T.~P., \&
  {Witzel}, A. 1995, \aap, 293, 479

\bibitem[{{Rickett} {et~al.}(2001){Rickett}, {Witzel}, {Kraus}, {Krichbaum}, \&
  {Qian}}]{Rickett_ea2001}
{Rickett}, B.~J., {Witzel}, A., {Kraus}, A., {Krichbaum}, T.~P., \& {Qian},
  S.~J. 2001, \apjl, 550, L11

\bibitem[{{Sawada-Satoh} {et~al.}(2000){Sawada-Satoh}, {Inoue}, {Shibata},
  {Kameno}, {Migenes}, {Nakai}, \& {Diamond}}]{Sawada_Satoh_ea2000}
{Sawada-Satoh}, S., {Inoue}, M., {Shibata}, K.~M., {Kameno}, S., {Migenes}, V.,
  {Nakai}, N., \& {Diamond}, P.~J. 2000, \pasj, 52, 421

\bibitem[{{Simien} \& {Prugniel}(2002)}]{Simien_P2002}
{Simien}, F., \& {Prugniel}, P. 2002, \aap, 384, 371

\bibitem[{{Simonetti} {et~al.}(1985){Simonetti}, {Cordes}, \&
  {Heeschen}}]{Simonetti_CH1985}
{Simonetti}, J.~H., {Cordes}, J.~M., \& {Heeschen}, D.~S. 1985, \apj, 296, 46

\bibitem[{{Sincell} \& {Krolik}(1994)}]{Sincell_K1994}
{Sincell}, M.~W., \& {Krolik}, J.~H. 1994, \apj, 430, 550

\bibitem[{{Solanes} {et~al.}(2002){Solanes}, {Sanchis}, {Salvador-Sol{\' e}},
  {Giovanelli}, \& {Haynes}}]{Solanes_ea2002}
{Solanes}, J.~M., {Sanchis}, T., {Salvador-Sol{\' e}}, E., {Giovanelli}, R., \&
  {Haynes}, M.~P. 2002, \aj, 124, 2440

\bibitem[{{Taylor} \& {Cordes}(1993)}]{Taylor_C1993}
{Taylor}, J.~H., \& {Cordes}, J.~M. 1993, \apj, 411, 674

\bibitem[{{Terashima} \& {Wilson}(2003)}]{Terashima_W2003}
{Terashima}, Y., \& {Wilson}, A.~S. 2003, \apj, 583, 145

\bibitem[{{Terry} {et~al.}(2002){Terry}, {Paturel}, \& {Ekholm}}]{Terry_PE2002}
{Terry}, J.~N., {Paturel}, G., \& {Ekholm}, T. 2002, \aap, 393, 57

\bibitem[{{Thompson} {et~al.}(1980){Thompson}, {Clark}, {Wade}, \&
  {Napier}}]{Thompson_ea1980}
{Thompson}, A.~R., {Clark}, B.~G., {Wade}, C.~M., \& {Napier}, P.~J. 1980,
  \apjs, 44, 151

\bibitem[{{Tonry} {et~al.}(2001){Tonry}, {Dressler}, {Blakeslee}, {Ajhar},
  {Fletcher}, {Luppino}, {Metzger}, \& {Moore}}]{Tonry_ea2001}
{Tonry}, J.~L., {Dressler}, A., {Blakeslee}, J.~P., {Ajhar}, E.~A., {Fletcher},
  A.~B., {Luppino}, G.~A., {Metzger}, M.~R., \& {Moore}, C.~B. 2001, \apj, 546,
  681

\bibitem[{{Tremaine} {et~al.}(2002)}]{Tremaine_ea2002}
{Tremaine}, S., {et~al.} 2002, \apj, 574, 740

\bibitem[{{Trotter} {et~al.}(1998){Trotter}, {Greenhill}, {Moran}, {Reid},
  {Irwin}, \& {Lo}}]{Trotter_ea1998}
{Trotter}, A.~S., {Greenhill}, L.~J., {Moran}, J.~M., {Reid}, M.~J., {Irwin},
  J.~A., \& {Lo}, K. 1998, \apj, 495, 740

\bibitem[{{Tully}(1988)}]{Tully_1988}
{Tully}, R.~B. 1988, {Nearby galaxies catalog} ({Cambridge}: Cambridge
  University Press)

\bibitem[{{Tully} {et~al.}(1992){Tully}, {Shaya}, \& {Pierce}}]{Tully_SP1992}
{Tully}, R.~B., {Shaya}, E.~J., \& {Pierce}, M.~J. 1992, \apjs, 80, 479

\bibitem[{{Tutui} \& {Sofue}(1997)}]{Tutui_S1997}
{Tutui}, Y., \& {Sofue}, Y. 1997, \aap, 326, 915

\bibitem[{{Ulrich} {et~al.}(1997){Ulrich}, {Maraschi}, \&
  {Urry}}]{Ulrich_MU1997}
{Ulrich}, M., {Maraschi}, L., \& {Urry}, C.~M. 1997, \araa, 35, 445

\bibitem[{{Ulvestad}(2003)}]{Ulvestad_2003}
{Ulvestad}, J.~S. 2003, in ASP Conf. Ser. 300: Radio Astronomy at the Fringe,
  ed. J.~A. {Zensus}, M.~H. {Cohen}, \& E.~{Ros} (San Francisco: ASP), 97

\bibitem[{{Ulvestad} {et~al.}(2004){Ulvestad}, {Antonucci}, \&
  {Barvainis}}]{Ulvestad_AB2004}
{Ulvestad}, J.~S., {Antonucci}, R.~R.~J., \& {Barvainis}, R. 2004, \apj,
  submitted

\bibitem[{{Ulvestad} \& {Ho}(2001{\natexlab{a}})}]{Ulvestad_H2001.0}
{Ulvestad}, J.~S., \& {Ho}, L.~C. 2001{\natexlab{a}}, \apj, 558, 561

\bibitem[{{Ulvestad} \& {Ho}(2001{\natexlab{b}})}]{Ulvestad_H2001.1}
---. 2001{\natexlab{b}}, \apjl, 562, L133

\bibitem[{{van Moorsel} {et~al.}(1996){van Moorsel}, {Kemball}, \&
  {Greisen}}]{van_Moorsel_KG1996}
{van Moorsel}, G., {Kemball}, A., \& {Greisen}, E. 1996, in ASP Conf. Ser. 101:
  Astronomical Data Analysis Software and Systems V, ed. G.~H. {Jacoby} \&
  J.~{Barnes} (San Francisco: ASP), 37

\bibitem[{{Wagner} \& {Witzel}(1995)}]{Wagner_W1995}
{Wagner}, S.~J., \& {Witzel}, A. 1995, \araa, 33, 163

\bibitem[{{Walker}(1998)}]{Walker_1998}
{Walker}, M.~A. 1998, \mnras, 294, 307

\bibitem[{{Walker}(2001)}]{Walker_2001}
---. 2001, \mnras, 321, 176

\bibitem[{{Whitmore} {et~al.}(1985){Whitmore}, {McElroy}, \&
  {Tonry}}]{Whitmore_MT1985}
{Whitmore}, B.~C., {McElroy}, D.~B., \& {Tonry}, J.~L. 1985, \apjs, 59, 1

\bibitem[{{Wilson} {et~al.}(1998)}]{Wilson_ea1998}
{Wilson}, A.~S., {et~al.} 1998, \apj, 505, 587

\bibitem[{{Yuan} {et~al.}(2002{\natexlab{a}}){Yuan}, {Markoff}, \&
  {Falcke}}]{Yuan_MF2002}
{Yuan}, F., {Markoff}, S., \& {Falcke}, H. 2002{\natexlab{a}}, \aap, 383, 854

\bibitem[{{Yuan} {et~al.}(2002{\natexlab{b}}){Yuan}, {Markoff}, {Falcke}, \&
  {Biermann}}]{Yuan_ea2002}
{Yuan}, F., {Markoff}, S., {Falcke}, H., \& {Biermann}, P.~L.
  2002{\natexlab{b}}, \aap, 391, 139

\bibitem[{{Zensus}(1997)}]{Zensus_1997}
{Zensus}, J.~A. 1997, \araa, 35, 607

\end{thebibliography}

\bibliographystyle{apj}



\clearpage

\clearpage

\clearpage

\clearpage

\clearpage

\begin{figure}
\epsscale{0.8}
\figurenum{\ref{fig:tar_time_0}c}
\plotone{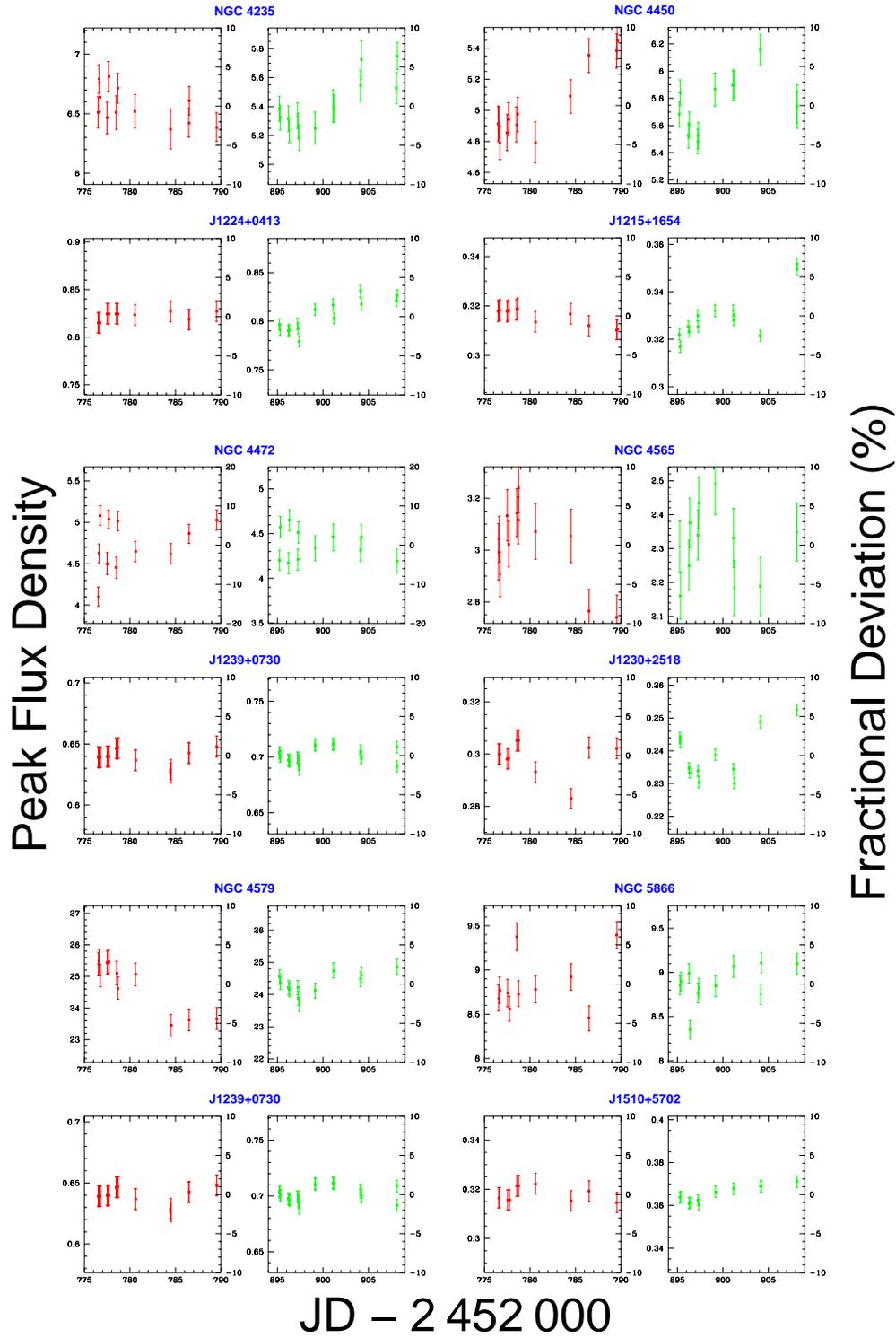}
\caption{Target time series for NGC~{4235}, NGC~{4450}, NGC~{4472},
  NGC~{4565}, NGC~{4579}, and NGC~{5866}.  Details are the same as
  Figure~2a.  The fractional scale extends from
  $-10$\% to $+10$\% for all objects except NGC~{4472}.
  \label{fig:tar_time_3} }
\end{figure}

\clearpage

\begin{figure}
\epsscale{1}
\plotone{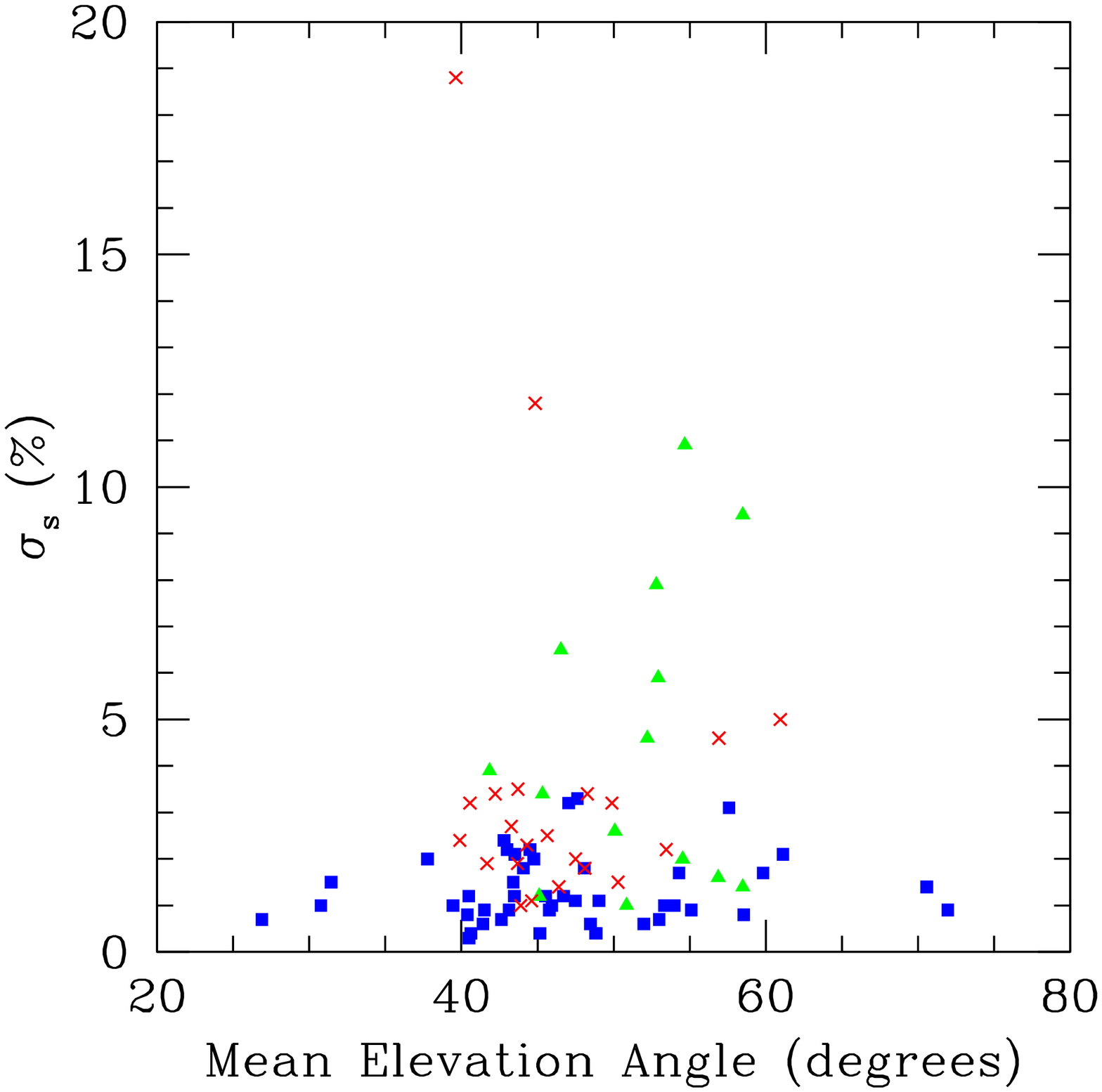}
\caption{RMS scatter versus mean elevation angle.  The fractional
  measured scatter of all our objects for the months of May and
  September is plotted against the mean elevation angle of the object
  for each month.  Calibrator objects are shown as (blue) squares.
  Extended target galaxies are shown as (green) triangles, and
  point-like target galaxies are shown as (red) crosses.  Objects
  observed at mostly low elevation angles do not exhibit higher levels
  of scatter, suggesting that self-calibration has been able to
  adequately compensate for variations in the atmosphere.  (See the
  on-line paper for a color version of this figure.)
  \label{fig:target_scatter_me} }
\end{figure}

\clearpage

\begin{figure}
\epsscale{1}
\plotone{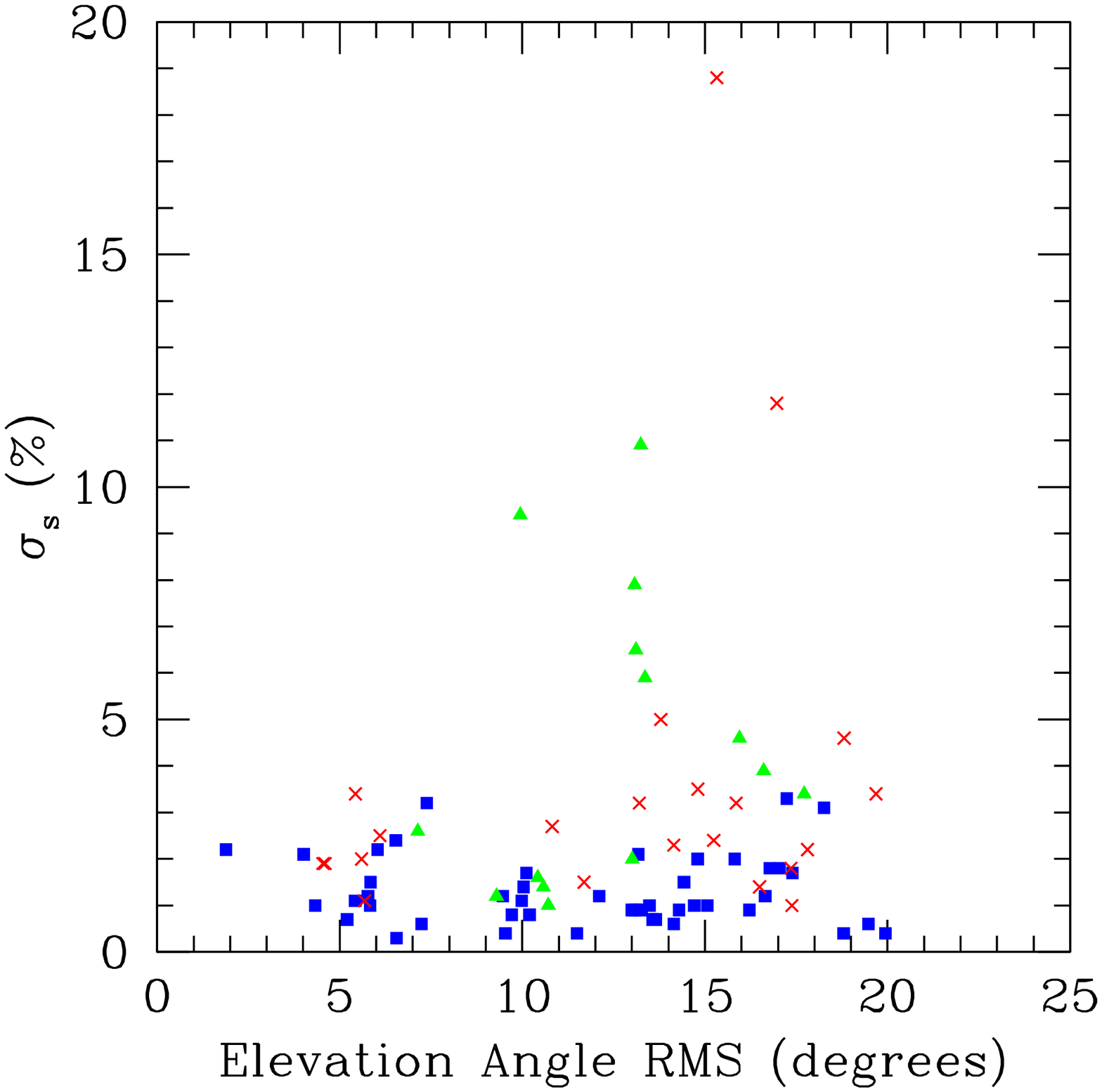}
\caption{RMS scatter versus RMS elevation angle.  The fractional
  measured scatter of all our objects for the months of May and
  September is plotted against the RMS variation in elevation angle of
  the object for each month.  Calibrator objects are shown as (blue)
  squares.  Extended target galaxies are shown as (green) triangles,
  and point-like target galaxies are shown as (red) crosses.  There is
  no significant trend for objects observed at more varied elevation
  angles (and therefore more substantially different $(u,v)$ coverage)
  to have larger amounts of scatter in their measured flux densities.
  (See the on-line paper for a color version of this figure.)
  \label{fig:target_scatter_se} }
\end{figure}

\clearpage

\begin{figure}
\epsscale{1}
\plotone{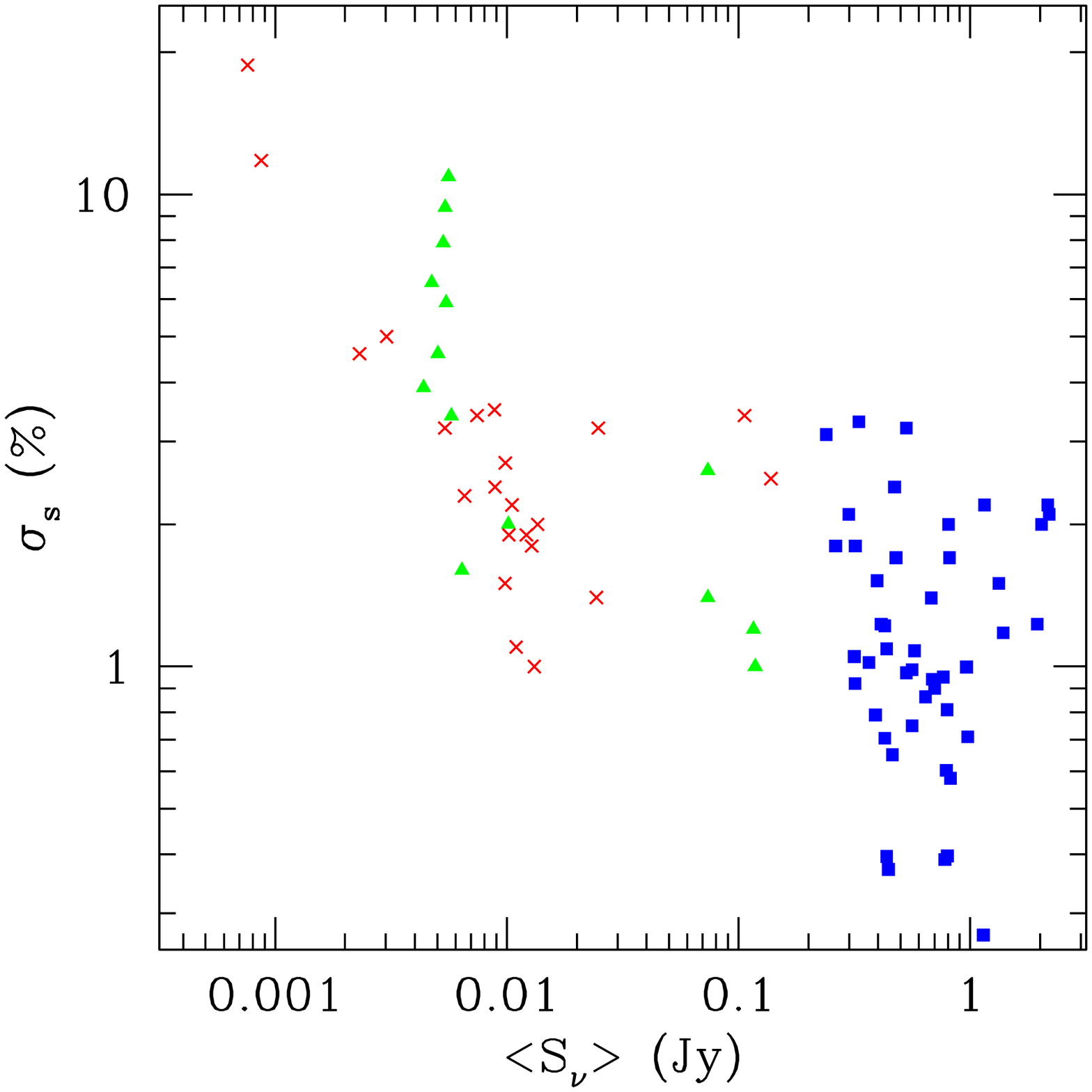}
\caption{RMS scatter versus mean flux density.  The fractional
  measured scatter of all our objects for the months of May and
  September is plotted against the mean flux density ($<S_\nu>$) for
  each month.  Calibrator objects are shown as (blue) squares.
  Extended target galaxies are shown as (green) triangles, and
  point-like target galaxies are shown as (red) crosses.  The objects
  with large variations in flux density all have mean flux densities
  less than about 8~mJy.  However, only the two weakest point-like
  target galaxies show scatter levels above 4\%, and it is unclear if
  this is a consequence of being unable to properly self-calibrate the
  weak targets NGC~777 and NGC~4565, or whether NGC~777 and NGC~4565
  are just far more variable that the other galaxies in this survey.
  (See the on-line paper for a color version of this figure.)
  \label{fig:target_scatter_fl} }
\end{figure}

\clearpage

\begin{figure}
\epsscale{0.93}
\vspace{-0.75in}
\plotone{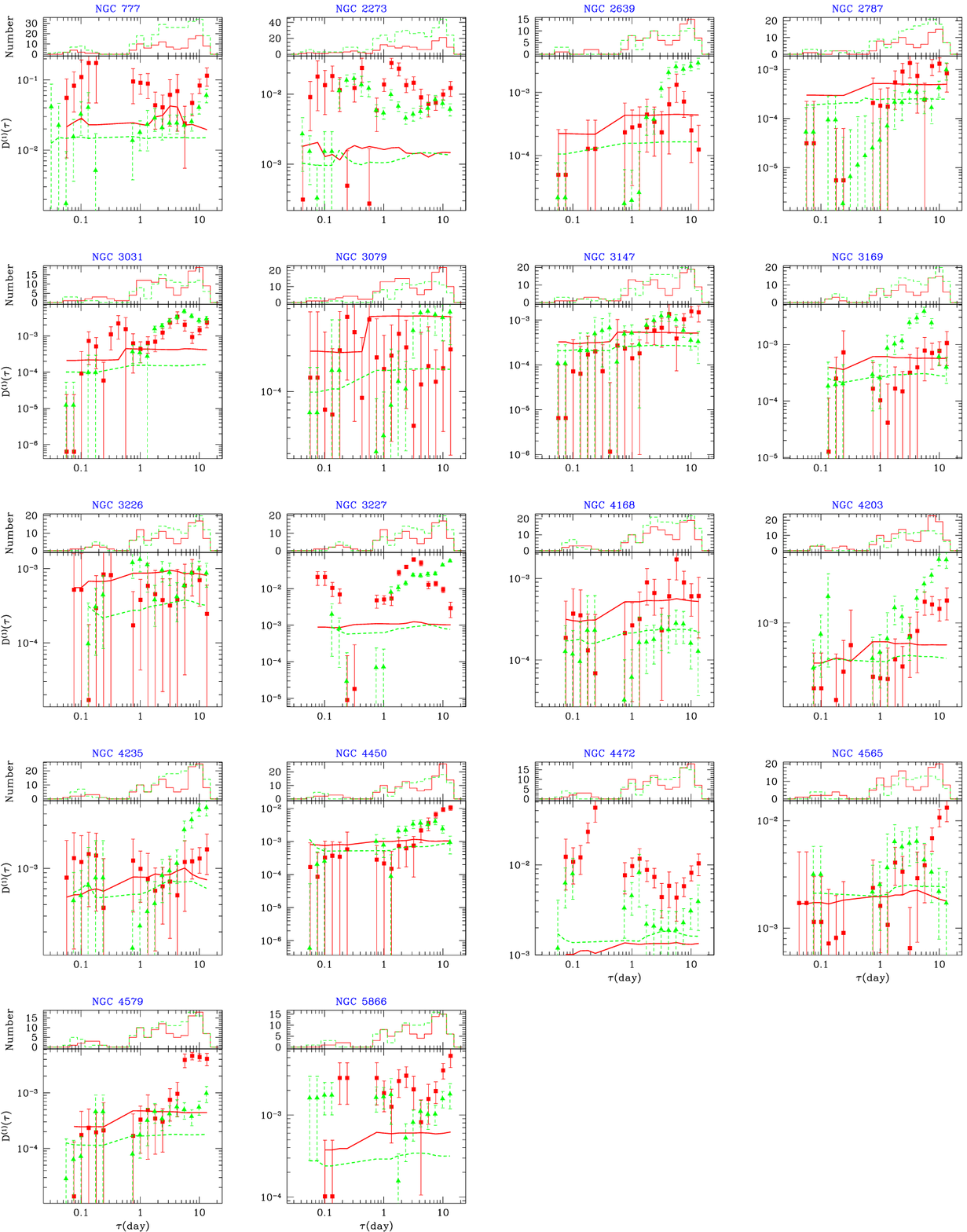}
\caption{Target galaxy variability structure functions.  The first
  order structure function is plotted for May (red squares) and
  September (green triangles) as a function of the lag time $\tau$.
  Error-bars indicate 1-$\sigma$ uncertainty levels.  The (mostly)
  horizontal solid (May, red) and dashed (September, green) lines in
  the structure function plot areas show the measurement error bias in
  the structure function, calculated from the estimated uncertainty
  levels of the individual data-points in each $\tau$ bin.  The
  measured structure functions have not been corrected for the
  measurement bias.  Just above the structure function plot is a
  histogram of the number of data point comparisons in each structure
  function point.  The May histogram is shown as the solid (red) line,
  while the September histogram is shown by the dashed (green) line.
  (See the on-line paper for a color version of this figure.)
  \label{fig:target_struc_func} }
\end{figure}

\clearpage

\begin{figure}
\plotone{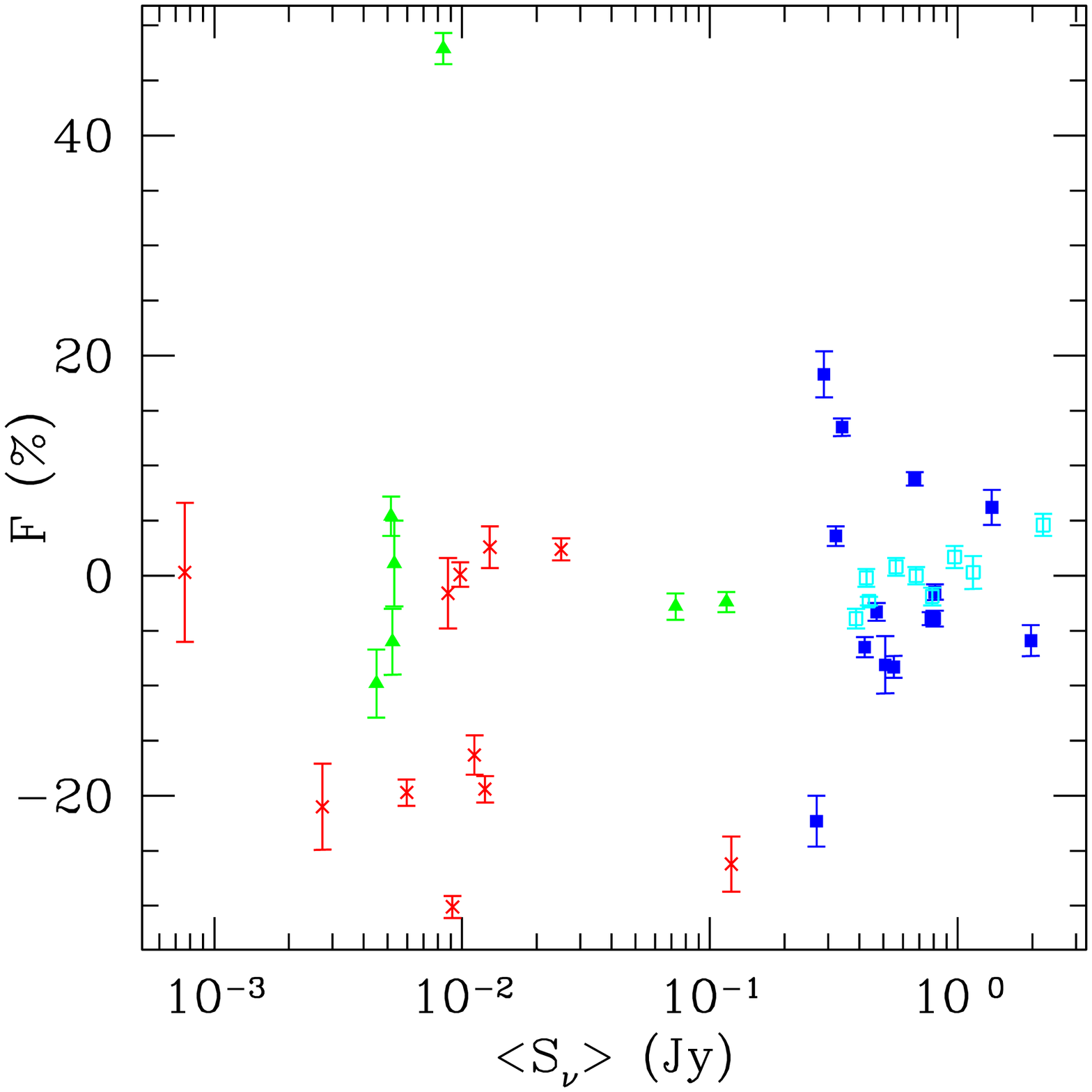}
\caption{The long-term fractional variation is plotted as a function
  of mean flux density.  Negative values indicate that the source was
  weaker in 2003 September than 2003 May.  Calibrator objects are
  shown as squares, with CSO calibrators indicated by open (cyan)
  squares and non-CSO calibrators by solid (blue) squares.  Extended
  target galaxies are shown as (green) triangles, and point-like
  target galaxies are shown as (red) crosses.  (See the on-line paper
  for a color version of this figure.)
  \label{fig:long_term_var_plot} }
\end{figure}
\clearpage

\begin{figure}
\plotone{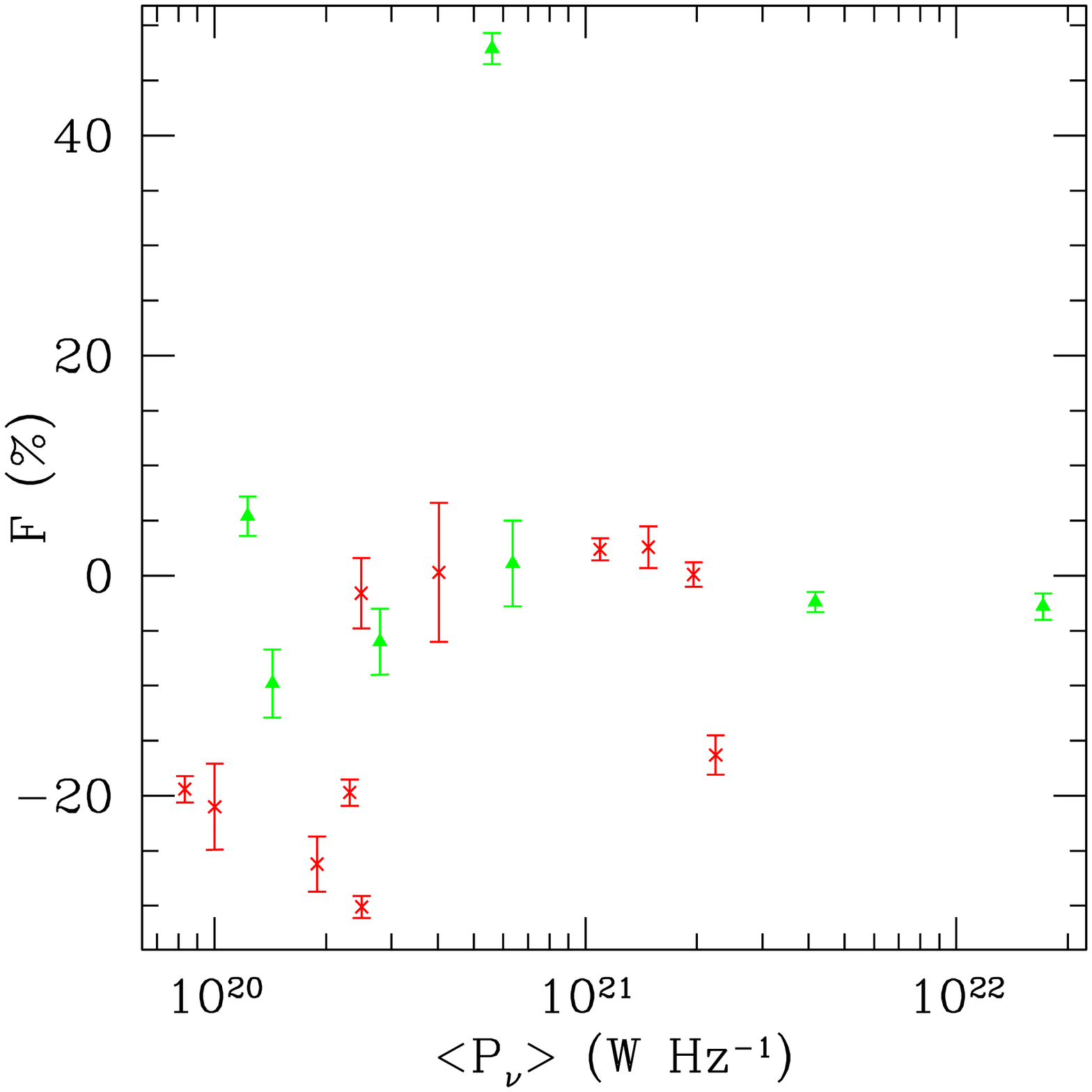}
\caption{The long-term fractional variation is plotted as a function
  of mean radio luminosity.  The radio luminosity ($P_\nu$) was
  calculated assuming isotropic radiation.  Extended target galaxies
  are shown as (green) triangles, and point-like target galaxies are
  shown as (red) crosses.  (See the on-line paper for a color version
  of this figure.)
  \label{fig:long_term_var_lum} }
\end{figure}

\clearpage

\begin{figure}
\plotone{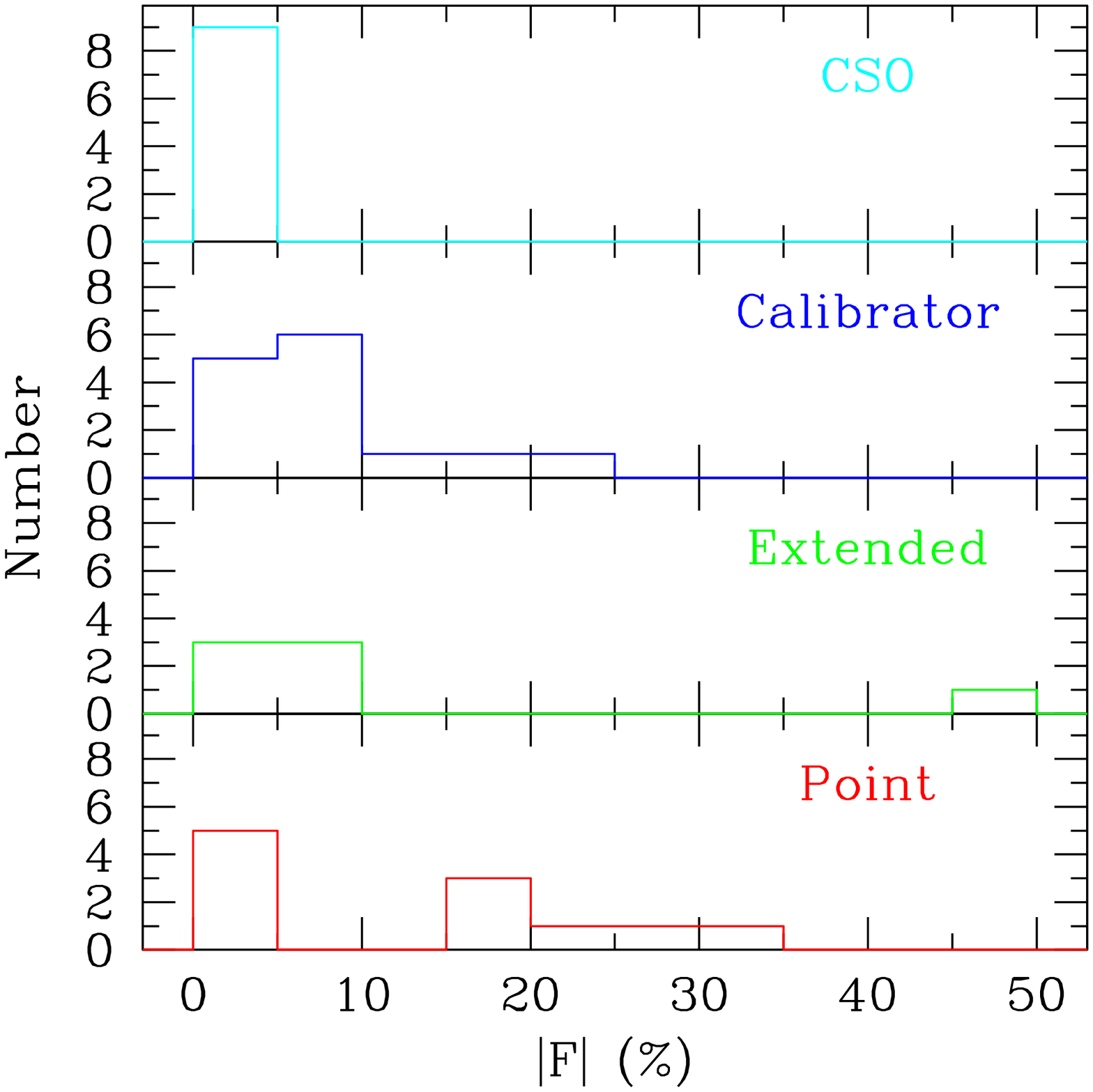}
\caption{Histogram of the number of sources as a function of the
  absolute long-term fractional variation.  CSO calibrators have been
  separated from non-CSO calibrators, and extended galaxies are
  separated from point-like galaxies.
  \label{fig:long_term_var_hist} }
\end{figure}

\clearpage

\begin{figure}
\epsscale{0.8}
\plotone{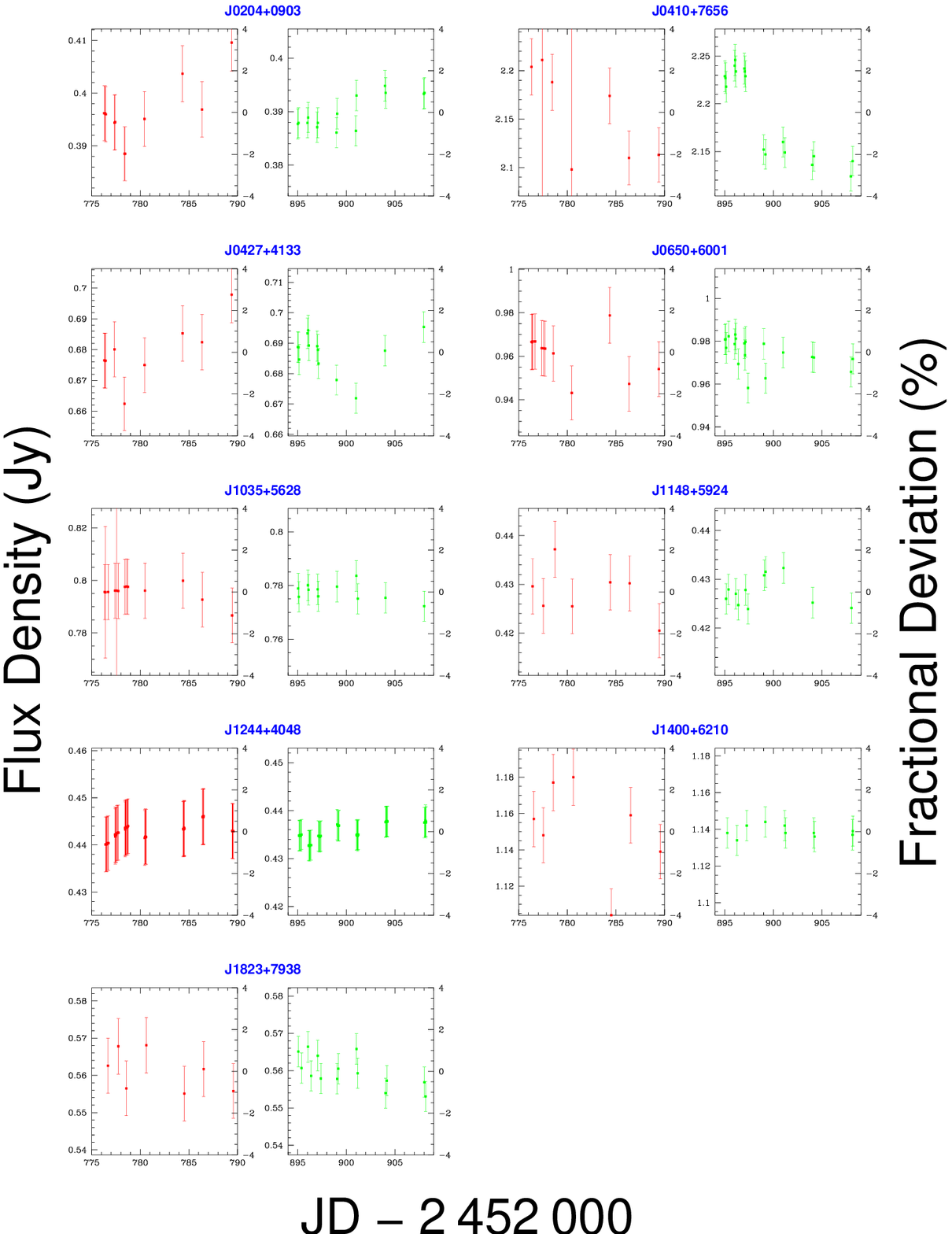}
\caption{CSO Time Series.  The calibrated flux densities for all 9 CSO
  calibrators are shown as a function of time.  Two plots are shown
  for each CSO.  The left-hand plots (red points in the online version)
  are for 2003 May, and the right-hand plots (green in the online
  version) are for 2003 September.  The horizontal axis indicates the
  number of days since Julian Date $2\,452\,000$.  The vertical axis
  on the left-hand side of each individual plot shows the flux density
  of the measurement.  The vertical axis on the right-hand side of
  each plot gives the relative difference from the mean value for the
  month.  The fractional scale is the same for all 9 CSOs.  Error-bars
  show the 1-$\sigma$ uncertainty in the measurements, including random
  errors and the estimated systematic error.  (See the on-line paper
  for a color version of this figure.)
  \label{fig:cso_var} }
\end{figure}

\clearpage

\begin{figure}
\plotone{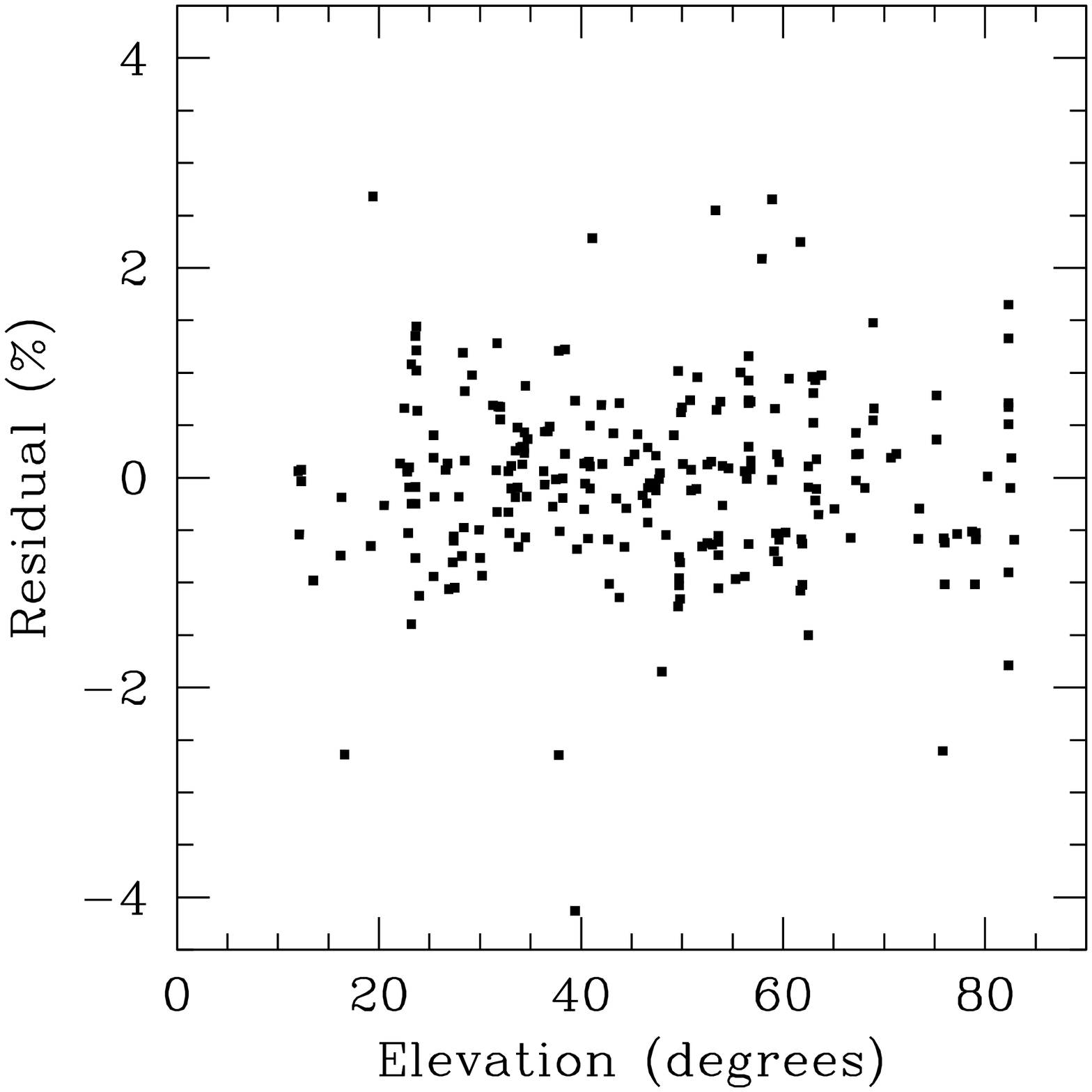}
\caption{The CSO residuals to the calibration fit are plotted as a
  function of elevation angle.  No significant trends are present in
  the data.
  \label{fig:cso_elev} }
\end{figure}

\clearpage

\begin{figure}
\plotone{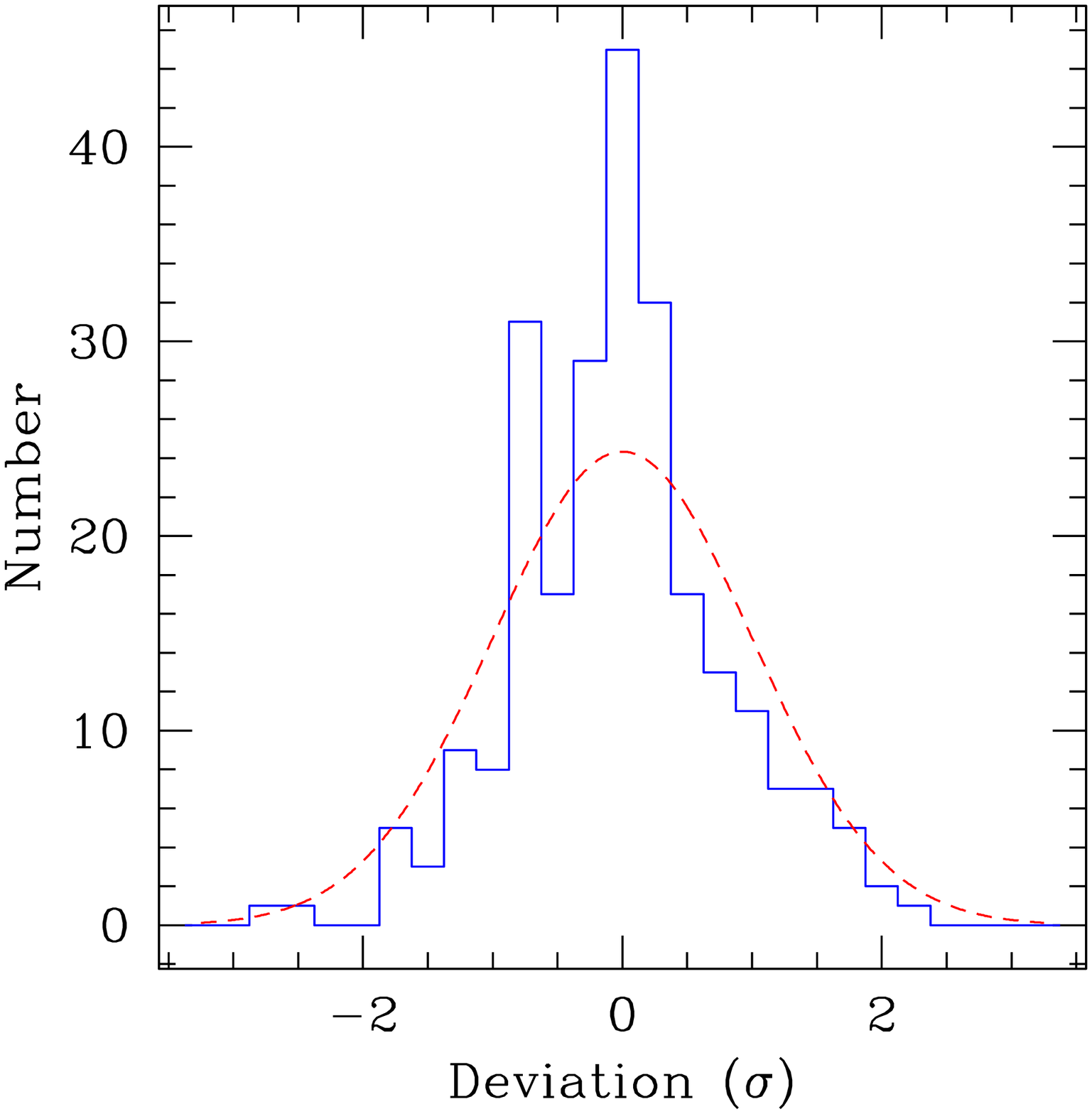}
\caption{Histogram of the CSO residuals to the calibration fit.  The
  solid (blue) line shows the deviation of each CSO measurement from
  the least squares calibration fit in units of the RMS scatter for
  the month of observation, with the May and September data combined.
  The dashed (red) line shows the corresponding Gaussian (normal)
  probability function.  (See the on-line paper for a color version of
  this figure.)
  \label{fig:cso_hist} }
\end{figure}

\clearpage

\begin{figure}
\plotone{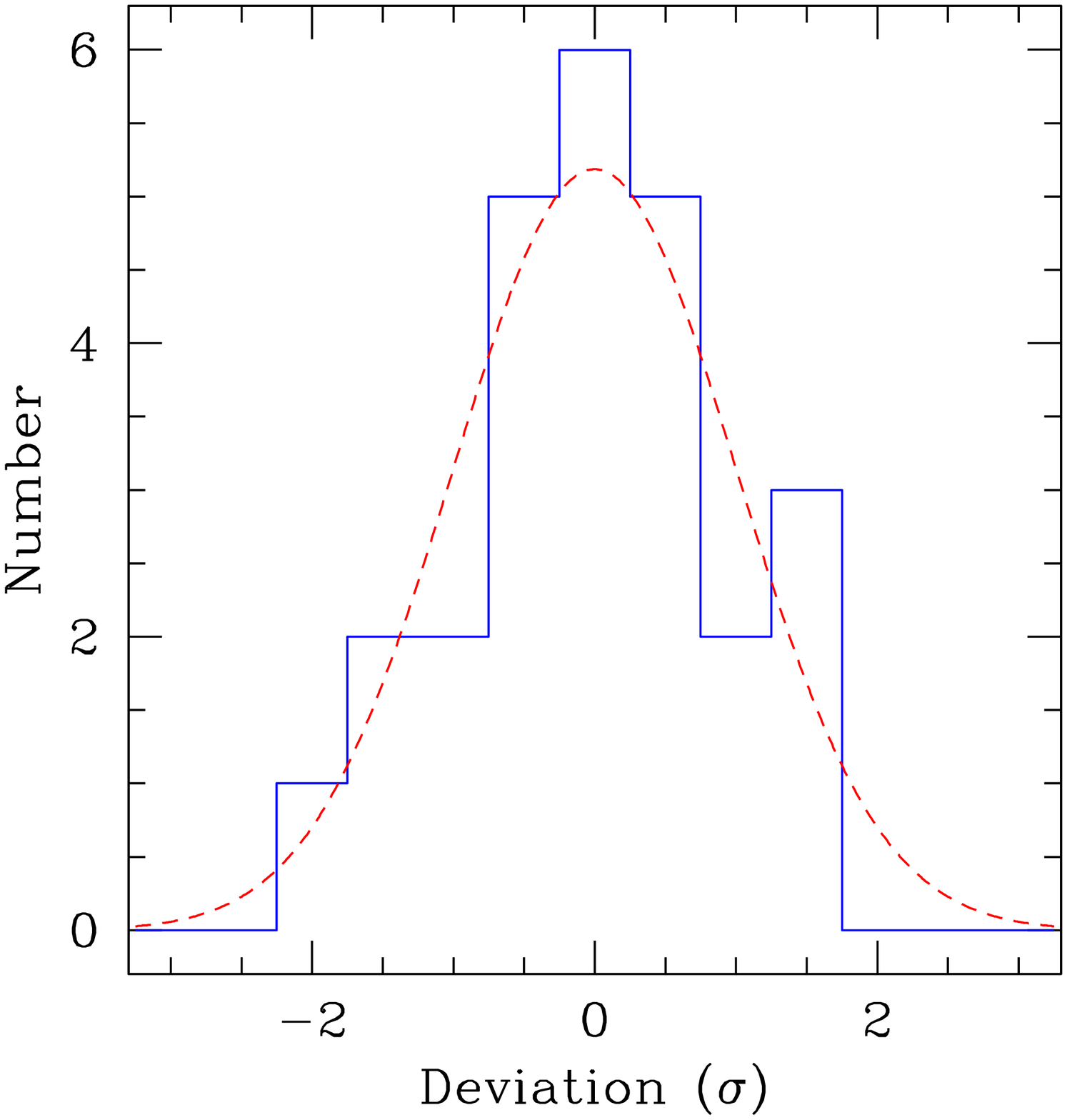}
\caption{NGC~4168 residual histogram.  The
  solid (blue) line shows a histogram of the residuals for the
  NGC~4168 measurements assuming a constant flux density, in units of
  the expected error level.  The dashed (red) line shows the
  corresponding Gaussian (normal) probability function.  (See the
  on-line paper for a color version of this figure.)
  \label{fig:NGC4168_hist} }
\end{figure}

\clearpage

\begin{figure}
\plotone{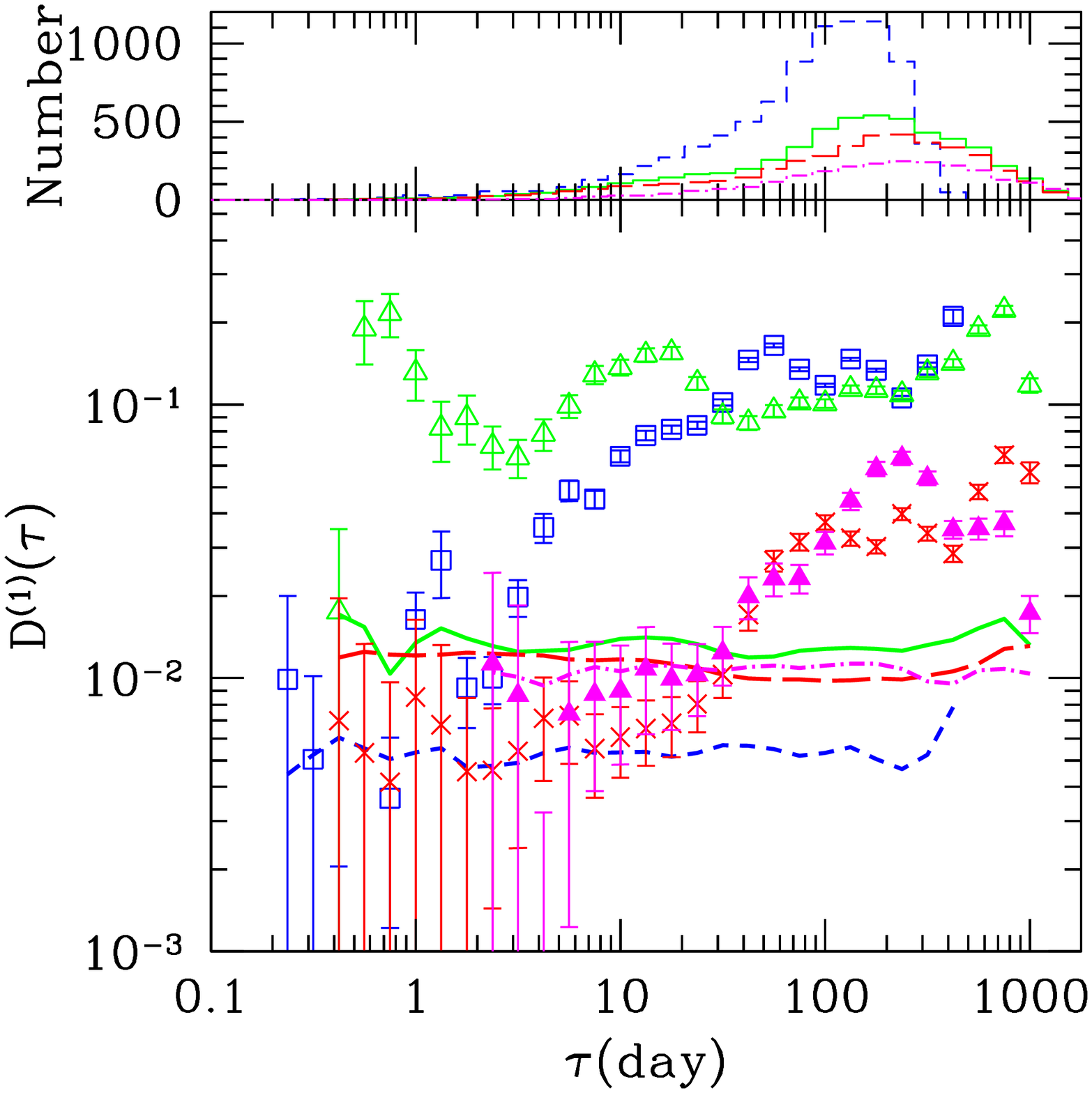}
\caption{Variability structure functions for NGC~3031 (M81) during
  1993--1997.  Flux density measurements of NGC~3031 following the
  supernova 1993J event taken with the VLA and the Ryle Telescope and
  presented in \citet{Ho_ea1999} have been used to generate structure
  functions covering time lags from less than one day to 1000~days.
  The plot information is similar to the plots in
  Figure~\ref{fig:target_struc_func}.  Data at 1.4~GHz are shown
  as solid (magenta) triangles and by the dot-dashed line.  Data at
  4.9~GHz are shown as (red) crosses and by the long-dashed lines.
  Data at 8.4~GHz are shown as the open (green) triangles and by the
  solid lines.  Data at 15.2~GHz are shown by the open (blue) squares
  and by the short-dashed lines.  (See the on-line paper for a color
  version of this figure.)
  \label{fig:M81_struc_func} }
\end{figure}

\clearpage

\begin{deluxetable}{lllllcccccccc}
  \rotate
  \tablecaption{Variability Sample\label{tab:sample}}
  \tabletypesize{\scriptsize}
  \tablewidth{0pt}
  \tablehead{
    \colhead{Galaxy} & \colhead{Hubble} & \colhead{AGN Type}
    & \colhead{RA} & \colhead{Dec} & \colhead{Ref}
    & \colhead{$M_\mathrm{BH}$} & \colhead{Ref}
    & \colhead{$D$} & \colhead{Ref}
    & \colhead{$\theta_{1000 R_\mathrm{S}}$}
    & \colhead{Phase} & \colhead{Offset}
    \\
    \colhead{} & \colhead{Type} & \colhead{}
    & \colhead{J2000} & \colhead{J2000} & \colhead{}
    & \colhead{(M$_\Sun$)} & \colhead{}
    & \colhead{(Mpc)} & \colhead{}
    & \colhead{($\mu$as)}
    & \colhead{Calibrator} & \colhead{(\degr)}
    \\
    \colhead{(1)} & \colhead{(2)} & \colhead{(3)} 
    & \colhead{(4)}  & \colhead{(5)} & \colhead{(6)}
    & \colhead{(7)}  & \colhead{(8)}
    & \colhead{(9)}  & \colhead{(10)}
    & \colhead{(11)}
    & \colhead{(12)} & \colhead{(13)}
    }
\startdata
\NGC{\phn 777} & E1          & S2/L2::  & 02 00 14.907 & 31 25 45.90 & 1 
  & $8.6\times 10^8$ & 6 & 66.5 & 14    & 255
  & \objectname[VLA_Cal]{J0205$+$3212} & 1.29
\\
\NGC{2273} & SB(r)a:     & S2       & 06 50 08.65752 & 60 50 44.9012&2
  & $2.0\times 10^7$ & 7 & 31.6 & 15    & \phn12
  & \objectname[VLA_Cal]{J0650$+$6001} & 0.82
\\
\NGC{2639} & (R)SA(r)a?  & S1.9     & 08 43 38.07792 & 50 12 20.0037&2
  & $1.1\times 10^8$ & 8 & 44.4 & 16    & \phn47
  & \objectname[VLA_Cal]{J0832$+$4913} & 2.07
\\
\NGC{2787} & SB(r)0$+$   & L1.9     & 09 19 18.60514 & 69 12 11.6465&2
  & $1.6\times 10^8$ & 9 & \phn7.5 & 17 &    433
  & \objectname[VLA_Cal]{J0903$+$6757} & 1.88
\\
\NGC{3031} & SA(s)ab     & S1.5     & 09 55 33.1731 & 69 03 55.061 & 4
  & $6.9\times 10^7$ & 6 & \phn3.6 & 18 &    381
  & \objectname[VLA_Cal]{J0958$+$6533} & 3.51
\\[1.5ex]
\NGC{3079} & SB(s)c spin & S2       & 10 01 57.8050 & 55 40 47.200 & 5
  & $4.2\times 10^7$ & 8 & 17.3 & 19    & \phn48
  & \objectname[VLA_Cal]{J1035$+$5628} & 4.69
\\
\NGC{3147} & SA(rs)bc    & S2       & 10 16 53.6503 & 73 24 02.696 & 3
  & $4.4\times 10^8$ &10 & 40.9 & 20    &    214
  & \objectname[VLA_Cal]{J1048$+$7143} & 2.89
\\
\NGC{3169} & SA(s)a pec  & L2       & 10 14 15.05027 & 03 27 57.8750&2
  & $7.2\times 10^7$ &11 & 40.8 & 21    & \phn35
  & \objectname[VLA_Cal]{J1016$+$0513} & 1.81
\\
\NGC{3226} & E2: pec     & L1.9     & 10 23 27.00837 & 19 53 54.6806&2
  & $1.4\times 10^8$ &12 & 23.6 & 17    &    116
  & \objectname[VLA_Cal]{J1016$+$2037} & 1.74
\\
\NGC{3227} & SAB(s)a pec & S1.5     & 10 23 30.579 & 19 51 54.18 & 1
  & $2.2\times 10^7$ & 7 & 21.1 & 15    & \phn21
  & \objectname[VLA_Cal]{J1016$+$2037} & 1.76
\\[1.5ex]
\NGC{4168} & E2          & S1.9:    & 12 12 17.2685 & 13 12 18.701 & 3
  & $1.0\times 10^8$ & 6 & 30.9 & 17    & \phn65
  & \objectname[VLA_Cal]{J1207$+$1211} & 1.60
\\
\NGC{4203} & SAB0$-$:    & L1.9     & 12 15 05.0554 & 33 11 50.382 & 3
  & $7.0\times 10^7$ & 9 & 15.1 & 17    & \phn92
  & \objectname[VLA_Cal]{J1215$+$3448} & 1.62
\\
\NGC{4235} & SA(s)a spin & S1.2     & 12 17 09.8818 & 07 11 29.670 & 3
  & $4.8\times 10^7$ &13 & 18.0 & 22    & \phn53
  & \objectname[VLA_Cal]{J1222$+$0413} & 3.24
\\
\NGC{4450} & SA(s)ab     & L1.9     & 12 28 29.5908 & 17 05 05.972 & 3
  & $2.1\times 10^7$ & 8 & 14.1 & 22    & \phn30
  & \objectname[VLA_Cal]{J1215$+$1654} & 3.21
\\
\NGC{4472} & E2          & S2::     & 12 29 46.76189 & 08 00 01.7129&2
  & $7.2\times 10^8$ & 6 & 16.3 & 17    &    869
  & \objectname[VLA_Cal]{J1239$+$0730} & 2.44
\\[1.5ex]
\NGC{4565} & SA(s)b? spin& S1.9     & 12 36 20.78023 & 25 59 15.6298&2
  & $2.9\times 10^7$ & 8 & 17.5 & 17    & \phn32
  & \objectname[VLA_Cal]{J1230$+$2518} & 1.54
\\
\NGC{4579} & SAB(rs)b    & S1.9/L1.9& 12 37 43.5223 & 11 49 05.488 & 3
  & $6.1\times 10^7$ & 6 & 19.1 & 22    & \phn63
  & \objectname[VLA_Cal]{J1239$+$0730} & 4.33
\\
\NGC{5866} & SA0 $+$ spin& T2       & 15 06 29.49889 & 55 45 47.5681&2
  & $5.2\times 10^7$ & 6 & 15.4 & 17    & \phn67
  & \objectname[VLA_Cal]{J1510$+$5702} & 1.37
\\
\enddata
\tablecomments{Column~(1) gives the galaxy name.  Column~(2) gives the
  Hubble classification and Column~(3) AGN classification are from
  \citet{Ho_FS1997.0}, with S representing Seyfert galaxies and L
  representing LINER galaxies.  Columns (4) and~(5) give the positions
  of the radio cores of the galaxies (units of right ascension are
  hours, minutes, and seconds, and units of declination are degrees,
  arcminutes, and arcseconds), and Column~(6) gives the reference for
  the positions.  Our VLA positions should be good to about 0\farcs04.
  Column~(7) gives the adopted mass of the central black hole, using
  velocity dispersion measurements taken from the references in
  Column~(8) and the $M_\mathrm{BH}$-$\sigma$ relation from
  \citet{Tremaine_ea2002}.  Columns (9) and~(10) give the adopted
  distance and reference for each galaxy.  For cases in which only
  recession velocities are known, we adopt a value of $H_0 =
  75$~km~s$^{-1}$~Mpc$^{-1}$.  Column~(11) gives the angular size in
  microarcseconds equivalent to 1000 Schwarzschild radii at the
  distance of the galaxy.  Column~(12) gives the phase calibrator used
  for our VLA variability study, and Column~(13) gives the angular
  separation between the galaxy and the calibrator.  }
\tablerefs{1. 8.5~GHz VLA observations, this paper; 2. 8.4~GHz VLBA
  observations, this paper; 3. ICRF \citealt{Ma_ea1998} (see the
  extension at \url{ftp://hpiers.obspm.fr/iers/icrf/iau/icrf-Ext.1/});
  4. \citealt[Fig.~4]{Trotter_ea1998}; 5. \citealt{Anderson_UH2004};
  6. \citealt{Prugniel_ea2001} (see the HyperLeda database at
  \url{http://www-obs.univ-lyon1.fr/hypercat/}); 7.
  \citealt{Nelson_W1995}; 8. \citealt{McElroy_1995}; 9.
  \citealt*{Barth_HS2002}; 10. \citealt*{Whitmore_MT1985}; 11.
  \citealt{Heraudeau_S1998}; 12. \citealt{Simien_P2002}; 13.
  \citealt{Jimenez-Benito_ea2000}; 14. \citealt{deVaucouleurs_ea1991};
  15 \citealt{Terry_PE2002}; 16. \citealt{Gallimore_ea1999}; 17.
  \citealt{Tonry_ea2001}; 18. \citealt{Freedman_ea2001}; 19.
  \citealt{Tully_SP1992}; 20. \citealt{Tully_1988}; 21.
  \citealt{Tutui_S1997}; 22. \citealt{Solanes_ea2002} }
\end{deluxetable}

\clearpage

\begin{deluxetable}{llccc}
  \tablecaption{VLBA Observation Details\label{tab:VLBA_obs}}
  \tablewidth{0pt}
  \tablehead{
    \colhead{Galaxy} & \colhead{UT Date} 
    & \colhead{Phase} & \colhead{Offset} 
    & \colhead{Check} \\
    \colhead{} & \colhead{} 
    & \colhead{Calibrator} & \colhead{(\degr)}
    & \colhead{(\degr)} \\
    \colhead{(1)} & \colhead{(2)} & \colhead{(3)} 
    & \colhead{(4)} & \colhead{(5)} 
    }
\startdata
\NGC{\phn 777} & 2003 Sep \hfill 15 & 
\objectname[ICRF]{J0205+3212} & 1.29 & 1.54 \\
\NGC{2273} & 2003 Sep \hfill 25 & 
\objectname[VLBA]{J0650+6001} & 0.82 & 1.64 \\
\NGC{2639} & 2003 Mar \hfill 03 & 
\objectname[ICRF]{J0832+4913} & 2.07 & 3.16 \\
\NGC{2787} & 2003 Mar \hfill 03 & 
\objectname[VLBA]{J0903+6757} & 1.88 & 1.14 \\
\NGC{3169} & 2003 Mar \hfill 03 & 
\objectname[VLBA]{J1016+0513} & 1.81 & 2.30 \\
\NGC{3226} & 2003 Oct \hfill 02 & 
\objectname[ICRF]{J1024+1912} & 0.76 & 2.36 \\
\NGC{3227} & 2003 Oct \hfill 02 & 
\objectname[ICRF]{J1024+1912} & 0.72 & 2.36 \\
\NGC{4472} & 2003 Sep \hfill 22 & 
\objectname[VLBA]{J1238+0723} & 2.14 & 2.70 \\
\NGC{4565} & 2003 Sep \hfill 25 & 
\objectname[VLBA]{J1230+2518} & 1.54 & 3.52 \\
\NGC{5866} & 2003 Mar \hfill 25 & 
\objectname[ICRF]{J1510+5702} & 1.37 & 5.77 \\
\enddata
\tablecomments{Column~(1): Galaxy name.  Column~(2): UT observing
  date.  Column~(3): Calibration source for phase referencing.
  Column~(4): Angular separation between the galaxy and phase calibrator.
  Column~(5): Angular separation between the phase calibrator and the
  check source.}
\end{deluxetable}

\clearpage

\begin{deluxetable}{lccccccc}
  \tabletypesize{\footnotesize}
  \tablecaption{VLBA Galaxy Attributes\label{tab:VLBA_att}}
  \tablewidth{0pt}
  \tablehead{
    \colhead{Galaxy} 
    & \colhead{$S^\mathrm{P}_\nu$} & \colhead{$S^\mathrm{I}_\nu$}
    & \colhead{RMS}
    & \colhead{coherence} 
    & \colhead{$\Theta_\mathrm{min}$} & \colhead{$\Theta_\mathrm{max}$}
    & \colhead{$T_\mathrm{b}$}
    \\
    \colhead{} 
    & \colhead{$\!\!\!\!$($\mathrm{mJy~beam^{-1}}$)} & \colhead{(mJy)}
    & \colhead{$\!\!\!\!$($\mathrm{\mu Jy~beam^{-1}}$)$\!\!\!\!$} & \colhead{ratio}
    & \colhead{(mas)} & \colhead{(mas)}
    & \colhead{(K)}
    \\
    \colhead{(1)} & \colhead{(2)} 
    & \colhead{(3)} & \colhead{(4)} 
    & \colhead{(5)} & \colhead{(6)}
    & \colhead{(7)} & \colhead{(8)} 
    }
\startdata
\NGC{\phn 777} 
   & \nodata & \nodata
   & 30
   & 1.185
   & \nodata & \nodata
   & \nodata
   \\
\NGC{2273} 
   & $0.44 \pm 0.04$ & $ 2.4 \pm 0.2$
   & 31
   & 1.020
   & \nodata & \nodata
   & $\phm{>}4\times 10^6$
   \\
\NGC{2639} 
   & $53 \pm 3$ & $ 66 \pm 3$
   & 50
   & 1.014
   & \nodata & \nodata
   & $>1\times 10^9$
   \\
\NGC{2787} 
   & $14.4 \pm 0.7$ & $14.4 \pm 0.7$
   & 37
   & 1.005
   &   0.15  &   0.23
   & $>1\times 10^9$
   \\
\NGC{3169} 
   & $8.6 \pm 0.4$ & $8.9 \pm 0.5$
   & 38
   & 1.005
   &   0.33  &   0.49
   & $>5\times 10^8$
   \\
\NGC{3226} 
   & $7.9 \pm 0.4$ & $9.8 \pm 0.5$
   & 37
   & 1.006
   & \nodata & \nodata
   & $>2\times 10^8$
   \\
\NGC{3227} 
   & \nodata          & \nodata          
   & 42
   & 1.219
   & \nodata    & \nodata
   & \nodata
   \\
\NGC{4472} 
   & $3.6 \pm 0.2$ & $3.8 \pm 0.2$
   & 32
   & 1.098
   &  0.00  &  0.73
   & $>2\times 10^8$
   \\
\NGC{4565} 
   & $1.9 \pm 0.1$ & $1.9 \pm 0.1$
   & 36
   & 1.062
   &  0.54  &  0.86
   & $>1\times 10^8$
   \\
\NGC{5866} 
   & $5.4 \pm 0.3$ & $7.1 \pm 0.4$
   & 40
   & 1.007
   &  1.92   &  1.99
   & $>1\times 10^8$
   \\
\enddata
\tablecomments{Column~(1) gives the galaxy name.  Column~(2) gives the
  peak flux density of the source from Gaussian fitting, corrected for
  decoherence, while Column~(3) gives the integrated flux density in
  the VLBA image.  Column~(4) gives the RMS noise in the image far
  away from the source.  Column~(5) gives the estimated increase in
  peak flux density by correcting for coherence losses based on
  self-calibration improvements in the check source.  Columns (6)
  and~(7) give respectively the 1-$\sigma$ lower and upper limits on
  the deconvolved major axis of the source sized obtained through
  Gaussian fitting of the image.  Column~(8) gives the lower limit to
  the brightness temperature.  For galaxies with no deconvolved size
  listed in Column~(7), details of the calculation procedure are given
  in Appendix~\ref{sec:sources}.  For the remaining galaxies, a beam
  size of the maximum of Column~(7) or one third of the synthesized
  beam width in both major and minor axes was assumed, since source
  sizes significantly smaller than the synthesized beam are unreliable
  even for our images with signal to noise ratios of several hundred.
}
\end{deluxetable}

\clearpage

\begin{deluxetable}{lccr}
  \tablecaption{VLA Observation Details\label{tab:VLA_obs}}
  \tablewidth{0pt}
  \tablecolumns{4}
  \tablehead{
    \colhead{UT Date} 
    & \colhead{LST Start} & \colhead{LST End} 
    & \colhead{Length} 
    \\
    \colhead{(1)} & \colhead{(2)} & \colhead{(3)} 
    & \colhead{(4)}  
    }
\startdata
\cutinhead{D$\rightarrow$A Configuration}
May 16 & 3:00 & 14:30 & 11.5 \\
May 17 & 4:30 & 14:30 & 10.0 \\
May 18 & 5:00 & 15:00 & 10.0 \\
May 20 & 7:00 & 12:00 &  5.0 \\
May 24 & 5:00 & 10:00 &  5.0 \\
May 26 & 5:00 & 10:00 &  5.0 \\
May 29 & 6:30 & 11:30 &  5.0 \\
\cutinhead{A$\rightarrow$BnA Configuration}
Sep 12 & 4:00 & 14:00 & 10.0 \\
Sep 13 & 4:00 & 14:00 & 10.0 \\
Sep 14 & 4:00 & 14:30 & 10.5 \\
Sep 16 & 4:00 & 10:30 &  6.5 \\
Sep 18 & 4:00 & 10:30 &  6.5 \\
Sep 21 & 4:00 & 10:30 &  6.5 \\
Sep 25 & 4:00 & 10:30 &  6.5 \\
\enddata
\tablecomments{Column~(1): UT observing date.  Column~(2) is the Local
  Sidereal Time (LST) at the start of observations, and Column~(3) is
  the LST at the end of observations in hours and minutes.  Column~(4)
  gives the total observing time in hours for the corresponding day.
}
\end{deluxetable}

\clearpage

\begin{deluxetable}{llclrrrlcclrrrlrr}
  \tablecaption{Variability Statistics\label{tab:var_stats}}
  \rotate
  \tabletypesize{\scriptsize}
  \tablecolumns{17}
  \tablewidth{0pt}
  \tablehead{
    \colhead{} & \colhead{}
    & \multicolumn{6}{c}{2003 May} 
    & \colhead{}
    & \multicolumn{6}{c}{2003 September} 
    & \colhead{} & \colhead{} \\
    \cline{3-8} \cline{10-15} \\
    \colhead{Name} & \colhead{Class}
    & \colhead{$N$} & \colhead{$<S_\nu>$}
    & \colhead{$\sigma_\mathrm{e}$} & \colhead{$\sigma_\mathrm{s}$}
    & \colhead{$\sigma_\mathrm{d}$} & \colhead{$P_{\chi^2}$}
    & \colhead{}
    & \colhead{$N$} & \colhead{$<S_\nu>$}
    & \colhead{$\sigma_\mathrm{e}$} & \colhead{$\sigma_\mathrm{s}$}
    & \colhead{$\sigma_\mathrm{d}$} & \colhead{$P_{\chi^2}$}
    & \colhead{$F$} & \colhead{$\delta F$}
    \\
    \colhead{} & \colhead{}
    & \colhead{}  & \colhead{(Jy~bm$^{-1}$)}
    & \colhead{(\%)} & \colhead{(\%)}
    & \colhead{(\%)} & \colhead{}
    & \colhead{\ }
    & \colhead{}  & \colhead{(Jy~bm$^{-1}$)}
    & \colhead{(\%)} & \colhead{(\%)}
    & \colhead{(\%)} & \colhead{}
    & \colhead{(\%)} & \colhead{(\%)}
    \\
    \colhead{(1)} & \colhead{(2)}
    & \colhead{(3)} & \colhead{(4)}  
    & \colhead{(5)} & \colhead{(6)}  
    & \colhead{(7)} & \colhead{(8)}  
    &\colhead{}
    & \colhead{(9)} & \colhead{(10)} 
    & \colhead{(11)}& \colhead{(12)} 
    & \colhead{(13)}& \colhead{(14)} 
    & \colhead{(15)}& \colhead{(16)}
}
\startdata
\cutinhead{Target Galaxies and Phase Calibrators}\\
\NGC{\phn 777}&P & 11& 0.00076& 11.4& 18.8& 14.9& $1.4 \times 10^{-4}$&& 17& 0.00087& 8.7& 11.8& 8.0& $9.4 \times 10^{-3}$& +0.3& 6.3\\*
\objectname[VLA_Cal]{J0205$+$3212}&C & 11& 2.03& 1.4& 2.0& 1.4& $2.0 \times 10^{-2}$&& 17& 1.943& 0.7& 1.2& 1.0& $5.6 \times 10^{-5}$& -5.9& 1.4\\*
\NGC{2273}&E & 12& 0.0053& 2.8& 7.9& 7.4& $1.1 \times 10^{-12}$&& 19& 0.00547& 2.5& 5.9& 5.3& $2.0 \times 10^{-13}$& +1.1& 3.9\\*
\objectname[VLA_Cal]{J0650$+$6001}&CS & 12& 0.962& 1.4& 1.0& \nodata& $8.8 \times 10^{-1}$&& 19& 0.975& 0.7& 0.7& 0.0& $4.4 \times 10^{-1}$& +1.7& 1.0\\*
\NGC{2639}&J & 10& 0.0737& 1.5& 1.4& \nodata& $5.0 \times 10^{-1}$&& 10& 0.0736& 0.9& 2.6& 2.5& $1.9 \times 10^{-12}$& -2.8& 1.2\\*
\objectname[VLA_Cal]{J0832$+$4913}&C & 10& 0.478& 1.4& 1.7& 0.9& $1.5 \times 10^{-1}$&& 10& 0.4613& 0.7& 0.6& \nodata& $5.9 \times 10^{-1}$& -3.3& 0.8\\[2ex]
\NGC{2787}&P & 10& 0.01357& 1.6& 2.0& 1.1& $1.1 \times 10^{-1}$&& 13& 0.01095& 1.1& 1.1& 0.2& $4.1 \times 10^{-1}$& -19.4& 1.2\\*
\objectname[VLA_Cal]{J0903$+$6757}&C & 10& 0.574& 1.4& 1.1& \nodata& $8.0 \times 10^{-1}$&& 13& 0.529& 0.7& 1.0& 0.7& $3.3 \times 10^{-2}$& -8.3& 1.0\\*
\NGC{3031}&P & 12& 0.1379& 1.5& 2.5& 2.0& $8.7 \times 10^{-4}$&& 11& 0.106& 0.9& 3.4& 3.3& $2.5 \times 10^{-28}$& -26.2& 2.5\\*
\objectname[VLA_Cal]{J0958$+$6533}&C & 12& 0.530& 1.4& 3.2& 2.9& $2.0 \times 10^{-9}$&& 11& 0.471& 0.7& 2.4& 2.3& $1.2 \times 10^{-20}$& -8.1& 2.6\\*
\NGC{3079}&EJ & 13& 0.1178& 1.5& 1.0& \nodata& $9.5 \times 10^{-1}$&& 11& 0.1157& 0.9& 1.2& 0.8& $4.2 \times 10^{-2}$& -2.4& 0.9\\*
\objectname[VLA_Cal]{J1035$+$5628}&CS & 13& 0.7956& 2.0& 0.4& \nodata& $1.0 \times 10^{0}$&& 11& 0.7776& 0.7& 0.4& \nodata& $9.8 \times 10^{-1}$& -1.9& 0.8\\[2ex]
\NGC{3147}&D & 12& 0.01213& 1.6& 1.9& 1.0& $1.4 \times 10^{-1}$&& 13& 0.01018& 1.2& 1.9& 1.5& $1.4 \times 10^{-3}$& -16.3& 1.8\\*
\objectname[VLA_Cal]{J1048$+$7143}&C & 12& 1.330& 1.4& 1.5& 0.6& $2.4 \times 10^{-1}$&& 13& 1.388& 0.7& 1.2& 0.9& $9.3 \times 10^{-4}$& +6.2& 1.6\\*
\NGC{3169}&D & 10& 0.00982& 1.7& 1.5& \nodata& $6.0 \times 10^{-1}$&& 12& 0.00983& 1.2& 2.7& 2.4& $4.4 \times 10^{-7}$& +0.1& 1.1\\*
\objectname[VLA_Cal]{J1016$+$0513}&C & 10& 0.435& 1.4& 1.1& \nodata& $7.8 \times 10^{-1}$&& 12& 0.412& 0.7& 1.2& 1.0& $6.7 \times 10^{-4}$& -6.5& 0.9\\*
\NGC{3226}&E & 11& 0.00639& 2.1& 1.6& \nodata& $8.6 \times 10^{-1}$&& 12& 0.01013& 1.3& 2.0& 1.5& $2.0 \times 10^{-3}$& +47.9& 1.4\\*
\objectname[VLA_Cal]{J1016$+$2037}&C & 22& 0.793& 1.4& 0.8& \nodata& $1.0 \times 10^{0}$&& 24& 0.765& 0.7& 0.9& 0.6& $1.3 \times 10^{-2}$& -3.9& 0.6\\[2ex]
\NGC{3227}&EJ & 11& 0.0054& 2.3& 9.4& 9.1& $2.0 \times 10^{-25}$&& 12& 0.0056& 2.0& 10.9& 10.7& $8.4 \times 10^{-69}$& -6.0& 3.0\\*
\objectname[VLA_Cal]{J1016$+$2037}&C & 22& 0.793& 1.4& 0.8& \nodata& $1.0 \times 10^{0}$&& 24& 0.765& 0.7& 0.9& 0.6& $1.3 \times 10^{-2}$& -3.9& 0.6\\*
\NGC{4168}&P & 12& 0.01278& 1.6& 1.8& 0.6& $3.6 \times 10^{-1}$&& 14& 0.01313& 1.1& 1.0& \nodata& $6.8 \times 10^{-1}$& +2.6& 1.9\\*
\objectname[VLA_Cal]{J1207$+$1211}&C & 12& 0.262& 1.4& 1.8& 1.1& $7.6 \times 10^{-2}$&& 14& 0.319& 0.7& 1.8& 1.6& $5.4 \times 10^{-12}$&+18.3& 2.1\\*
\NGC{4203}&P & 12& 0.01052& 1.7& 2.2& 1.4& $5.1 \times 10^{-2}$&& 11& 0.00743& 1.4& 3.4& 3.1& $8.4 \times 10^{-9}$& -30.1& 1.0\\*
\objectname[VLA_Cal]{J1215$+$3448}&C & 12& 0.814& 1.4& 1.7& 1.0& $9.4 \times 10^{-2}$&& 11& 0.789& 0.7& 0.6& \nodata& $7.2 \times 10^{-1}$& -1.5& 0.7\\[2ex]
\NGC{4235}&P & 12& 0.00656& 2.0& 2.3& 1.0& $2.0 \times 10^{-1}$&& 14& 0.00540& 1.8& 3.2& 2.7& $1.5 \times 10^{-4}$& -19.7& 1.2\\*
\objectname[VLA_Cal]{J1222$+$0413}&C & 12& 0.822& 1.4& 0.6& \nodata& $1.0 \times 10^{0}$&& 14& 0.805& 0.7& 2.0& 1.8& $2.9 \times 10^{-15}$& -3.9& 0.7\\*
\NGC{4450}&EJ & 12& 0.00503& 2.3& 4.6& 4.0& $5.3 \times 10^{-6}$&& 12& 0.00575& 1.9& 3.4& 2.9& $1.8 \times 10^{-5}$& +5.4& 1.8\\*
\objectname[VLA_Cal]{J1215$+$1654}&C & 12& 0.3159& 1.4& 1.0& \nodata& $8.4 \times 10^{-1}$&& 12& 0.330& 0.7& 3.3& 3.2& $1.1 \times 10^{-39}$& +3.6& 0.9\\*
\NGC{4472}&E & 11& 0.00473& 2.6& 6.5& 6.0& $1.2 \times 10^{-10}$&& 11& 0.00437& 2.9& 3.9& 2.6& $3.1 \times 10^{-2}$& -9.8& 3.1\\*
\objectname[VLA_Cal]{J1239$+$0730}&C & 22& 0.641& 1.4& 0.9& \nodata& $9.9 \times 10^{-1}$&& 23& 0.701& 0.7& 0.9& 0.5& $4.1 \times 10^{-2}$& +8.8& 0.6\\[2ex]
\NGC{4565}&P & 12& 0.00302& 3.1& 5.0& 3.8& $1.0 \times 10^{-3}$&& 11& 0.00231& 3.4& 4.6& 3.0& $6.1 \times 10^{-2}$& -21.0& 3.9\\*
\objectname[VLA_Cal]{J1230$+$2518}&C & 12& 0.299& 1.4& 2.1& 1.6& $4.6 \times 10^{-3}$&& 11& 0.239& 0.7& 3.1& 3.1& $1.4 \times 10^{-34}$& -22.3& 2.3\\*
\NGC{4579}&D & 11& 0.0248& 1.5& 3.2& 2.9& $5.7 \times 10^{-7}$&& 12& 0.0243& 0.9& 1.4& 1.1& $6.4 \times 10^{-3}$& +2.4& 1.0\\*
\objectname[VLA_Cal]{J1239$+$0730}&C & 22& 0.641& 1.4& 0.9& \nodata& $9.9 \times 10^{-1}$&& 23& 0.701& 0.7& 0.9& 0.5& $4.1 \times 10^{-2}$& +8.8& 0.6\\*
\NGC{5866}&P & 10& 0.00884& 1.7& 3.5& 3.1& $2.9 \times 10^{-5}$&& 11& 0.00887& 1.3& 2.4& 2.1& $2.1 \times 10^{-5}$& -1.6& 3.2\\*
\objectname[VLA_Cal]{J1510$+$5702}&C & 10& 0.3179& 1.4& 0.9& \nodata& $9.1 \times 10^{-1}$&& 11& 0.365& 0.7& 1.0& 0.7& $3.1 \times 10^{-2}$& +13.5& 0.8\\[8ex]
\cutinhead{CSO Calibrators}\\
\objectname[COINS]{J0204$+$0903}&S & 11& 0.396& 1.4& 1.5& 0.6& $2.9 \times 10^{-1}$&& 14& 0.3898& 0.7& 0.8& 0.3& $2.6 \times 10^{-1}$& -3.9& 0.9\\*
\objectname[COINS]{J0410$+$7656}&SE & 7& 2.16& 7.8& 2.2& \nodata& $1.9 \times 10^{-1}$&& 17& 2.19& 0.7& 2.1& 2.0& $2.4 \times 10^{-22}$& +4.6& 1.0\\*
\objectname[COINS]{J0427$+$4133}&S & 9& 0.679& 1.4& 1.4& 0.2& $4.2 \times 10^{-1}$&& 13& 0.687& 0.7& 0.9& 0.6& $4.5 \times 10^{-2}$& +0.0& 0.8\\*
\objectname[COINS]{J0650$+$6001}&CS & 12& 0.962& 1.4& 1.0& \nodata& $8.8 \times 10^{-1}$&& 19& 0.975& 0.7& 0.7& 0.0& $4.4 \times 10^{-1}$& +1.7& 1.0\\*
\objectname[COINS]{J1035$+$5628}&CS & 13& 0.7956& 2.0& 0.4& \nodata& $1.0 \times 10^{0}$&& 11& 0.7776& 0.7& 0.4& \nodata& $9.8 \times 10^{-1}$& -1.9& 0.8\\[2ex]
\objectname[COINS]{J1148$+$5924}&S & 7& 0.428& 1.4& 1.2& \nodata& $5.8 \times 10^{-1}$&& 11& 0.4274& 0.7& 0.7& \nodata& $4.7 \times 10^{-1}$& -0.2& 0.8\\*
\objectname[COINS]{J1244$+$4048}&S & 44& 0.4428& 1.4& 0.4& \nodata& $1.0 \times 10^{0}$&& 42& 0.4356& 0.7& 0.4& \nodata& $1.0 \times 10^{0}$& -2.3& 0.4\\*
\objectname[COINS]{J1400$+$6210}&S & 7& 1.152& 1.4& 2.2& 1.8& $1.1 \times 10^{-2}$&& 10& 1.1387& 0.7& 0.3& \nodata& $1.0 \times 10^{0}$& +0.3& 1.5\\*
\objectname[COINS]{J1823$+$7938}&S & 7& 0.561& 1.4& 1.0& \nodata& $8.0 \times 10^{-1}$&& 14& 0.560& 0.7& 0.7& 0.2& $3.6 \times 10^{-1}$& +0.8& 0.8\\
\enddata
\tablecomments{Simple variability statistics are given for the target
  galaxies and calibration sources in this study.  Column~(1) gives
  the name of the object.  The phase calibrator used for each target
  galaxy is listed immediately below the target information, so some
  calibrators are listed multiple times.  Column~(2) gives the object
  classification.  P indicates a point source, D indicates a galaxy
  dominated by a point source containing at least 80\% of the flux, E
  indicates an extended source, and J indicates the presence of a jet
  feature.  C indicates that the source was used as a phase
  calibrator, and S indicates that the object is a CSO.  Columns (3)
  and~(9) give the number of independent observations for May and
  September, respectively.  Columns (4) and~(10) give the mean peak
  flux density.  Columns (5) and~(11) give the mean expected scatter
  in the data based on the measurement noise and expected statistical
  errors.  Columns (6) and~(12) give the actual RMS scatter of the
  data.  Columns (7) and~(13) give the predicted de-biased scatter in
  the data, correcting for measurement uncertainty.  Columns (8)
  and~(14) give the $\chi^2$ probability that a constant brightness
  object would have an RMS scatter at least as large as actually
  observed given the expected measurement errors for the given number
  of data-points ($N-1$ degrees of freedom).  Column~(15) gives the
  fractional variation from May to September, and Column~(16) gives the
  uncertainty in that value.  The fractional variation was calculated
  from the mean of the last three days of May and the first three days
  of September in order to provide the most similar $(u,v)$ coverage.
  See \S~\ref{sec:gross_stat} and \S~\ref{sec:long-term} for more
  details.  }
\end{deluxetable}

\clearpage

\begin{deluxetable}{lrrrrr}
  \tablecaption{Galaxy Variability Fractions\label{tab:var_fraction}}
  \tablewidth{0pt}
  \tablecolumns{6}
  \tabletypesize{\footnotesize}
  \tablehead{
    \colhead{} 
    & \multicolumn{2}{c}{Reliable} 
    & \colhead{}
    & \multicolumn{2}{c}{Reliable $+$ Tentative} 
    \\
    \cline{2-3} \cline{5-6} 
    \colhead{Class} 
    & \colhead{Number} 
    & \colhead{Fraction}
    & \colhead{}
    & \colhead{Number}
    & \colhead{Fraction} 
    \\
    \colhead{} 
    & \colhead{} 
    & \colhead{(\%)}
    & \colhead{}
    & \colhead{}
    & \colhead{(\%)}
    \\
    \colhead{(1)} 
    & \colhead{(2)} 
    & \colhead{(3)}
    & \colhead{}
    & \colhead{(4)}  
    & \colhead{(5)} 
  }
\startdata
\cutinhead{Galaxies with $P_{\chi^2} < 0.01$}
E$+$J         & $1/\phn4$   & $25\pm 22$ && $3/\phn6$ & $50\pm 20$ \\
D             & $1/\phn2$   & $50\pm 35$ && $4/\phn6$ & $67\pm 19$ \\
P             & $8/14$   & $57\pm 13$ && $9/16$ & $56\pm 12$ \\
P$+$D          & $9/16$   & $56\pm 12$ && $13/22$ & $59\pm 10$ \\
P$+$D$+$E$+$J  & $10/20$   & $50\pm 11$ && $16/28$ & $57\pm \phn9$ \\
\cutinhead{Galaxies with $P_{\chi^2} < 0.001$}
P$+$D$+$E$+$J  & $7/20$   & $35\pm 11$ && $12/28$ & $43\pm \phn9$ \\
\cutinhead{Galaxies with $P_{\chi^2} < 0.0005$}
P$+$D$+$E$+$J  & $5/20$   & $25\pm 10$ && $10/28$ & $36\pm \phn9$ \\
\cutinhead{Galaxies with $P_{\chi^2} < 0.01$ and $\sigma_\mathrm{s} > 4$\%}
P$+$D$+$E$+$J  & $3/20$   & $15\pm \phn8$ && $4/28$ & $14\pm \phn7$ \\
\cutinhead{Galaxies with $P_{\chi^2} < 0.01$ and $\sigma_\mathrm{d} > 4$\%}
P$+$D$+$E$+$J  & $2/20$   & $10\pm \phn7$ && $3/28$ & $11\pm \phn6$ \\
\enddata
\tablecomments{Column~(1) indicates the object classes studied, using
  the classification from \S~\ref{sec:gross_stat}.  Columns (2)
  and~(4) give the number counts for the number of variable sources
  and the total number of sources.  Columns (3) and~(5) give the
  fraction of source-months which show variability.  Columns (2)
  and~(3) give information for only datasets classified as
  ``reliable'', while Columns (4) and~(5) also include ``tentative''
  datasets.  The galaxies with known $(u,v)$ problems (NGC~2273,
  NGC~2639, NGC~3227, and NGC~4472) are not included in these
  statistics.  }
\end{deluxetable}

\clearpage

\begin{deluxetable}{llcrrrrrr}
  \rotate
  \tablecaption{Variability Categorization\label{tab:var_cat}}
  \tablewidth{0pt}
  \tabletypesize{\footnotesize}
  \tablehead{
    \colhead{Galaxy} 
    & \colhead{Class} & \colhead{Var.}
    & \colhead{$T_\mathrm{b,O,1}$}
    & \colhead{$T_\mathrm{b,O,4}$}
    & \colhead{$T_\mathrm{b,M,50}$}
    & \colhead{$T_\mathrm{b,M,90}$}
    & \colhead{$T_\mathrm{b,R,50}$}
    & \colhead{$T_\mathrm{b,R,90}$}
    \\
    \colhead{} 
    & \colhead{} & \colhead{} 
    & \colhead{$\!\!\!$(\scriptsize{$\log_{10}(K)$})$\!\!\!$} & \colhead{$\!\!\!$(\scriptsize{$\log_{10}(K)$})$\!\!\!$}
    & \colhead{$\!\!\!$(\scriptsize{$\log_{10}(K)$})$\!\!\!$} & \colhead{$\!\!\!$(\scriptsize{$\log_{10}(K)$})$\!\!\!$} 
    & \colhead{$\!\!\!$(\scriptsize{$\log_{10}(K)$})$\!\!\!$} & \colhead{$\!\!\!$(\scriptsize{$\log_{10}(K)$})$\!\!\!$} 
    \\
    \colhead{(1)} 
    & \colhead{(2)} & \colhead{(3)} 
    & \colhead{(4)}  & \colhead{(5)} & \colhead{(6)}
    & \colhead{(7)}  & \colhead{(8)} & \colhead{(9)}
    }
\startdata
\NGC{\phn 777} &P&  var      & 13.7 & 11.6 & 10.0&  9.8& 10.8& 10.0                   \\
\NGC{2787} &P&  const    &  8.5 &  7.5 &  7.6&\nodata &  7.9&\nodata                  \\
\NGC{3031} &P&  var      & 11.2 &  9.5 &  9.1&  8.9&  9.5&  9.3                       \\
\NGC{3079} &EJ& const    & 10.2 &  7.4 &\nodata &\nodata &\nodata &\nodata            \\
\NGC{3147} &D&  var      & 10.8 & 10.3 &  9.9&  9.8& 10.3& 10.0                       \\[1.5ex]
\NGC{3169} &D&  var?     & 10.7 & 10.4 & 10.1&  9.9& 10.5& 10.2                       \\
\NGC{3226} &E&  var      & 10.9 &  9.8 &  9.3&  8.9&  9.8&  9.1                       \\
\NGC{4168} &P&  const    &  7.2 &  7.1 &\nodata &\nodata &\nodata &\nodata            \\
\NGC{4203} &P&  var?     &  9.8 &  9.0 &  8.8&  8.4&  9.1&  8.7                       \\
\NGC{4235} &P&  var      &  9.1 &  8.7 &  8.5&  8.2&  8.7&  8.4                       \\[1.5ex]
\NGC{4450} &EJ& var?     &  9.6 &  9.0 &  8.9&  8.4&  9.1&  8.9                       \\
\NGC{4565} &P&  var      &  9.7 &  9.2 &  8.8&  8.4&  9.2&  8.6                       \\
\NGC{4579} &D&  var?     & 10.1 &  9.7 &  9.4&  9.0&  9.8&  9.3                       \\
\NGC{5866} &P&  var?     & 12.4 & 10.5 &  9.5&  9.3& 12.2& 10.4                       \\
\enddata
\tablecomments{Column~(1) gives the galaxy name.  Column~(2) gives the
  image classification from Table~\ref{tab:var_stats}.  Column~(3)
  gives our variability classification for the target galaxy;
  ``const'' indicates that the source appears to be constant, and
  ``var'' indicates that the source is variable.  Galaxies which have
  slightly questionable variability are designated by ``var?''.
  Columns (4) and~(5) give the highest and fourth highest apparent
  brightness temperatures measured directly from the target dataset.
  Columns (6) and~(7) give the 50\% (best) and 90\% (minimum)
  confidence estimates of the variability brightness temperature,
  respectively, for our first Monte Carlo simulation.  Columns (8)
  and~(9) give the results for the random brightness temperature
  simulation, which is our best attempt to account for all biases.
  See \S~\ref{sec:T_b} for details.  }
\end{deluxetable}

\clearpage

\begin{deluxetable}{lrrrrrrrrrrrrrrrrrr}
  \tablecaption{Intraday Scintillation Results\label{tab:var_scintillation}}
  \tablewidth{0pt}
  \tablecolumns{19}
  \rotate
  \tabletypesize{\scriptsize}
  \tablehead{
    \colhead{} &\colhead{} &\colhead{} &\colhead{} 
    & \multicolumn{7}{c}{2003 May} 
    & \colhead{}
    & \multicolumn{7}{c}{2003 September} 
    \\
    \cline{5-11} \cline{13-19} \\
    \colhead{Galaxy} 
    & \colhead{$\nu_0$} 
    & \colhead{$\theta_\mathrm{F}$} & \colhead{$t_\mathrm{F}$}
    & \colhead{$\tau_\mathrm{max}$} & \colhead{$\theta_\tau$} 
    & \colhead{$m$} & \colhead{$\theta_m$} 
    & \colhead{$R$} & \colhead{$T_{\mathrm{b}\,m}$}
    & \colhead{$\tau_m$}
    & \colhead{~}
    & \colhead{$\tau_\mathrm{max}$}& \colhead{$\theta_\tau$} 
    & \colhead{$m$} & \colhead{$\theta_m$} 
    & \colhead{$R$} & \colhead{$T_{\mathrm{b}\,m}$}
    & \colhead{$\tau_m$}
    \\
    \colhead{} 
    & \colhead{(GHz)} 
    & \colhead{($\mu$as)} & \colhead{(day)}
    & \colhead{(day)} & \colhead{($\mu$as)} 
    & \colhead{(\%)}  & \colhead{($\mu$as)}
    & \colhead{($R_\mathrm{S}$)} & \colhead{$\!\!\!\!\!\!\!$($\log_{10}(K)$)$\!\!\!$}
    & \colhead{(day)} 
    & \colhead{}
    & \colhead{(day)}& \colhead{($\mu$as)} 
    & \colhead{(\%)}  & \colhead{($\mu$as)}
    & \colhead{($R_\mathrm{S}$)} & \colhead{$\!\!\!\!\!\!\!$($\log_{10}(K)$)$\!\!\!$}
    & \colhead{(day)} 
    \\
    \colhead{(1)} 
    & \colhead{(2)} 
    & \colhead{(3)} & \colhead{(4)}  
    & \colhead{(5)} & \colhead{(6)}
    & \colhead{(7)} & \colhead{(8)} 
    & \colhead{(9)} & \colhead{(10)}
    & \colhead{(11)}
    & \colhead{}
    & \colhead{(12)} & \colhead{(13)}
    & \colhead{(14)} & \colhead{(15)}
    & \colhead{(16)} & \colhead{(17)}
    & \colhead{(18)}
    \\
    }
\startdata
\NGC{\phn 777}& 11.2&    2.7 &  0.12&0.5    &  11   &14.9   &19     &38     & 10.6  &0.9    && 0.13  & 3     &8.0    & 33    &   65  & 10.2  &    1.5\\
\NGC{2787}&  9.8&    3.0 &  0.14&\nodata&\nodata&\nodata&\nodata&\nodata&\nodata&\nodata&&\nodata&\nodata&\nodata&\nodata&\nodata&\nodata&\nodata\\
\NGC{3031}&  9.4&    3.1 &  0.14&0.4    &9      &2.0    &101    &  130  & 11.4  &    4.5&& 6     &133    &3.3    &66     &  86   & 11.7  &    3.0\\
\NGC{3079}&  8.8&    3.4 &  0.14&\nodata&\nodata&\nodata&\nodata&\nodata&\nodata&\nodata&&\nodata&\nodata&\nodata&\nodata&\nodata&\nodata&\nodata\\
\NGC{3147}&  9.5&    3.0 &  0.14&\nodata&\nodata&\nodata&\nodata&\nodata&\nodata&\nodata&& 3     &64     &1.5    &126    &  300  & 10.1  &    5.9\\[1.5ex]
\NGC{3169}&  9.1&    3.2 &  0.14&\nodata&\nodata&\nodata&\nodata&\nodata&\nodata&\nodata&& 6.5\tablenotemark{a}   &149    &2.4    &86     & 1200  & 10.4  &    3.7\\
\NGC{3226}&  8.5&    3.6 &  0.15&\nodata&\nodata&\nodata&\nodata&\nodata&\nodata&\nodata&& 1.0   &24     &1.5    &132    &  570  & 10.0  &    5.5\\
\NGC{4168}&  7.5&    3.5 &  0.18&\nodata&\nodata&\nodata&\nodata&\nodata&\nodata&\nodata&&\nodata&\nodata&\nodata&\nodata&\nodata&\nodata&\nodata\\
\NGC{4203}&  6.4&    3.1 &  0.28&\nodata&\nodata&\nodata&\nodata&\nodata&\nodata&\nodata&&$\ge 10$&$\ge 111$&3.1 &  43   &   440 & 10.9  &    3.9\\
\NGC{4235}&  7.7&    3.5 &  0.17&\nodata&\nodata&\nodata&\nodata&\nodata&\nodata&\nodata&&$\ge 10$&$\ge 206$&2.7 &  69   &   650 & 10.3  &    3.4\\[1.5ex]
\NGC{4450}&  7.1&    3.4 &  0.21&$\ge 10$\tablenotemark{a}&$\ge 162$&4.0 & 43    &  732  & 10.7  &    2.7&& 3.5\tablenotemark{a}   & 57    &2.9    &  57   &   970 & 10.5  &    3.5\\
\NGC{4565}&  6.4&    3.1 &  0.27&1.7    &20     &3.8    & 36    &  560  & 10.6  &    3.2&&\nodata&\nodata&\nodata&\nodata&\nodata&\nodata&\nodata\\
\NGC{4579}&  7.5&    3.5 &  0.18&5.5\tablenotemark{b}    &107    &2.9    &63     &  500  & 11.1  &   3.2 && 4\tablenotemark{a}     &78     &1.1    & 144   &  1200 & 10.4  &    7.4\\
\NGC{5866}&  9.9&    3.5 &  0.12&0.2    &  6    &3.1    &83     &  620  & 10.4  &   2.9 && 0.1   &3      &2.1    & 116   &   860 & 10.1  &    4.0\\
\enddata
\tablenotetext{a}{The variability estimate for this target-month may
  be influenced by calibration errors.}
\tablenotetext{b}{Small $(u,v)$ effects may be present for this
  target-month.}
\tablecomments{Column~(1) gives the galaxy name.  Column~(2) gives the
  transition frequency between diffractive and refractive interstellar
  scattering.  Column~(3) gives the angular size of the first Fresnel
  zone, and Column~(4) gives the scintillation time for a point
  source.  Columns (2)--(4) were calculated using the NE2001 software
  package from \citet{Cordes_L2002}.  Columns (5) and~(12) give the
  estimated timescale for the first peak in the structure functions
  plotted in Figure~\ref{fig:target_struc_func}.  For structure
  functions which are continuing to rise past a timescale of 10~days,
  the column is marked $\ge 10$.  Columns (6) and~(13) give the
  predicted angular size of the source based on the variability
  timescale using Equation~10 of \citet{Walker_1998}.  Columns (7)
  and~(14) give the modulation index, which is just the de-biased RMS
  from Table~\ref{tab:var_stats}.  Columns (8) and~(15) give the
  predicted angular size of the source based on the modulation index
  using Equation~9 of \citet{Walker_1998}.  Columns (9) and~(16) give
  the equivalent linear \emph{radius} of the modulation-based source
  size, in units of the Schwarzschild radius of the galaxy's black
  hole.  Columns (10) and~(17) give the equivalent brightness
  temperature for the source angular size $\theta_m$ and the mean flux
  density in Table~\ref{tab:var_stats}.  Finally, Columns (11)
  and~(18) give the equivalent variability timescale which should have
  been observed if the modulation was caused by refractive
  interstellar scintillation.  Columns (5) to~(11) give results for
  the 2003~May observations, and Columns (12) to~(18) give results for
  2003 September.  Data values are only shown for those target
  galaxies with $P_{\chi^2}$ values less than $0.01$ in
  Table~\ref{tab:var_stats} which are not known to have $(u,v)$
  problems.  }
\end{deluxetable}

\clearpage

\begin{deluxetable}{lccc}
  \tablecaption{CSO Calibrator List\label{tab:CSO_list}}
  \tablewidth{0pt}
  \tablehead{
    \colhead{CSO Name} & \colhead{RA}  & \colhead{Dec}
    & \colhead{RMS}
    \\
    \colhead{} & \colhead{(J2000)} & \colhead{(J2000)} 
    & \colhead{(\%)}
    \\
    \colhead{(1)} & \colhead{(2)} & \colhead{(3)} 
    & \colhead{(4)}
    }
\startdata
\objectname[COINS]{J0204$+$0903} & 02 04 34.7589 & 09 03 49.248 & 1.24
\\
\objectname[COINS]{J0410$+$7656} & 04 10 45.6057 & 76 56 45.301 & \nodata
\\
\objectname[COINS]{J0427$+$4133} & 04 27 46.0455 & 41 33 01.099 & 1.18
\\
\objectname[COINS]{J0650$+$6001}\tablenotemark{a} & 06 50 31.2543 & 60
01 44.555 & 0.91
\\
\objectname[COINS]{J1035$+$5628}\tablenotemark{b} & 10 35 07.0399 & 56
28 46.792 & 0.41
\\
\objectname[COINS]{J1148$+$5924} & 11 48 50.3582 & 59 24 56.382 & 0.91
\\
\objectname[COINS]{J1244$+$4048} & 12 44 49.1872 & 40 48 06.153 & 0.47
\\
\objectname[COINS]{J1400$+$6210} & 14 00 28.6526 & 62 10 38.526 & 1.40
\\
\objectname[COINS]{J1823$+$7938} & 18 23 14.1087 & 79 38 49.002 & 0.85
\\
\enddata
\tablenotetext{a}{Phase calibrator for \NGC{2273}}
\tablenotetext{b}{Phase calibrator for \NGC{3079}}
\tablecomments{Column~(1) gives the J2000 names of the CSOs.  Columns
  (2) and~(3) give the positions of the CSOs (units of right ascension
  are hours, minutes, and seconds, and units of declination are
  degrees, arcminutes, and arcseconds).  Column~(4) gives the RMS
  residual level for each CSO, combining both May and September data.  }
\end{deluxetable}

\clearpage

\begin{deluxetable}{lrrrrr}
  \tablecaption{NGC~3031 Long-Term Variability Statistics\label{tab:M81}}
  \tablewidth{0pt}
  \tablehead{
    \colhead{Frequency} 
    & \colhead{$\sigma_\mathrm{e}$}
    & \colhead{$\sigma_\mathrm{s}$}
    & \colhead{$\sigma_\mathrm{d}$}
    & \colhead{$T_\mathrm{b,R,50}$}
    & \colhead{$T_\mathrm{b,R,90}$}
    \\
    \colhead{(GHz)} 
    & \colhead{(\%)}
    & \colhead{(\%)}
    & \colhead{(\%)}
    & \colhead{($\log_{10}(K)$)} & \colhead{($\log_{10}(K)$)} 
    \\
    \colhead{(1)} 
    & \colhead{(2)} & \colhead{(3)} 
    & \colhead{(4)}  & \colhead{(5)} & \colhead{(6)}
    }
\startdata
 1.4 &  7.3 & 13.6 & 11.4 & \nodata & \nodata \\
 4.9 &  7.3 & 12.4 & 10.0 & \nodata & \nodata \\
 8.4 &  8.1 & 24.8 & 23.4 & 11.6 & 10.8 \\
15.2 &  5.1 & 24.6 & 24.1 &  9.2 &  9.0 \\
\enddata
\tablecomments{The long term variability statistics for NGC~3031 are
  presented from the data in \citet{Ho_ea1999}.  Column~(1) gives the
  observed frequency.  Columns (2)--(4) give the mean expected scatter
  (measurement error), the RMS observed scatter, and the de-biased
  scatter, respectively, similar to Table~\ref{tab:var_stats}.
  Columns (5) and~(6) give the minimum brightness temperature of the
  variable component of the emission at a 50\% and 90\% confidence
  level, respectively, as in Table~\ref{tab:var_cat}.  The 1.4 and
  4.9~GHz variability data are both consistent with brightness
  temperatures below $10^5$~K.}
\end{deluxetable}


\end{document}